\documentclass[sigconf]{acmart}

\AtBeginDocument{%
  \providecommand\BibTeX{{%
    Bib\TeX}}}

\setcopyright{acmcopyright}
\copyrightyear{2023}
\acmYear{2023}
\acmDOI{XXXXXXX.XXXXXXX}

\acmConference[Conference acronym 'XX]{Make sure to enter the correct
  conference title from your rights confirmation emai}{June 03--05,
  2018}{Woodstock, NY}
\acmPrice{15.00}
\acmISBN{978-1-4503-XXXX-X/18/06}

\usepackage{algpseudocode}
\usepackage{algorithm}
\algnewcommand\algorithmicforeach{\textbf{for each}}
\algdef{S}[FOR]{ForEach}[1]{\algorithmicforeach\ #1\ \algorithmicdo}

\usepackage{graphicx}
\usepackage{textcomp}
\usepackage{xcolor}
\def\BibTeX{{\rm B\kern-.05em{\sc i\kern-.025em b}\kern-.08em
    T\kern-.1667em\lower.7ex\hbox{E}\kern-.125emX}}

\usepackage[utf8]{inputenc} % allow utf-8 input
\usepackage[T1]{fontenc}    % use 8-bit T1 fonts
\usepackage{hyperref}       % hyperlinks
\usepackage{url}            % simple URL typesetting
\usepackage{booktabs}       % professional-quality tables
\usepackage{amsfonts}       % blackboard math symbols
\usepackage{nicefrac}       % compact symbols for 1/2, etc.

\usepackage[utf8]{inputenc} % allow utf-8 input
\usepackage[T1]{fontenc}    % use 8-bit T1 fonts

\usepackage{subfig}
\usepackage{tabularx}
\usepackage{graphicx}

\begin{document}

\title{On the Discredibility of Membership Inference Attacks}

\author{Shahbaz Rezaei}
\email{srezaei@ucdavis.edu}
\orcid{0000-0003-1583-0114}
\affiliation{
  \institution{University of California}
  \city{Davis}
  \state{CA}
  \country{USA}
}

\author{Xin Liu}
\email{xinliu@ucdavis.edu}
\affiliation{%
  \institution{University of California}
  \city{Davis}
  \state{CA}
  \country{USA}
}

% \renewcommand{\shortauthors}{Rezaei et al.}

\begin{abstract}
With the wide-spread application of machine learning models, it has become critical to study the potential data leakage of models trained on sensitive data. Recently, various membership inference (MI) attacks are proposed to determine if a sample was part of the training set or not. The question is whether these attacks can be reliably used in practice. We show that MI models frequently misclassify neighboring nonmember samples of a member sample as members. In other words, they have a high false positive rate on the subpopulations of the exact member samples that they can identify. We then showcase a practical application of MI attacks where this issue has a real-world repercussion. Here, MI attacks are used by an external auditor (investigator) to show to a judge/jury that an auditee unlawfully used sensitive data. Due to the high false positive rate of MI attacks on member's subpopulations, auditee challenges the credibility of the auditor by revealing the performance of the MI attacks on these subpopulations. We argue that current membership inference attacks can identify memorized subpopulations, but they cannot reliably identify which exact sample in the subpopulation was used during the training.
\end{abstract}





\maketitle

\section{Introduction}
\label{sec-introduction}

The wide-spread deployment of machine learning in various applications that deal with sensitive data, such as health records and personal information, has raised concerns about the leakage of sensitive training data post-deployment. Recently, a few studies suggest that machine learning  models memorize the training data \cite{yeom2018privacy} and, consequently, various attacks, called membership inference (MI), have been proposed to identify the training samples \cite{shokri2017membership, sablayrolles2019white, jayaraman2020revisiting, salem2018ml, liu2019socinf, song2019privacy, long2017towards, truex2019demystifying, long2018understanding, carlini2022membership, rezaei2022efficient, rezaei2021accuracy, li2022user}. Due to its simplicity, membership inference attacks have become a standard way to evaluate the privacy risk of machine learning models \cite{carlini2022membership, murakonda2020ml}.

Recent studies have shown that the evaluation of such models using average-case success metrics is misleading \cite{rezaei2020towards}. Specifically, a trivial random guess adjusted using the generalization gap, called \textit{gap attack} \cite{choo2020label} or \textit{naive attack} \cite{rezaei2020towards, leino2020stolen}, has shown to achieve similar performance as many membership inference attacks. Moreover, as argued in \cite{carlini2022membership} and \cite{ long2020pragmatic}, privacy is not an average case metric and a pragmatic approach should avoid relying on such metrics. To better demonstrate the privacy risk of a model, true positive rate at low false positive rate is suggested in \cite{carlini2022membership} as used in various areas of computer security \cite{ho2017detecting, kantchelian2015better, kolter2006learning, metsis2006spam}. Using the true positive rate at low false positive rate has revealed that many de facto membership inference attacks, such as \cite{shokri2017membership, yeom2018privacy, jayaraman2020revisiting}, catastrophically fail. Only the state-of-the-art MI attacks that use some form of sample difficulty calibration \cite{watson2021importance}, such as \cite{sablayrolles2019white, carlini2022membership, watson2021importance, rezaei2022efficient}, can identify some training samples at low false positive rates.

\textbf{Contributions.} In this paper, we aim to answer the following question: \textit{Can membership inference attacks with low false positive be reliably used in practice?} We show that despite their low false positive when evaluated on the entire dataset, they have a very high false positive when evaluated on nonmember samples belonging to the exact subpopulations that identified member samples are coming from. The reason this issue has not been realized in previous experimental evaluations with common datasets, such as CIFAR10, was that identified member samples were often outliers (w.r.t to the dataset in hand) and there has not been enough samples from the same subpopulation to investigate it. Moreover, we show that the membership score of two samples are correlated with how semantically close they are. This reveals the inaccuracy of the current MI attacks for \textit{record-level membership inference}. Our findings suggest that the current MI attacks may be better suited for \textit{subpopulation-based membership inference}, where the attacker concludes that \textit{a sample} from the subpopulation is used during the training, but \textit{the exact sample is unknown}.

To manifest this problem in a real world application, we introduce a potential application of MI attacks for the purpose of external auditing. In this application scenario, MI attacks are used as an auditing tool to investigate unlawful use of sensitive data. Here, an auditor aims to show to the judge/jury that private data has been unlawfully used by the auditee under investigation. The auditor uses an MI attack, and reports the performance of the MI attack model along the samples labeled as members (at low false positive rate) to the judge. For the claimed member samples provided by the auditor, auditee can provide an unlimited number of non-member samples \textit{from the claimed member samples' subpopulation} to the judge for which the MI attack constantly fails. We call this process \textit{discredibility}. 

Discredibility allows the auditee to seriously damage the credibility of MI attack models on the claimed member samples and, hence, get the case dismissed. Consequently, we argue that current MI attacks should not be used alone as a golden standard for prosecution, like DNA matching, for record-level MI. It can, however, be used during an investigation phase to enable the collection of further evidence.
% lead and collect more evidence because it has a lower bar in comparison with the prosecution phase of criminal-justice process \cite{lander2016forensic}. 
Also, it could be used when further evidence (i.e. information about the prior) is available to safely exclude all other semantically similar samples from the membership analysis.

To show the high false positive of MI attacks on those subpopulations, we need to find them first. Unfortunately, due to the small number of samples available in a common evaluation dataset, like CIFAR10, it is unlikely to find enough sample, if any, for a meaningful evaluation. We propose three algorithms to create these subpopulations, refereed to as \textit{discrediting dataset} in the auditing example:
%, a dataset for which the false positive of MI attacks are significantly large:
1) searching through another public dataset, 2) crafting semantically similar samples to the target sample using a generative model, and 3) adversarially perturbing a non-member sample to embody the semantic representation of a member sample. We demonstrate that
the false positive rate on these samples are up to several thousand times more than the false positive evaluated on the entire dataset.

%We systematically evaluate our discrediting algorithms over multiple datasets and models. We demonstrate that our approach can increase the false positive rate up to several thousand times more than the claimed low rate for SOTA algorithms. 
%Our algorithms can even increase the false positive rates of older approaches, such as \cite{shokri2017membership, yeom2018privacy}. Nevertheless, the results are less significant because they cannot achieve low false positive rate in the first place to start with. 
%Our results show that even SOTA algorithms are all prone to the discredibility problem. 

\textbf{New Insights.} 
% We analyze the two hypotheses implicitly used as building blocks of our discrediting algorithms. 
We investigate two hypotheses that establish a positive correlation between the membership score of a member sample and its neighboring samples, and also a positive correlation between the semantic closeness of two neighbors and their membership scores. 
% Although we start with a potential application of MI attacks in the auditing scenario, these two hypotheses are valid regardless of the application scenario.
Consequently, MI attacks are prone to incorrectly classifying nonmember samples in the neighborhood of member samples as members. Furthermore, these two findings suggest that the current MI attacks are more reliable in identifying \textit{memorized subpopulations} than individual samples.

% Not only does this attests the effectiveness of our discrediting algorithms, but also it shows the unreliability of \textit{record-level} (sample-level) MI attacks in general. 

\textbf{Implications on the Application of MI Attacks.} The new insight, that current MI attacks are identifying memorized subpopulations, undermines the reliability of using MI in real applications. However, this insight implies a new potential direction for MI attacks. It suggests that current \textit{"record-level" MI attacks} are in fact better at \textit{"subpopulation-level" membership inference}. For example, in face recognition where each subpopulation likely represents a user, current MI attacks may achieve better user-level membership inference than record-level membership inference. This new direction of MI attacks needs further investigation.

% \textbf{Legal Consideration: } The criminal-justice process mainly consists of two stages \cite{lander2016forensic}: investigation and prosecution which has higher standard than investigation. We argue that although MI methodologies can be a useful tool during investigation phase, it should not be considered as a \textit{golden standard} for a prosecution in court at this stage. It can still be a useful tool to corroborate other evidences.

\textbf{Implications on the evaluation of MI Attacks.} If a false positive rate on member's subpopulation is often larger than other subpopulations, as we show in this paper, we may need to find a better way to evaluate record-level MI attacks. We argue that for each sample identified as member by an MI attack, it is more reasonable to evaluate the attack performance on the same subpopulation. Let's consider a face recognition model under investigation. If we probe the model with an image from Amazonian indigenous tribes and identify it as a member, we argue that we cannot use the false positive on white Americans as a baseline. It is crucial to study the MI attack behavior on other samples from Amazonian indigenous tribes to see if it can distinguish them. In fact, adjusting the evaluation criteria based on the target sample is done in practice. For example, in DNA analysis, random match probability is calculated using an appropriate statistical formula that takes population substructure of the case at hand into account because the frequency of genetic variants varies among ethnic groups \cite{balding1994dna}. 

To show the importance of evaluation set from another point of view, let's think about a thought experiment where we replace all/some member samples with their semantically similar neighbors. In fact, for any real world application, it is highly unlikely for an MI attacker to have access to all exact member samples during the evaluation. In this thought experiment, the false positive rate increases significantly. The practice of including all member samples during evaluation phase is refereed to as a \textbf{closed-set} experimental design. Such closed-set designs are known to underestimate the false-positive rate \cite{monson2022planning}. For example, firearm matching in forensic science used to have the closed-set experimental design. In firearm matching, examiners are given a set of samples and asked to find the gun the ammunition had been fired from. In a closed-set design, the source gun is always present. When a similar study without the closed-set assumption has been conducted, the false positive rate jumped from $0.02\%$ (closed-set) to $1.5\%$ \cite{baldwin2014study}. Most recent practices in forensic science nowadays follow an open-set experimental design to measure a more accurate false positive \cite{monson2022planning, duez2018development, chapnick2021results, mattijssen2020validity, pauw2013faid2009}. We believe that although reporting the true positive at low false positive rate has been a great progress in MI attack evaluation, more investigation is needed for more reliable evaluation.

\section{Related Work}

Membership inference aims to identify samples used during the training of a target model, referred to as a \textit{victim model}. Samples that have been used during the training are reffered to as \textit{members} or \textit{train} samples, and other samples as \textit{non-members}, \textit{non-train} or \textit{test} samples. First generation of membership inference attacks were built upon the intuition that the confidence output of a victim model exhibits different distribution between train and non-train samples \cite{rezaei2020towards}. Simply put, the victim model is more confident on train samples than on non-train samples. Hence, the first membership inference attack on deep models was proposed in \cite{shokri2017membership} using this idea. They train a membership inference attack model that takes the confidence output of a model as an input and predicts its membership status.
%The training dataset of the MI attack model is constructed by training \textit{shadow models} and taking their confidence output. Shadow models are trained on the same task as the victim model, but with different dataset. Since the training set of the shadow models is known to the attacker, the attacker can easily construct the labels for the MI attack dataset. 
Many papers use the same idea with different variations of less restrictive assumptions \cite{salem2018ml, liu2019socinf, song2019privacy, long2017towards, truex2019demystifying, long2018understanding, yeom2018privacy, rezaei2020towards, zou2020privacy, li2020label}. 

The effectiveness of the first generation of MI attack has been seriously challenged when it has been shown that they can barely outperform a trivial baseline, called \textit{gap attack} \cite{choo2020label} or \textit{naive attack} \cite{rezaei2020towards, leino2020stolen}. Gap attack labels a sample as a member if it is correctly classified by the victim model, and non-member otherwise. In \cite{rezaei2020towards}, the authors go one step further and show that the seemingly intuitive assumption that was the basis of these attacks generally do not hold. In other words, the distribution of confidence output of member and non-member samples are not significantly different, particularly when correctly classified samples are looked upon seperately which constitute the majority of samples. Furthermore, Carlini et al. \cite{carlini2022membership} argue that using average-case metric is not suitable for security-related applications and suggest using the true positive rate at a low false positive rate.

The main challenge for the first generation of membership inference attacks was distinguishing between hard member samples (for which the confidence is low) from easy non-member samples (for which the confidence is high). As suggested in \cite{watson2021importance}, MI attack should have an adaptable reference point to which it compares the confidence of the target sample, called sample calibration. Most SOTA MI attacks that can perform well in a low false positive rate solve this issue by calibrating the confidence so that it takes the difficulty of the target sample into account \cite{sablayrolles2019white, watson2021importance, carlini2022membership, rezaei2022efficient, ye2021enhanced}. In the \textbf{Watson attack}, the attacker excludes the target sample from the training set, and then train multiple shadow models. As a result, the attacker can obtain the average confidence output of a model in the absence of the target sample in the training data as baseline. 
%By comparing the confidence output of the victim model with the average confidence output of the shadow models, the attacker can predict the membership status of the sample. 
In \cite{sablayrolles2019white, carlini2022membership}, a variation of this idea was used with one main difference. These attacks include two set of shadow models: one where the training set excludes the target sample and one where the training set includes the target sample. In \cite{rezaei2022efficient}, a slightly different and more efficient calibration has been proposed where it does not require training shadow models. In this attack, the attacker uses a BiGAN architecture to craft samples from the same subpopulation as the target sample. Then, the attacks compares the confidence output of the target sample versus the subpopulation. 
% If the target sample has higher confidence than its subpopulation, it is an indication of a member sample. This attack achieves similar accuracy as other calibration-based attacks, while decreasing the training computation cost of shadow models significantly.

% Majority of membership inference attacks in literature use confidence output of the victim model as the main feature. There are, however, a few novel approaches that utilize other features \cite{rezaei2020towards, choo2020label, rahimian2020sampling}. Assuming a white-box access to the victim model, Rezaei et al. \cite{rezaei2020towards} launch a set of attack using distance to the decision boundary, gradient w.r.t model weight, and gradient w.r.t input.  However, none of the features significantly outperform the confidence output. A similar MI attack approach based on distance to the decision boundary has also appeared in \cite{choo2020label} in the black box setting. Another MI attack approach is to compare the prediction of the target sample with the prediction label of its transformed versions \cite{choo2020label} or randomly perturbed versions \cite{rahimian2020sampling, jayaraman2020revisiting}. The intuition is that deep models are more robust to training samples. Hence, the transformed/perturbed sample is less likely to be mislabeled by the victim model. Despite their moderate accuracy and black-box nature of some of these novel MI attacks, they perform poorly in terms of achieving low false positive rates.

There have been a few novel approaches that utilize other features than confidence output \cite{rezaei2020towards, choo2020label, rahimian2020sampling, jayaraman2020revisiting}. However, they perform poorly at low false positive regime. In summary, for practical reasons, we mainly focus on the SOTA MI attacks that perform well on low false positive rates \cite{watson2021importance, carlini2022membership, rezaei2022efficient}. We also report the performance of Shokri \cite{shokri2017membership} and Yeom \cite{yeom2018privacy} attacks because traditionally they have been a default baseline for comparison.

\section{Threat Model}
To better manifest the potential application of membership inference in practice, we showcase a scenario in a trial, where MI is used as an auditing tool to demonstrate the unlawful use of private data. We note that our findings applies to all other applications. Our threat model consists of three actors: an external \textbf{auditor}, or attacker in the MI literature, an \textbf{auditee}, or MI defender whose model is under MI attack, and a \textbf{judge} (or juries), who examines if the auditor's claim is credible enough. Unlike previous defense papers in literature where the goal is to reduce MI effectiveness, either by confidence masking or more private training, we focus on a case where the auditee (defender) can discredit the auditor's (attacker) claim post-attack. This threat model is fundamentally different from the literature and makes known MI attacks ineffective even against already trained or public models.

\subsection{Auditor (MI Attacker or Investigator)}
\textbf{Objectives:} The goal of the auditor is to use membership inference attacks on the auditee's model to find potential training samples that are private. To do this reliably, we assume that MI attacks are set to perform in the low false positive regime. The auditor then reports the potential members to the judge. We call these samples \textit{claimed member list}. As a proof of low false positive rate, the auditor needs to privately disclose its own training/validation data to the judge such that it can be confirmed. This data is not available to the auditee or any other actor.

\textbf{Assumptions:} The auditor in our threat model has the highest advantage it could have. It has a white-box access to the auditee's model with unlimited query. It has the capability to train multiple models if needed. It has access to a dataset coming from the same distribution as the auditee's dataset. It has access to a set of data points some of which have been potentially used as auditee's training data. To identify the member samples, auditor uses MI attacks.
%However, the auditor has no way to prove the membership of this data beyond using membership inference attacks. In other words, if the MI attack fails or is getting discredited, the auditor's claim about this portion of the data is as credible as any other data and is dismissed by the judge.

\subsection{Judge}
\textbf{Objectives:} The goal of the judge is to examine if auditor's claims are reasonable, i.e. high true positive at low false positive on the auditor dataset. If so, the judge will give the auditee a chance to challenge the auditor's claim. Here, if the auditee can successfully discredit the auditor's method (i.e., the MI attack), the judge will dismiss the case.

\subsection{Auditee (Defender)}
\textbf{Objectives:} The goal of the auditee is to discredit the MI method used by the auditor. To do so, the auditee aims to find a procedure by which it can craft/find unlimited number of non-member samples which the auditor's MI method likely mislabel as members. We call these samples \textit{discrediting samples} and the corresponding dataset \textit{discrediting dataset}.
In other words, the auditee tries to discredit the auditor by showing that his/her low false positive claim was in fact fallacious, and, thereby, every statement using this MI method is unreliable. Note that the non-membership status of discrediting samples should be agreed by all actors beyond reasonable doubt. Otherwise it cannot be used to discredit the auditor's MI attack. To fulfil this criterion, the samples can come from the sources becoming available only after the model is trained, can be randomly generated on-the-fly in the court, or can be crafted by adding small perturbation to samples that have already been labeled by the auditor as non-member.

\textbf{Assumptions:} The auditee has no information about the MI method deployed by the auditor, the auditor's dataset, or his/her capabilities. In other words, from the perspective of the auditee, the auditor's MI model is a black-box with no online query access to. The only information the auditor has is the claimed member list that the auditor claims to be a part of auditee's training set, which is then given to the auditee by the judge. These are the samples with highest membership score according to the MI attack used by the auditor.

\subsection{Discredibility Pipeline}

Given that the auditee's model is trained and publicly available, the trial's pipeline is as follows: 
\begin{enumerate}
    \item Using an MI attack, the auditor provides the claimed member list, a list of samples with highest membership score, to the judge stating that they are unlawfully used during training. To demonstrate the reliability of the MI attack, the auditor privately disclose the attack information and the training/validation dataset to prove the low false positive rate.
    \item The judge examines the claim. If the low false positive rate satisfies the low false positive threshold required, the judge gives the claimed member list to the auditee and asks if he/she challenges the claim.
    % \item The auditee uses a procedure to find/generate a large number of nonmember samples (discrediting samples), using methods in Section \ref{sec-disc-methods}, that are likely to be mislabeled by the auditor's MI model as members. The auditee, then, gives these discrediting samples to the judge and asks the judge to evaluate the performance of the MI method on.
    \item The auditee uses a procedure to find/generate a large number of nonmember samples (discrediting samples), using methods in Section \ref{sec-disc-methods}, that are coming from the same subpopulation as of claimed member list. The auditee, then, gives these discrediting samples to the judge and asks the judge to evaluate the performance of the MI method on.
    \item If the false positive rate of the auditor's MI attack on discrediting samples are significantly larger than what is claimed earlier by the auditor, the judge dismisses the case and consider the auditor's MI attack unreliable.

\end{enumerate}

\subsection{Discredibility Criteria}
We note that auditee must provide a discrediting dataset with certain distribution to be able to discredit the MI attack. Certainly, there are numerous unnatural random inputs that can trigger false positive which obviously cannot be used for discredibility purpose. Here, the discrediting dataset should either follow the same distribution as of auditee's training data (which is often not known by all parties) or the same subpopulation as of the claimed member list. In our discrediting algorithms, we mainly focus on the latter because it is known to all parties. Moreover, as we discussed earlier in Section \ref{sec-introduction}, it is more aligned with practical practices in forensic investigation. For instance, if the auditor identify an image from Amazonian indigenous tribes as member, it is more reasonable to credit/discredit the MI model by evaluate the performance on a set of another images from Amazonian indigenous tribes rather than white Americans.

\section{Discredibility Mechanisms}
\label{sec-disc-methods}

\subsection{Problem Statement}
As stated earlier, the goal of the auditee is to find a set of non-member samples from the subpopulation of the claimed member list. Let $Y(\cdot)$ and $E(\cdot)$ be the auditee's model under investigation, and the encoder part of the auditee's model, respectively. Similar to \cite{rezaei2022efficient}, encoder here refers to the output of the last fully connected layer before the softmax, also known as the latent representation. Let's denote the last layer operation of the auditee model by $l(\cdot)$. In other words, $Y(x)=l(E(x))$. We denote the auditor's MI attack model by $M(\cdot)$. Moreover, let $D_c$, $D_p$, and $D_d$ be the claimed member list provided by the auditor, public dataset agreed by all parties to only contain non-members, and the discrediting dataset, respectively. 

Formally speaking, the goal is to find a set of samples in $D_p$, labeled as $D_d$, such  that for each $x \in D_p$, there is a another sample $x' \in D_c$ for which $E(x) \approx E(x')$. Interestingly, this process is independent of the MI attack model, $M(\cdot)$. Hence, auditee can discredit the auditor without any knowledge about the attack.
% the goal is to find a mapping from $D_c$ to a subset in $D_p$ to be mislabeled as member by $M$ with high probability. 
% The challenge to find such a mapping is that the MI attack is a complete black-box and it is not even allowed to be queried. Hence, the only information the auditee has about the MI attack is the samples identified as members ($D_c$) with high membership score. 
%Later in Section \ref{sec-analysis}, we show that $D_c$ is not necessary for the discrediting purpose. In this case, auditee can safely use the mapping with the actual training data (members), which is known to the model trainer, instead of $D_c$. 

% In this section, we will show three algorithms to generate/find nonmember samples on which the MI attack catastrophically fails. Interestingly, these samples happen to come from the same subpopulation as the $D_c$ which is aligned with the evaluation practice criteria used in forensic science.

\subsection{Mapping and Intuition}
\label{sec-intuition}

To simply put, the mapping consists of finding/generating samples that has similar latent representation as the samples in $D_c$. Auditee uses the encoder, $E(\cdot)$, to find the latent representation to which he/she has white-box access. For a member sample $x$ marked by auditor with a high membership score, auditee's discredibility algorithm aims to find a non-member samples $x'$, where $E(x) \approx E(x')$. The intuition as of why this causes the current MI attacks to misclassify can be analyzed by considering neural networks as deterministic functions with certain properties.
%in two different perspectives:

As a deterministic function, a single layer ReLU network has shown to be locally linear. In fact, the entire multi-layer ReLU network is a piece-wise linear function \cite{pascanu2013number}. Because $l(\cdot)$ is a single-layer ReLU function, if $E(x) \approx E(x')$, then $l(E(x)) \approx l(E(x'))$ or $Y(x) \approx Y(x')$. Therefore, any MI attack that only takes the output of $Y(\cdot)$ as a feature fails to distinguish between $x$ and $x'$. This is particularly an issue for older generation of MI attacks, such as Shokri \cite{yeom2018privacy} and Yeom \cite{yeom2018privacy}.

The contemporary MI attacks often use the output of a set of extra models on a target sample, such as \cite{watson2021importance, sablayrolles2019white, carlini2022membership}. There are two challenges when it comes to applying the same argument here. First, the extra models the MI attackers use may be different when probing $x$ versus $x'$. For example, in \cite{carlini2022membership}, half of the extra models include the target sample in the training set and the other half excludes the target sample from the training set. As a result, when probing $x$ and $x'$, the extra models are not necessarily the same. However, we argue that since both samples belong to the same subpopulation, including or excluding either of them results in a similar behaviour from the final model's perspective on that subpopulation. 

The second challenge is that even if we assume the extra models are the same when probing two different samples, the encoder part of them are not the same as the encoder of the auditee's model, $E(\cdot)$. Let's denote the encoder part of an extra model by $E_e(\cdot)$. It has been empirically shown in \cite{rezaei2022efficient} that the $dist(E(x), E(x')) \approx dist(E_e(x), E_e(x'))$ despite $E(x)$ not necessarily being close to $E_e(x)$. This suggests that membership inference score of $x$ and $x'$ is likely to be similar even with respect to the new generation of MI attacks. We show the correlation between the closeness of two samples and their membership scores in Section \ref{sec-analysis} to provide an empirical evidence.

\begin{algorithm} 
\caption{Finding Discrediting Samples by Search}
\begin{algorithmic}[1]
\Require Encoder $E(\cdot)$; Claimed member list $D_c$; Non-member public dataset $D_p$; the number of neighbors to pull from $D_p$ per sample in $D_c$, denoted by $n_n$; a sample and the associated label (x, y); the number of samples from $D_c$ with highest membership score to find neighbors for, denoted by $n_c$
\State Initialize discrediting dataset $D_d = \{\}$
\State $D_c \gets Sort\_by\_membership\_score(D_c)$
\State $D_c \gets D_c[-n_c:]$
\ForEach {$(x,y) \in \mathcal D_c $}
    \ForEach {$(x',y') \in \mathcal D_p $}
        \If{$y=y'$}
            \State $d[x,x'] \gets dist(E(x), E(x'))$
        \EndIf
    \EndFor
    \State $d_s \gets argsort(d[x,:])$
    \State $D_d \gets D_d \cup d_s[:n_n]$
\EndFor
\State \Return $D_d$
\end{algorithmic}
\label{algo-search}
\end{algorithm}

\subsection{Discredibility Methods}

As discussed in Section \ref{sec-intuition}, discredibility is performed by sampling from $D_p$ provided that the samples belong to the same subpopulation as samples in $D_c$. Here, we use the latent representation of the auditee's model, $E(\cdot)$, to find subpopulations. Formally, we consider two samples, $x$ and $x'$, from the same subpopulation if $dist(E(x), E(x')) \leq \epsilon$. For the distance measure, the two prominent choices are Cosine distance (Cosine loss) and $L2$ norm (MSE loss). As shown in \cite{rezaei2022efficient}, there is not much difference between these to metrics when it comes to measuring sample similarities. Hence, we mainly use Cosine loss in this paper.

\begin{algorithm} 
\caption{Finding Discrediting Samples by Sample Generation}
\begin{algorithmic}[1]
\Require Auditee's model $Y(\cdot)$; Encoder $E(\cdot)$; Claimed member list $D_c$; Generator of the Rezaei's BiGAN architecture $G(\cdot)$; the number of samples to pull from $D_p$ per sample in $D_c$, denoted by $n_n$; a Gaussian random noise generator $\mathcal{N}(\mu,\,\sigma^{2})$; the number of samples from $D_c$ with highest membership score to craft samples from, denoted by $n_c$

\State Initialize discrediting dataset $D_d = \{\}$
\State $D_c \gets Sort\_by\_membership\_score(D_c)$
\State $D_c \gets D_c[-n_c:]$
\ForEach {$(x,y) \in \mathcal D_c $}
    \For {i = $0$ to $n_n$}
        \State $\epsilon \sim \mathcal{N}(\mu,\,\sigma^{2})$
        \State $x' \gets G(E(x) + \epsilon)$
        \If{$y=Y(x')$}
            \State $D_d \gets D_d \cup \{x'\}$
        \EndIf
    \EndFor
\EndFor
\State \Return $D_d$
\end{algorithmic}
\label{algo-bigan}
\end{algorithm}

\begin{algorithm} 
\caption{Finding Discrediting Samples by Adversarial Perturbation}
\begin{algorithmic}[1]
\Require Encoder $E(\cdot)$; Claimed member list $D_c$; the number of adversarial samples per sample in $D_c$, denoted by $n_n$; a targeted adversarial attack $adv\_attack(x, x', F(\cdot), dist(\cdot))$ that perturbs $x'$ such that $dist(F(x), F(x')) \thickapprox 0$; the number of steps to run the adversarial attack, denoted by $n_{adv}$; the number of samples from $D_c$ with highest membership score to find adversarially perturbed neighbors for, denoted by $n_c$; a function returning a non-member neighbor sample with the same class label as the input $S(\cdot)$ 

\State Initialize discrediting dataset $D_d = \{\}$
\State $D_c \gets Sort\_by\_membership\_score(D_c)$
\State $D_c \gets D_c[-n_c:]$
\ForEach {$(x,y) \in \mathcal D_c $}
    \For {i = $0$ to $n_n$}
        % \State $x' \gets$ pick a random non-member sample s.t. $y=y'$
        \State $(x', y) \gets S(x, y)$
        \For {j = $0$ to $n_{adv}$}
            \State $x' \gets adv\_attack(x, x', E(\cdot), MSE(\cdot))$
        \EndFor
        \State $D_d \gets D_d \cup \{x'\}$
    \EndFor
\EndFor
\State \Return $D_d$
\end{algorithmic}
\label{algo-adv}
\end{algorithm}

In this paper, we propose three methods to find/generate samples from the same subpopulation:

\textbf{1. Using a Large Public Dataset:} If a large dataset, disjoint from the train set, is available to sample from, auditee can use it to create discrediting dataset, $D_d$. The procedure is straightforward, as shown in Algorithm \ref{algo-search}. Note that there is no guarantee that a sample with subpopulation constraint, i.e. $dist(E(x), E(x')) \leq \epsilon$, exists in $D_p$. For simplicity, we discard this condition and we add the closest $n_n$ samples to discrediting dataset although they may not necessarily be from the same subpopulation. The only criterion is that their class labels should match (line 6). Otherwise, they obviously do not belong to the same subpopulation. The empirical results in Section \ref{sec-exp-natural} shows that the discrediting samples are good enough for the purpose of discrediting the auditor. Hence, the challenge of defining $\epsilon$ is not crucial for the discrediting purpose and, hence, it is ignored in this paper.

\textbf{2. Using Generative Model:} We can use generative models to craft new samples. However, unconditional sample generation is an extremely inefficient exercise as it may take millions of queries for the model to generate a sample from the same subpopulation. In this paper, we use the BiGAN architecture proposed in \cite{rezaei2022efficient} to craft new samples. The generator in their architecture take the latent representation as an input and generate a sample accordingly. As shown in Algorithm \ref{algo-bigan}, we add a small random noise to the latent representation of a target sample and use it to generate a new sample from the BiGAN.

\textbf{3. Using Adversarial Perturbation:} In this method, we take a non-member sample that belongs to the same class as the target sample does, and we add a small adversarial perturbation such that the latent representation of the two samples approaches the same value. The algorithm is shown in Algorithm \ref{algo-adv}. Here, $x$ and $x'$ should belong to the same class, otherwise the auditor can easily tell the adversarial nature of it because it will be misclassified by the model. Although we can start the adversarial perturbation on any non-member sample ($x'$), we use a function ($S(\cdot)$) to find the closest neighbor with the same class label to increase the chance of reaching the same latent representation.

\section{Experimental Results}

\subsection{Evaluation Metrics}
As suggested in \cite{carlini2022membership}, it is more practical to use membership inference attack at the low false positive regime. Hence, in this paper, we mainly focus on true positive at a low false positive rate. For the sake of completeness, we also report the AUC of all MI attacks.

The second evaluation metric that we use in this paper is \textit{false positive to false positive} plot or ratio. This measures the false positive of an MI attack on an auditor's dataset in comparison with the discrediting dataset that auditee provides. Here, we disregard the true positive rate because positive samples include all training member samples on both cases. Thus, this set is assumed to be fixed in both auditor dataset and the discrediting dataset. 
% The goal of the auditee is to come up with a samples generation/look-up scheme that outputs negative (non-member) samples that are labeled as positive with a much larger false positive rate than acceptable. As a result, 
Hence, we only measure the false positive difference between these two datasets. In other words, the true positive is the same regardless of the evaluating dataset.

\begin{table}
%\scriptsize
  \caption{Accuracy of the auditee's models}
  \label{tbl-bal-acc-auditee}
  \centering
  \begin{tabular}{llll}
    \toprule
    \midrule
    Dataset & Model & Train accuracy & Test accuracy \\
    MNIST & MLP & 100\% & 97.71\% \\
    FMNIST & MLP & 100\% & 88.62\% \\
    SVHN & LeNet & 99.99\% & 87.72\% \\
    Cifar10 & LeNet & 97.13\% & 58.22\% \\
    Cifar10 & ResNet20 & 98.48\% & 74.58\% \\
    Cifar100 & LeNet & 98.27\% & 22.61\% \\
    Cifar100 & ResNet20 & 100.00\% & 33.30\% \\
    \bottomrule
  \end{tabular}
\end{table}

\begin{table}
%\scriptsize
  \caption{Comparison of AUC of prior membership inference attacks. S, Y, W, C, and R stands for Shokri, Yeom, Watson, Carlini, and Rezaei attacks, respectively.}
  \label{tbl-auc}
  \centering
  \begin{tabular}{llllll}
    Dataset/Model & S \cite{shokri2017membership} & Y \cite{yeom2018privacy} & W \cite{watson2021importance} & C \cite{carlini2022membership} & R \cite{rezaei2022efficient} \\
    \midrule
    MNIST/MLP & 52.43\% & 50.95\% & 53.53\% & 56.17\% & 51.11\%  \\
    FMNIST/MLP & 59.62\% & 56.38\% & 57.54\% & 58.55\% & 54.87\% \\
    SVHN/LeNet & 57.60\% & 57.88\% & 60.24\% & 69.94\% & 58.64\%\\
    C-10/LeNet & 72.62\% & 78.15\% & 73.77\% & 79.55\% & 76.12\%\\
    C-10/ResNet20 & 74.52\% & 70.75\% & 65.06\% & 72.19\% & 68.74\%\\
    C-100/LeNet & 82.03\% & 91.96\% & 90.19\% & 94.30\% & 93.83\% \\
    C-100/ResNet20 & 91.17\% & 92.81\% & 81.27\% & 93.39\% & 91.98\%\\
    \bottomrule
  \end{tabular}
\end{table}

\begin{table*}
%\scriptsize
  \caption{Comparison of prior membership inference attacks at low false positive rate. S, Y, W, C, and R stands for Shokri, Yeom, Watson, Carlini, and Rezaei attacks, respectively. Since the exact false positive is not always possible to achieve, we choose the lowest false positive between the stated ranges of $(0.01\%, 0.03\%)$ and $(1.0\%, 3.0\%)$. }
  \label{tbl-tp-at-lfp}
  \centering
  \begin{tabular}{llllllllllll}
    Dataset & Model & \multicolumn{5}{c}{TPR @ (0.01\%, 0.03\%) FPR} & \multicolumn{5}{c}{TPR @ (1.0\%, 3.0\%) FPR} \\
    \cmidrule(r){3-7}
    \cmidrule(r){8-12}
    % \toprule
    -  & - & S \cite{shokri2017membership} & Y \cite{yeom2018privacy} & W \cite{watson2021importance} & C \cite{carlini2022membership} & R \cite{rezaei2022efficient} & S \cite{shokri2017membership} & Y \cite{yeom2018privacy} & W \cite{watson2021importance} & C \cite{carlini2022membership} & R \cite{rezaei2022efficient} \\
    \midrule

    MNIST & MLP & 0.00\% & 0.00\% & 0.10\% & 0.00\% & 0.01\% & 0.00\% & 0.00\% & 2.40\%& 2.47\% & 0.47\% \\
    FMNIST & MLP & 0.00\% & 0.00\% & 0.39\% & 3.26\% & 0.08\% & 2.67\% & 0.00\% & 3.25\%& 5.39\% & 1.06\% \\
    SVHN & LeNet & 0.00\% & 0.00\% & 0.63\% & 0.00\% & 0.11\% & 0.00\% & 0.00\% & 5.20\%& 6.52\% & 2.61\% \\
    Cifar10 & LeNet & 0.00\% & 0.00\% & 0.52\% & 0.00\% & 0.28\% & 2.57\% & 0.00\% & 7.71\%& 10.37\% & 4.76\% \\
    Cifar10 & ResNet20 & 0.00\% & 0.00\% & 0.54\% & 0.57\% & 0.06\% & 3.55\% & 0.00\% & 5.80\%& 10.97\% & 5.10\% \\
    Cifar100 & LeNet & 0.06\% & 0.00\% & 1.68\% & 0.01\% & 0.17\% & 3.44\% & 0.00\% & 18.66\%& 18.71\% & 19.73\% \\
    Cifar100 & ResNet20 & 0.00\% & 0.00\% & 1.76\% & 16.26\% & 1.46\% & 4.64\% & 4.56\% & 14.20\%& 37.12\% & 26.71\% \\
    \bottomrule
  \end{tabular}
\end{table*}

\subsection{Experimental Setup}

We conduct experiments on a number of image classification benchmarks traditionally used for membership inference attack evaluation, including MNIST \cite{lecun1998gradient}, FMNIST \cite{fmnist}, SVHN \cite{netzer2011reading}, and CIFAR-10/CIFAR-100 \cite{krizhevsky2009learning}. For Algorithm \ref{algo-search} to work, we need a large public dataset to search from. For CIFAR-10, we use the CINIC dataset \cite{darlow2018cinic} as a public dataset, and for SVHN, we use the \textit{extra} portion of the dataset as a public dataset. For other datasets, we could not find a large public dataset to search through, and, hence, we only perform the second and third algorithms on them.

We divide the training set of these datasets into two parts: auditor and auditee training dataset. Similar to \cite{rezaei2022efficient}, we choose multi-layer perceptron (MLP) with 4 hidden layers for MNIST and FMNIST classification. For SVHN, we choose LeNet. For CIFAR-10 and CIFAR-100 we use both LeNet and ResNet20. We use SGD with a learning rate of $0.1$ to train all models. We decrease the learning rate by a factor of 10 at epoch 50 and 75. The performance of the auditee models is shown in Table \ref{tbl-bal-acc-auditee}.

We evaluate our discredibility methods on five MI attacks, namely Shokri \cite{shokri2017membership}, Yeom \cite{yeom2018privacy}, Watson \cite{watson2021importance}, Carlini \cite{carlini2022membership}, and Rezaei \cite{rezaei2022efficient}. Unless specified, we follow the same hyper-parameters to train MI attack models as suggested in their original papers. For the Shokri attack, we train 50 shadow models for all datasets. For the Watson and Rezaei attacks, we use the loss function as the base membership score before calibration. We use the same BiGAN architecture proposed in \cite{rezaei2022efficient} for both Rezaei's attack and Algorithm \ref{algo-bigan}.

Table \ref{tbl-auc} shows the AUC of the MI attacks. In Table \ref{tbl-tp-at-lfp}, the true positive rates of MI attacks at $0.01\%$ and $1.0\%$ false positive is presented. As also shown in \cite{carlini2022membership}, Shokri \cite{shokri2017membership} and Yeom \cite{yeom2018privacy} attacks does not work well in the low false positive regime. We omit other membership inference attacks in this study, such as Jayaraman \cite{jayaraman2020revisiting} and Song \cite{song2021systematic} due to the poor performance at low false positive rate \cite{carlini2022membership}.

% \begin{table*}
% %\scriptsize
%   \caption{Comparison of prior membership inference attacks at low false positive rate}
%   \label{tbl-tp-at-lfp}
%   \centering
%   \begin{tabular}{llllllllllll}
%     Dataset & Model & \multicolumn{5}{c}{TPR @ (0.001\%, 0.005\%) FPR} & \multicolumn{5}{c}{TPR @ (0.1\%, 0.5\%) FPR} \\
%     \cmidrule(r){3-7}
%     \cmidrule(r){8-12}
%     % \toprule
%     -  & - & S \cite{shokri2017membership} & Y \cite{yeom2018privacy} & W \cite{watson2021importance} & C \cite{carlini2022membership} & R \cite{rezaei2022efficient} & S \cite{shokri2017membership} & Y \cite{yeom2018privacy} & W \cite{watson2021importance} & C \cite{carlini2022membership} & R \cite{rezaei2022efficient} \\
%     \midrule
%     MNIST & MLP & 0.00\% & 0.00\% & 0.00\% & 0.00\% & 0.00\% & 0.00\% & 0.00\% & 0.49\%& 1.59\% & 0.03\% \\
%     FMNIST & MLP & 0.00\% & 0.00\% & 0.00\% & 0.00\% & 0.00\% & 0.00\% & 0.00\% & 0.81\%& 0.00\% & 0.29\% \\
%     SVHN & LeNet & 0.00\% & 0.00\% & 0.23\% & 0.00\% & 0.01\% & 0.00\% & 0.00\% & 1.76\%& 1.95\% & 0.56\% \\
%     Cifar10 & LeNet & 0.00\% & 0.00\% & 0.28\% & 0.00\% & 0.17\% & 0.00\% & 0.00\% & 2.02\%& 5.01\% & 1.04\% \\
%     Cifar10 & ResNet20 & 0.00\% & 0.00\% & 0.13\% & 0.00\% & 0.01\% & 0.32\% & 0.00\% & 1.59\%& 1.98\% & 1.01\% \\
%     Cifar100 & LeNet & 0.00\% & 0.00\% & 0.00\% & 0.00\% & 0.04\% & 0.49\% & 0.00\% & 3.78\%& 0.00\% & 2.68\% \\
%     Cifar100 & ResNet20 & 0.00\% & 0.00\% & 0.00\% & 0.00\% & 0.36\% & 2.76\% & 3.79\% & 3.94\%& 0.00\% & 8.50\% \\
%     \bottomrule
%   \end{tabular}
% \end{table*}

\begin{figure}
\centering
\includegraphics[width = 0.6\linewidth]{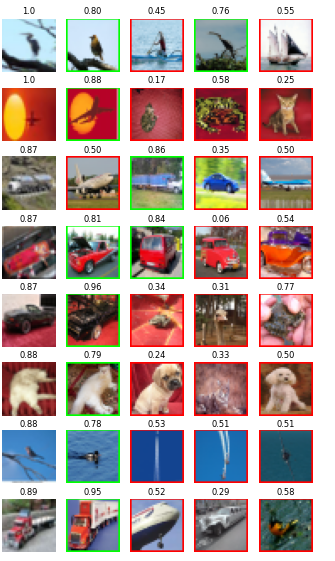}
\caption{The first columns shows member samples from CIFAR-10 dataset. The next four columns show the closest samples from the CINIC dataset to the sample in the first column. The value on top of each image shows the normalized Watson attack membership score. The neighboring samples that have the same label as the original sample is indicated by the green boarder. The boarder is red otherwise. Membership score of non-member neighboring samples of the member samples that belong to the same class often have high membership score.}
\label{fig-natural-samples}
\end{figure}

\begin{figure*}
\centering
\centering
\begin{tabular}{ccc}
\subfloat[FPR/FPR plot]{\includegraphics[width=0.35\linewidth]{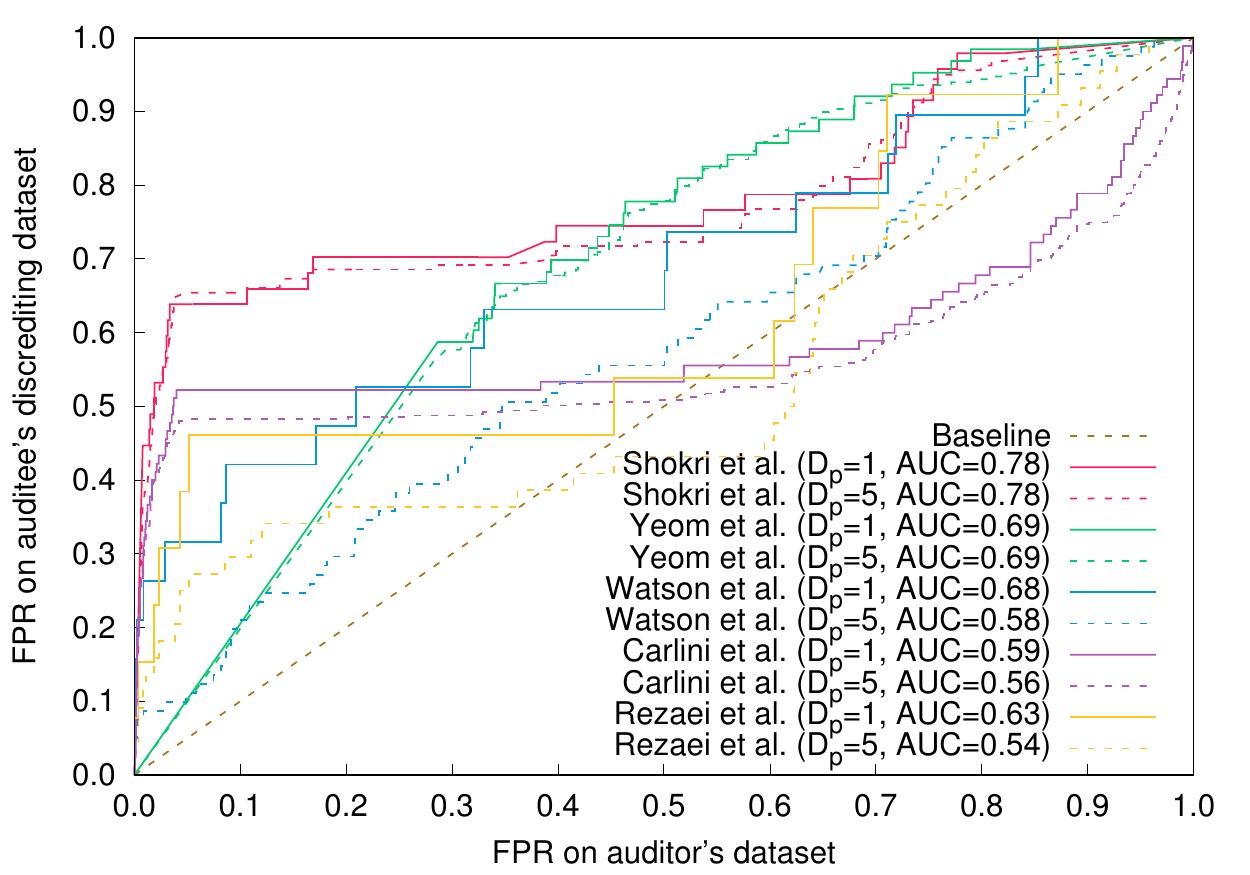}} 
& \subfloat[FPR/FPR logscale plot]{\includegraphics[width=0.35\linewidth]{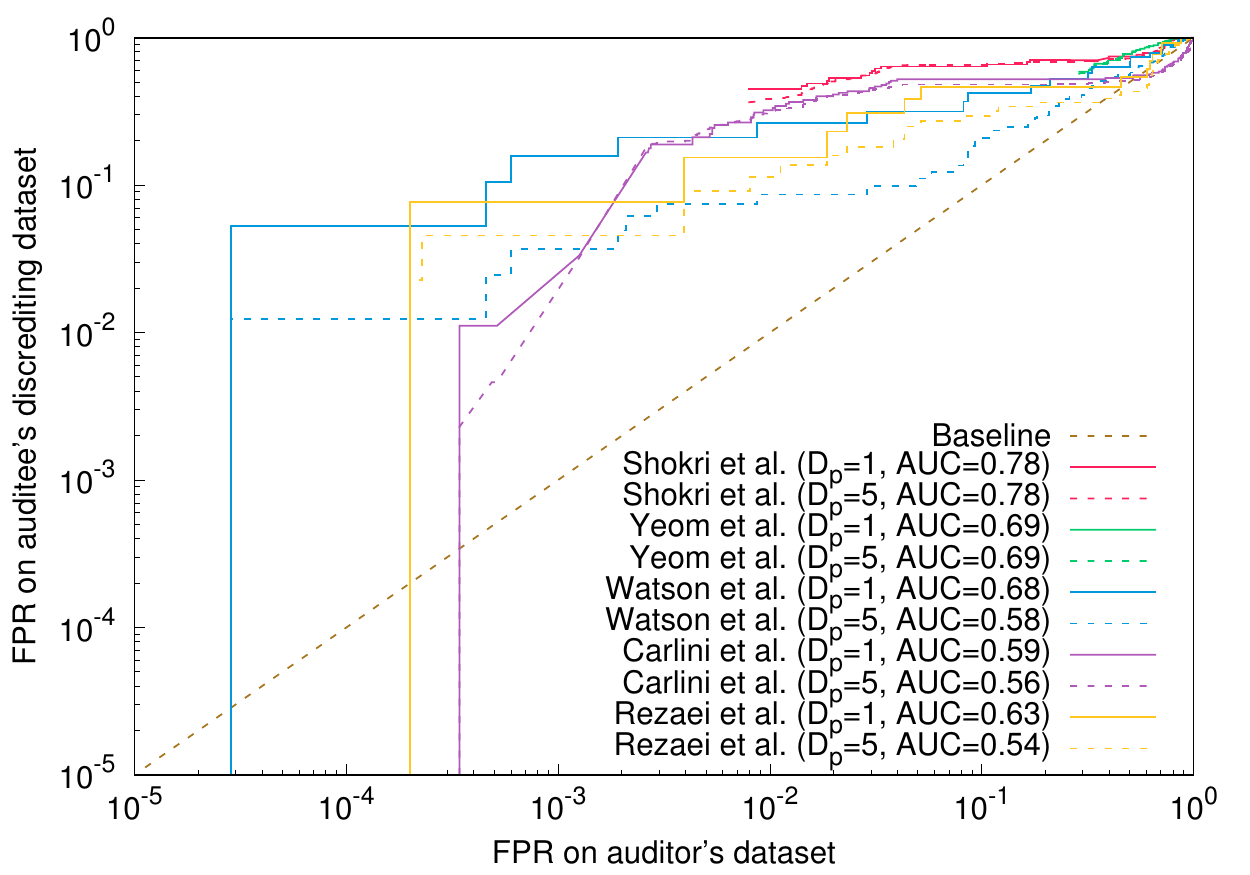}}\\
\end{tabular}
\caption{CIFAR-10/LeNet model. Discrediting algorithm \ref{algo-search} using CINIC dataset.}
\label{fig-fpr-fpr-c10-lenet-natural}
\end{figure*}

\begin{table*}
%\scriptsize
  \caption{Lowest false positive value on the auditor dataset. The numbers in parenthesis show the ratio of the false positive on discrediting dataset over the false positive on auditor dataset when Algorithm \ref{algo-search} is used for discrediting.}
  \label{tbl-natural}
  \centering
  \begin{tabular}{lllllll}
    Dataset & Model & Shokri \cite{shokri2017membership} & Yeom \cite{yeom2018privacy} & Watson \cite{watson2021importance} & Carlini \cite{carlini2022membership} & Rezeai \cite{rezaei2022efficient} \\
    \midrule
    SVHN & LeNet & 3.449\% ($\times$25.6 {\color{blue}$\uparrow$}) & 67.730\% ($\times$1.3 {\color{blue}$\uparrow$}) & 0.142\% ($\times$88.0 {\color{blue}$\uparrow$}) & 1.283\% ($\times$3.1 {\color{blue}$\uparrow$}) & 0.013\% ($\times$326.4 {\color{blue}$\uparrow$}) \\
    CIFAR-10 & LeNet & 0.791\% ($\times$56.5 {\color{blue}$\uparrow$}) & 28.631\% ($\times$2.1 {\color{blue}$\uparrow$}) & 0.003\% ($\times$1842.1 {\color{blue}$\uparrow$}) & 0.034\% ($\times$32.4 {\color{blue}$\uparrow$}) & 0.020\% ($\times$384.6 {\color{blue}$\uparrow$}) \\
    CIFAR-10 & ResNet20 & 0.049\% ($\times$363.3 {\color{blue}$\uparrow$}) & 17.469\% ($\times$4.3 {\color{blue}$\uparrow$}) & 0.029\% ($\times$116.7 {\color{blue}$\uparrow$}) & 0.011\% ($\times$102.9 {\color{blue}$\uparrow$}) & 0.009\% ($\times$364.6 {\color{blue}$\uparrow$}) \\
    
    \bottomrule
  \end{tabular}
\end{table*}

\subsection{Natural Subpopulation}
\label{sec-exp-natural}
Our first method to produce discrediting samples rely on searching samples in a large public dataset. The details of the algorithm is shown in Algorithm \ref{algo-search}. The only datasets for which we can find a large public dataset with similar classes are CIFAR-10 and SVHN. Figure \ref{fig-natural-samples} shows a few examples of member samples and their closest neighbors. In contrast with Algorithm \ref{algo-search}, in Figure \ref{fig-natural-samples} we show all neighboring samples, even the ones with different class labels for comprehension. It is worth mentioning that not all neighboring samples belong to the same class and, interestingly, the membership score of the neighboring samples with different class label are often significantly lower and should be discarded. 
%It is clear from this figure that the neighboring samples are not close pixel-wise to their corresponding member sample. Hence, a credible membership inference attack should differentiate the membership status of them.

The false positive to false positive plot of MI attacks for a LeNet model trained on CIFAR-10 is shown in Figure \ref{fig-fpr-fpr-c10-lenet-natural}. Here the x-axis shows the false positive on auditor's dataset for a given threshold and the y-axis shows the false positive on the discrediting dataset. Any region over the baseline indicates that the auditee successfully presents a dataset with a larger false positive. For a practical membership inference attack, the false positive should be small. Hence, we mostly focus on the log plot (Figure \ref{fig-fpr-fpr-c10-lenet-natural} (b)) where we can better study the behavior on low false positive rates. It is clear that the false positive is over hundreds to thousands times larger on discrediting dataset for Watson, Carlini, and Rezaei attacks. In addition, the lowest false positive for Shokri and Yoem attacks are too large for any practical usage. Nevertheless, our discredibility method still increases the false positive even further.

\begin{figure*}[h]
\centering
\centering
\begin{tabular}{ccc}
\subfloat[FPR/FPR plot]{\includegraphics[width=0.35\linewidth]{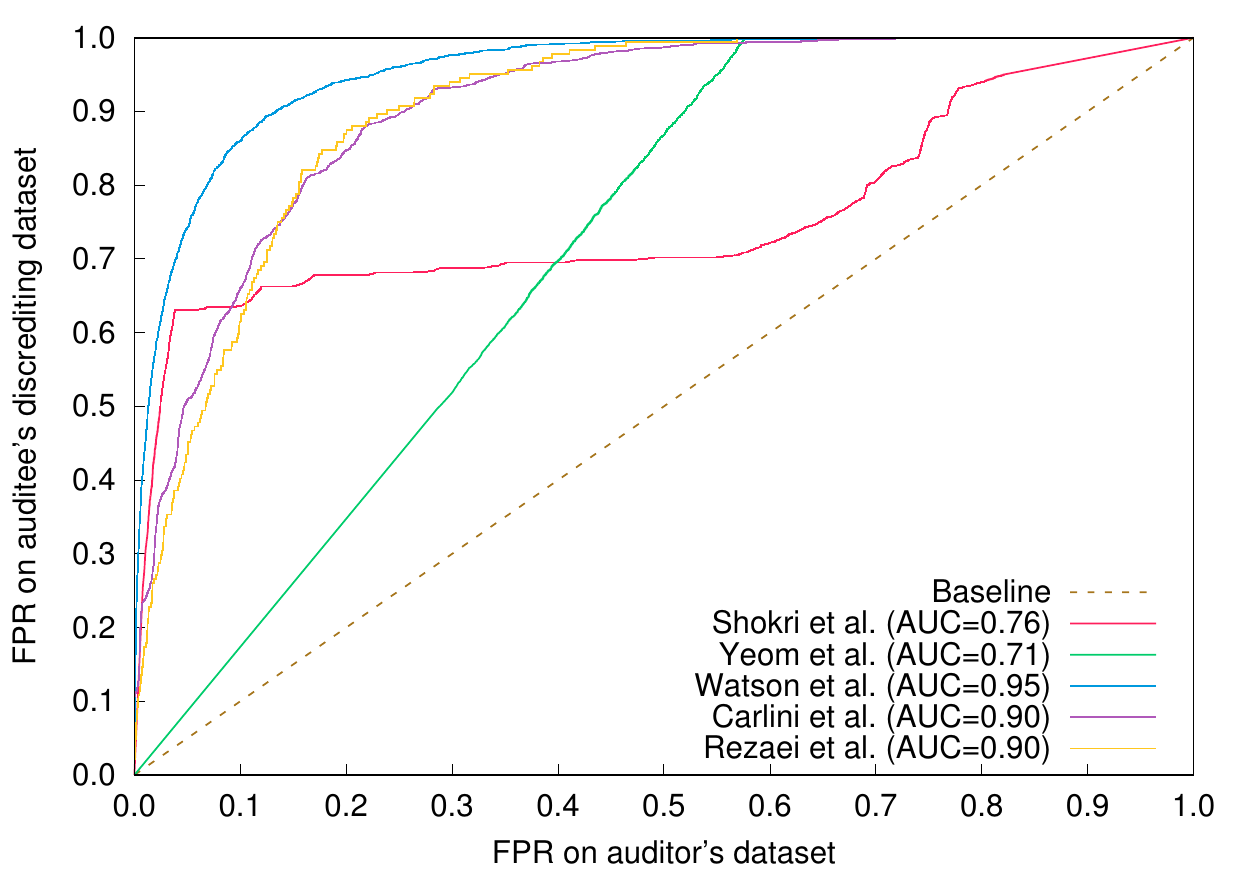}} 
& \subfloat[FPR/FPR logscale plot]{\includegraphics[width=0.35\linewidth]{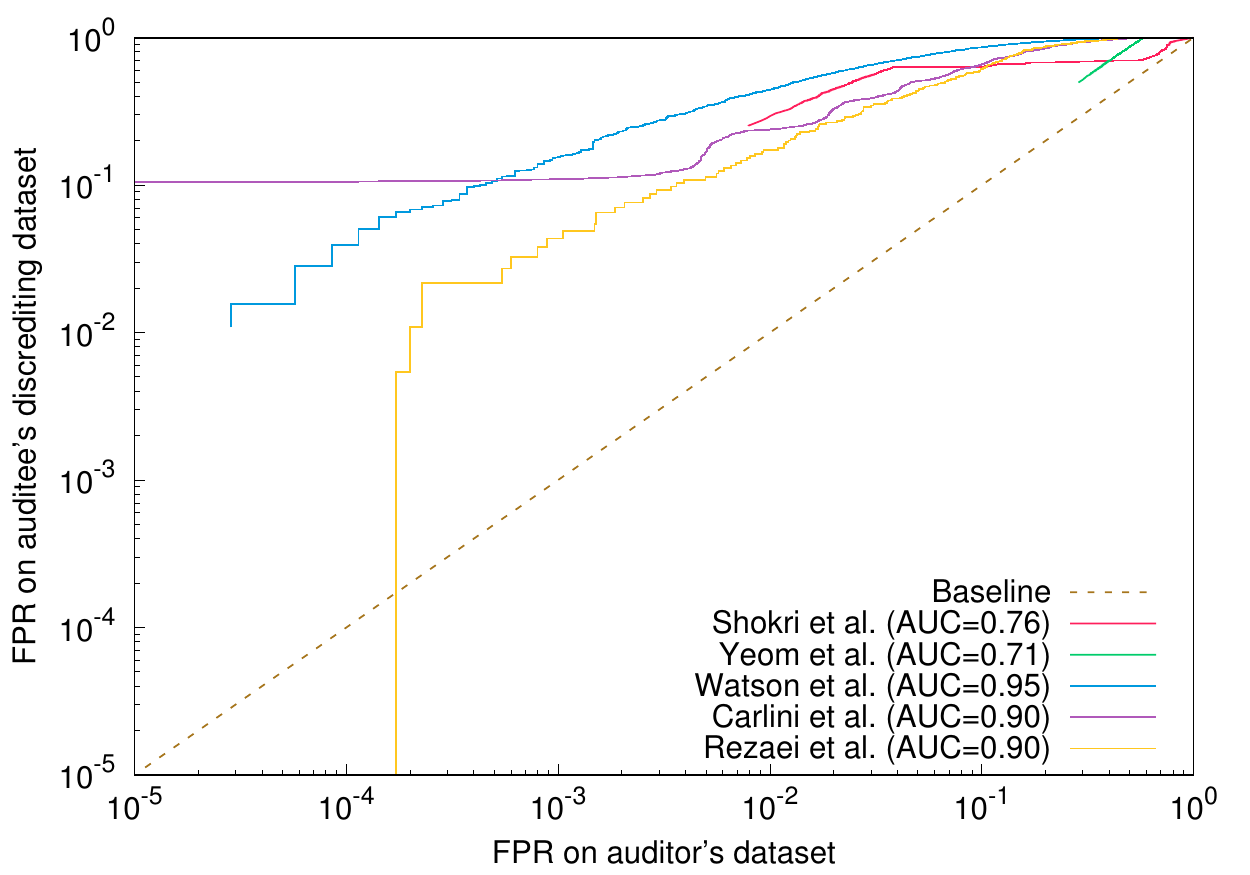}}\\
\end{tabular}
\caption{Cifar10/LeNet model. Discrediting algorithm \ref{algo-bigan}.}
\label{fig-fpr-fpr-c10-lenet-bigan}
\end{figure*}

\begin{table*}
%\scriptsize
  \caption{Lowest false positive value on the auditor dataset. The numbers in parenthesis show the ratio of the false positive on discrediting dataset over the false positive on auditor dataset when Algorithm \ref{algo-bigan} is used for discrediting.}
  \label{tbl-crafted}
  \centering
  \begin{tabular}{lllllll}
    Dataset & Model & Shokri \cite{shokri2017membership} & Yeom \cite{yeom2018privacy} & Watson \cite{watson2021importance} & Carlini \cite{carlini2022membership} & Rezeai \cite{rezaei2022efficient} \\
    \midrule
    MNIST & MLP & 4.750\% ($\times$14.3 {\color{blue}$\uparrow$}) & 77.400\% ($\times$1.0 {\color{blue}$\uparrow$}) & 0.010\% ($\times$97.9 {\color{blue}$\uparrow$}) & 2.740\% ($\times$2.9 {\color{blue}$\uparrow$}) & 0.003\% ($\times$37.5 {\color{blue}$\uparrow$}) \\
    FMNIST & MLP & 2.350\% ($\times$32.2 {\color{blue}$\uparrow$}) & 65.910\% ($\times$1.3 {\color{blue}$\uparrow$}) & 0.010\% ($\times$366.4 {\color{blue}$\uparrow$}) & 0.780\% ($\times$15.3 {\color{blue}$\uparrow$}) & 4.755\% ($\times$0.0 {\color{blue}$\uparrow$}) \\
    SVHN & LeNet & 3.449\% ($\times$17.5 {\color{blue}$\uparrow$}) & 67.730\% ($\times$1.2 {\color{blue}$\uparrow$}) & 0.002\% ($\times$5952.8 {\color{blue}$\uparrow$}) & 0.798\% ($\times$1.0 {\color{blue}$\uparrow$}) & 0.005\% ($\times$151.4 {\color{blue}$\uparrow$}) \\
    Cifar10 & LeNet & 0.791\% ($\times$31.9 {\color{blue}$\uparrow$}) & 28.631\% ($\times$1.7 {\color{blue}$\uparrow$}) & 0.003\% ($\times$379.1 {\color{blue}$\uparrow$}) & 0.003\% ($\times$3668.3 {\color{blue}$\uparrow$}) & 0.017\% ($\times$31.7 {\color{blue}$\uparrow$}) \\
    Cifar10 & ResNet20 & 0.049\% ($\times$17.5 {\color{blue}$\uparrow$}) & 17.469\% ($\times$1.2 {\color{blue}$\uparrow$}) & 0.003\% ($\times$101.7 {\color{blue}$\uparrow$}) & 0.063\% ($\times$1.5 {\color{blue}$\uparrow$}) & 0.002\% (-) \\
    Cifar100 & LeNet & 0.020\% ($\times$8.5 {\color{blue}$\uparrow$}) & 6.110\% ($\times$1.1 {\color{blue}$\uparrow$}) & 0.020\% ($\times$17.5 {\color{blue}$\uparrow$}) & 0.010\% ($\times$1.0 {\color{blue}$\uparrow$}) & 0.002\% (-) \\
    Cifar100 & ResNet20 & 0.630\% ($\times$6.7 {\color{blue}$\uparrow$}) & 0.890\% ($\times$2.9 {\color{blue}$\uparrow$}) & 0.010\% ($\times$45.0 {\color{blue}$\uparrow$}) & 0.020\% ($\times$2.5 {\color{blue}$\uparrow$}) & 0.002\% (-) \\
    \bottomrule
  \end{tabular}
\end{table*}

In Figure \ref{fig-fpr-fpr-c10-lenet-natural}, $D_p$ represents the number of neighboring samples from the same class we used to construct the discrediting dataset. As expected, $D_p=1$ slightly outperforms $D_p=5$ case potentially because the further away the samples is from the target sample, the less likely it is labeled the same way as the target sample, with respect to membership inference. In Section \ref{sec-analysis}, we analyze the correlation between distance and membership score in more depth. Due to the lack of space, we present the false positive to false positive plots of other dataset/models in Appendix \ref{appendix-natural}.

Table \ref{tbl-natural} shows the lowest possible false positive a membership inference can achieve on the auditor's dataset and the ratio of the false positive on discrediting dataset over the false positive on auditor dataset. 
%This significant increase clearly shows that the discrediting Algorithm \ref{algo-search} can be used when a large non-member dataset is available to search samples from.

% \clearpage
\subsection{Crafted Subpopulation}
In this section, we evaluate the effectiveness of Algorithm \ref{algo-bigan} to craft discrediting samples. 
%This method assumes that a generator model is available that takes a latent representation and crafts samples corresponding to the representation. We use the same BiGAN architecture proposed in \cite{rezaei2022efficient} to train such a generator. Since the generator is trained by the auditee, we directly use the auditee's model under investigation in the BiGAN architecture. All training hyper-parameters and details are adopted from the \cite{rezaei2022efficient}.
The false positive to false positive plot for a LeNet model trained on the CIFAR-10 is shown in Figure \ref{fig-fpr-fpr-c10-lenet-bigan}. In comparison with using real samples by Algorithm \ref{algo-search}, the effectiveness of this method varies accross different attacks/models/datasets. Nevertheless, it still increases the false positive rate more than 10 times for most MI attacks. Interestingly, Rezaei attack \cite{rezaei2022efficient} seems to be more immune to discrediting based on the BiGAN approach. The reason lays on how this attack works. Rezaei attack uses the same BiGAN architecture to craft similar samples to the target sample. Then, it uses the difference between the target sample's loss and the loss of average samples from the same subpopulation as the membership score. We find that the average loss difference between two crafted samples are often smaller than a natural sample and a crafted sample. Consequently, the membership scores of auditor samples (which are natural) are on average larger than the discrediting samples (which are crafted). In other words, the Rezaei attack \cite{rezaei2022efficient} is immune to this discrediting method because it can distinguish between crafted and natural samples, and not because it can identify member samples versus non-member samples. This occurs mainly because the BiGAN architecture is not good enough to generate indistinguishable natural samples. In Section \ref{sec-analysis}, we analyze this method in more depth. Nevertheless, the auditee can still use the other two methods to safely discredit the auditor if he/she uses Rezaei's MI attack.

Table \ref{tbl-crafted} shows the full results of all membership inference attacks for the lowest false positive. Clearly, the BiGAN approach of crafting discrediting samples does not work as effective on harder classification tasks, such as CIFAR-10 and CIFAR-100. This probably stems from the difficulty in training a high quality BiGAN to craft natural samples for these datasets. Further research is needed to see if this problem can be solved by using stronger generator trained on larger datasets.

\begin{figure*}[h]
\centering
\centering
\begin{tabular}{ccc}
\subfloat[Original images]{\includegraphics[width=0.28\linewidth]{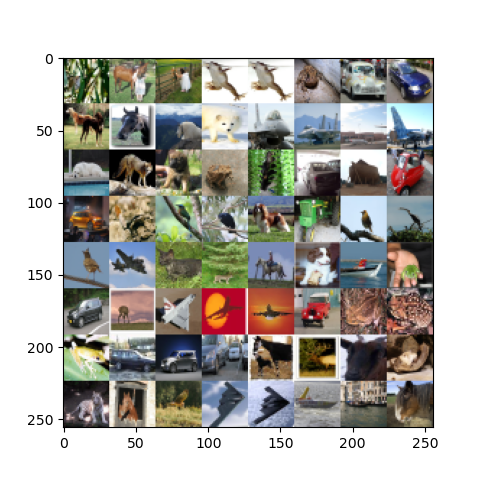}} 
& \subfloat[Adversarially tuned images ($\epsilon=0.01$)]{\includegraphics[width=0.28\linewidth]{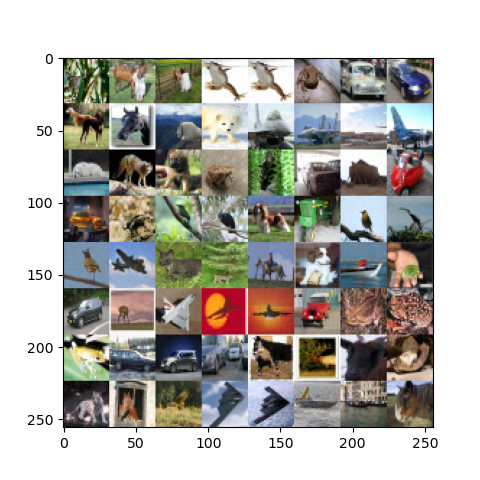}}
& \subfloat[Adversarially tuned images ($\epsilon=0.05$)]{\includegraphics[width=0.28\linewidth]{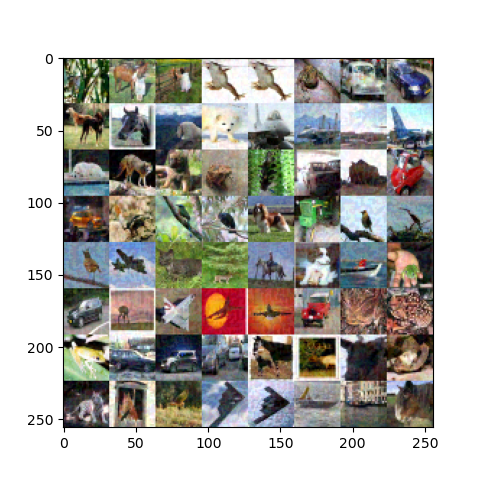}}\\
\end{tabular}
\caption{Natural samples versus the corresponding adversarially perturbed versions.}
\label{fig-adv-examples}
\end{figure*}

\begin{figure*}[h]
\centering
\centering
\begin{tabular}{ccc}
\subfloat[FPR/FPR plot]{\includegraphics[width=0.35\linewidth]{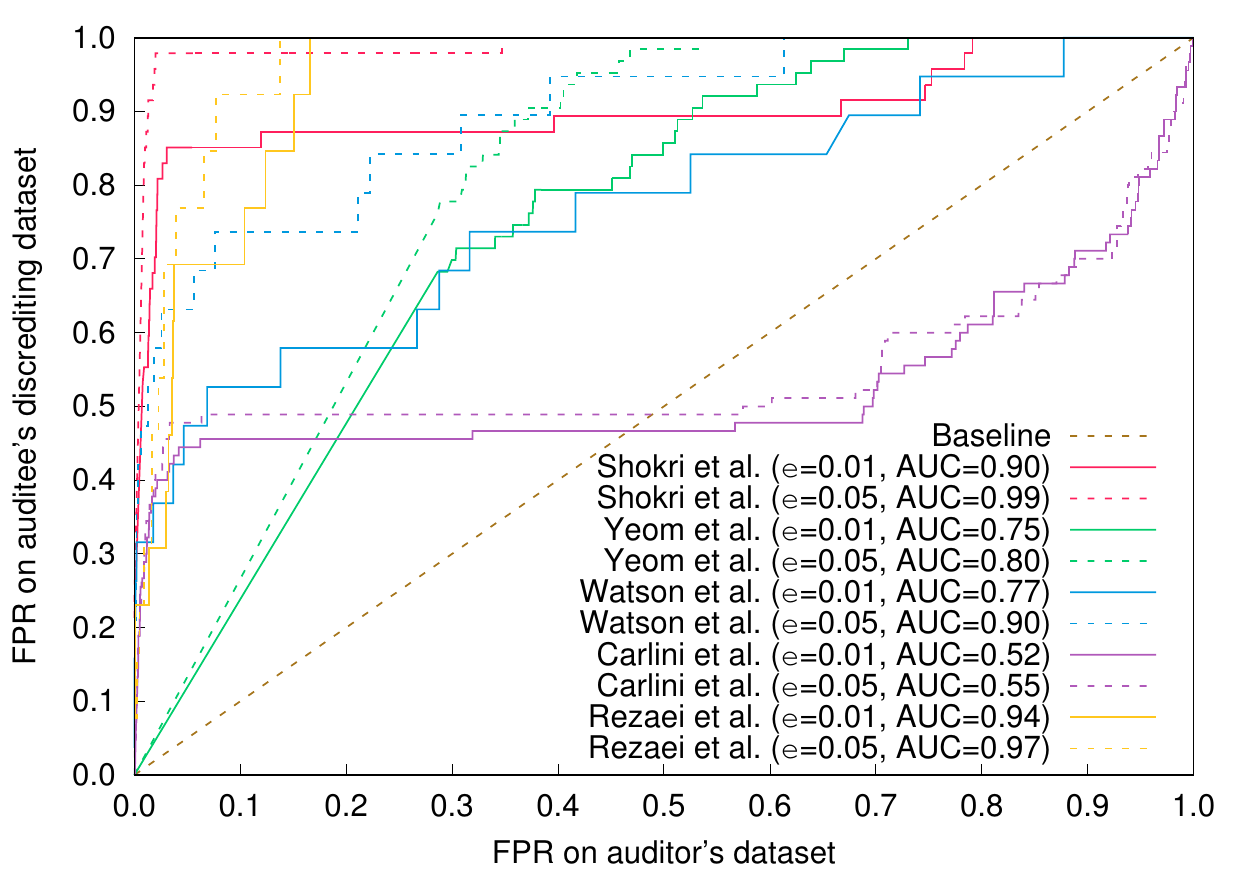}} 
& \subfloat[FPR/FPR logscale plot]{\includegraphics[width=0.35\linewidth]{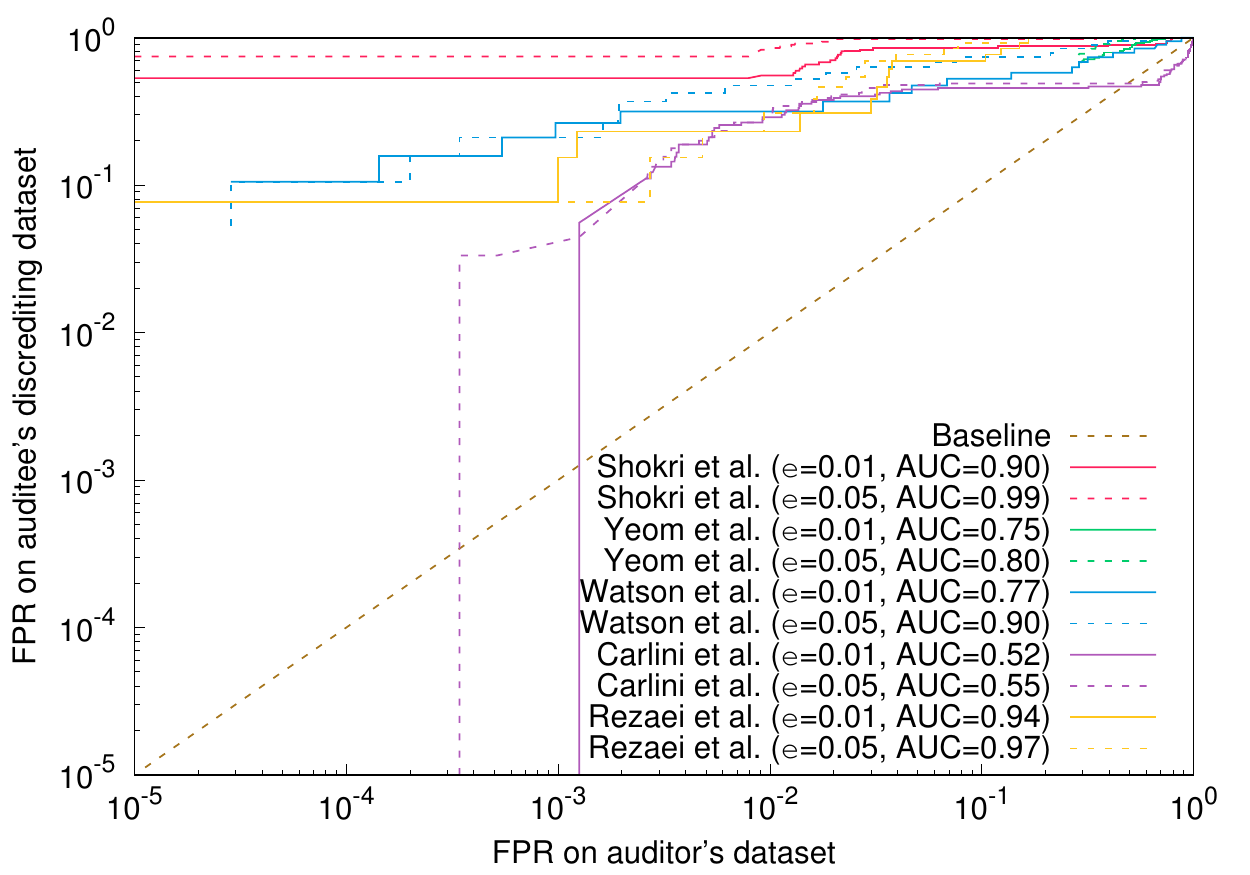}}\\
\end{tabular}
\caption{Cifar10/LeNet model. Discrediting algorithm \ref{algo-adv}.}
\label{fig-fpr-fpr-c10-lenet-adv}
\end{figure*}

\begin{table*}[h]
%\scriptsize
  \caption{Lowest false positive value on the auditor dataset. The numbers in parenthesis show the ratio of the false positive on discrediting dataset over the false positive on auditor dataset when Algorithm \ref{algo-adv} ($\epsilon=0.05$) is used for discrediting.}
  \label{tbl-adv}
  \centering
  \begin{tabular}{lllllll}
    Dataset & Model & Shokri \cite{shokri2017membership} & Yeom \cite{yeom2018privacy} & Watson \cite{watson2021importance} & Carlini \cite{carlini2022membership} & Rezeai \cite{rezaei2022efficient} \\
    \midrule
    MNIST & MLP & 4.750\% ($\times$21.1 {\color{blue}$\uparrow$}) & 77.400\% ($\times$1.3 {\color{blue}$\uparrow$}) & 0.010\% ($\times$138.9 {\color{blue}$\uparrow$}) & 0.050\% ($\times$20.0 {\color{blue}$\uparrow$}) & 0.062\% ($\times$18.4 {\color{blue}$\uparrow$}) \\
    FMNIST & MLP & 2.350\% ($\times$41.2 {\color{blue}$\uparrow$}) & 65.910\% ($\times$1.5 {\color{blue}$\uparrow$}) & 0.020\% ($\times$113.6 {\color{blue}$\uparrow$}) & 0.060\% ($\times$17.7 {\color{blue}$\uparrow$}) & 0.003\% ($\times$615.4 {\color{blue}$\uparrow$}) \\
    SVHN & LeNet & 3.449\% ($\times$27.7 {\color{blue}$\uparrow$}) & 67.730\% ($\times$1.4 {\color{blue}$\uparrow$}) & 0.252\% ($\times$49.6 {\color{blue}$\uparrow$}) & 0.238\% ($\times$8.4 {\color{blue}$\uparrow$}) & 0.003\% ($\times$1305.4 {\color{blue}$\uparrow$}) \\
    Cifar10 & LeNet & 0.791\% ($\times$94.1 {\color{blue}$\uparrow$}) & 28.631\% ($\times$2.7 {\color{blue}$\uparrow$}) & 0.003\% ($\times$1842.1 {\color{blue}$\uparrow$}) & 0.034\% ($\times$97.2 {\color{blue}$\uparrow$}) & 0.271\% ($\times$28.3 {\color{blue}$\uparrow$}) \\
    Cifar10 & ResNet20 & 0.049\% ($\times$302.8 {\color{blue}$\uparrow$}) & 17.469\% ($\times$5.3 {\color{blue}$\uparrow$}) & 0.003\% ($\times$1166.7 {\color{blue}$\uparrow$}) & 0.046\% ($\times$25.7 {\color{blue}$\uparrow$}) & 0.011\% ($\times$273.4 {\color{blue}$\uparrow$}) \\
    Cifar100 & LeNet & 0.030\% ($\times$1333.3 {\color{blue}$\uparrow$}) & 6.110\% ($\times$13.2 {\color{blue}$\uparrow$}) & 0.020\% ($\times$1363.6 {\color{blue}$\uparrow$}) & 0.430\% ($\times$11.6 {\color{blue}$\uparrow$}) & 0.009\% ($\times$972.2 {\color{blue}$\uparrow$}) \\
    Cifar100 & ResNet20 & 0.630\% ($\times$79.4 {\color{blue}$\uparrow$}) & 0.890\% ($\times$39.2 {\color{blue}$\uparrow$}) & 0.010\% ($\times$2222.2 {\color{blue}$\uparrow$}) & 0.030\% ($\times$370.4 {\color{blue}$\uparrow$}) & 0.003\% ($\times$5833.3 {\color{blue}$\uparrow$}) \\
    
    \bottomrule
  \end{tabular}
\end{table*}

% \clearpage
\subsection{Adversarially Tuned Subpopulation}
In this section, we evaluate Algorithm \ref{algo-adv} effectiveness in producing discrediting samples. This method requires an adversarial attack algorithm to perturb the input such that its latent representation of the sample converges to the latent representation of the target sample. We use projected gradient descent\footnote{We use the public implementation by Cleverhans lab at https://github.com/cleverhans-lab/cleverhans} (PGD) algorithm with step size 0.001 for 100 iterations. We try $\epsilon=0.01$ and $\epsilon=0.05$ to assess different perturbation budget. Figure \ref{fig-adv-examples} shows several natural samples from CIFAR-10 and the corresponding adversarially perturbed versions. Samples with perturbation of $\epsilon=0.01$ are imperceptible to human eyes from the natural samples. Perturbation of $\epsilon=0.05$, however, leaves visible footprint on otherwise natural samples.

\begin{figure*}[h]
\centering
\centering
\begin{tabular}{cccc}
\subfloat[Random samples from CINIC]{\includegraphics[width=0.23\linewidth]{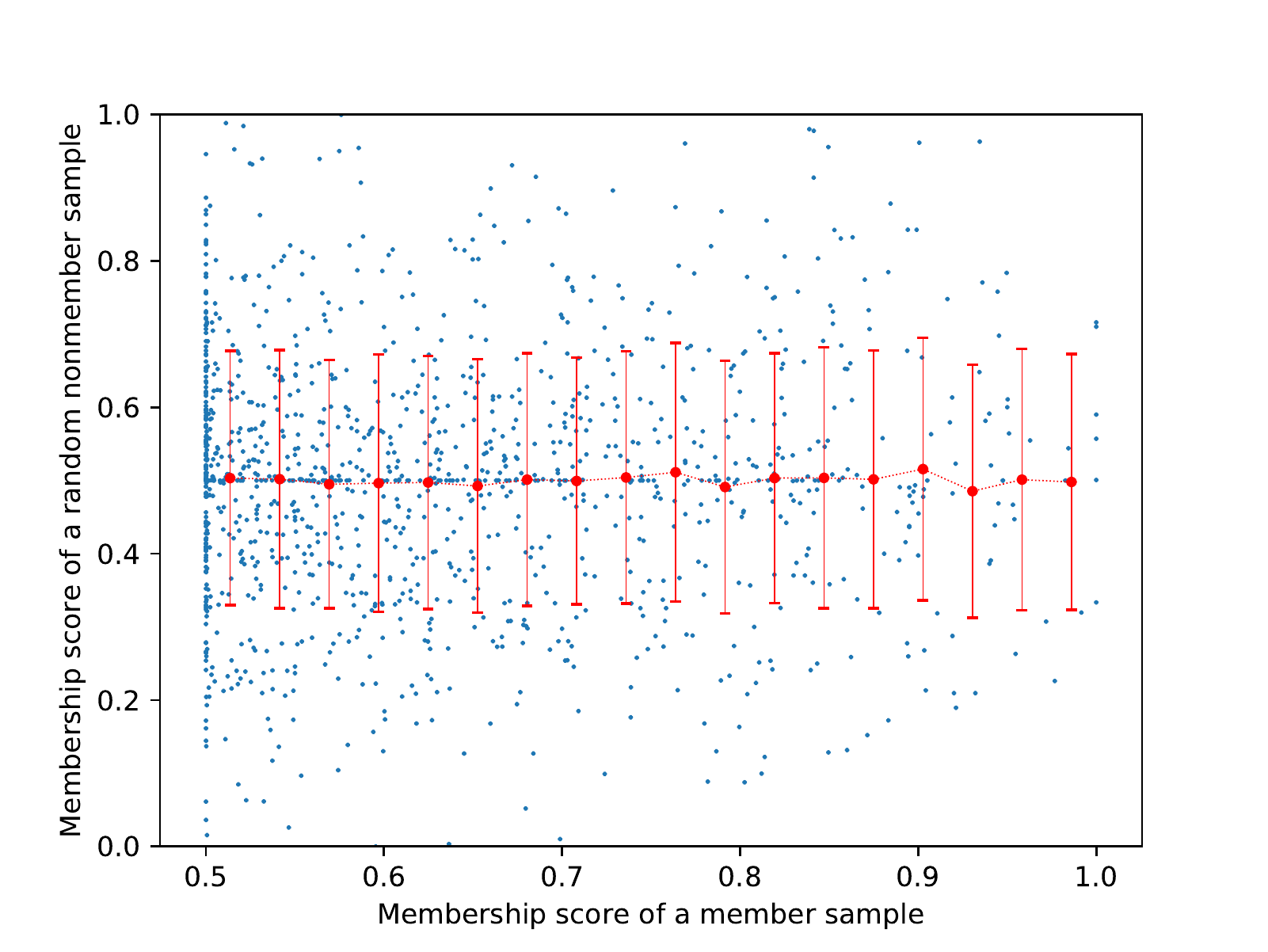}} 
& \subfloat[Closest sample from CINIC (Algorithm \ref{algo-search})]{\includegraphics[width=0.23\linewidth]{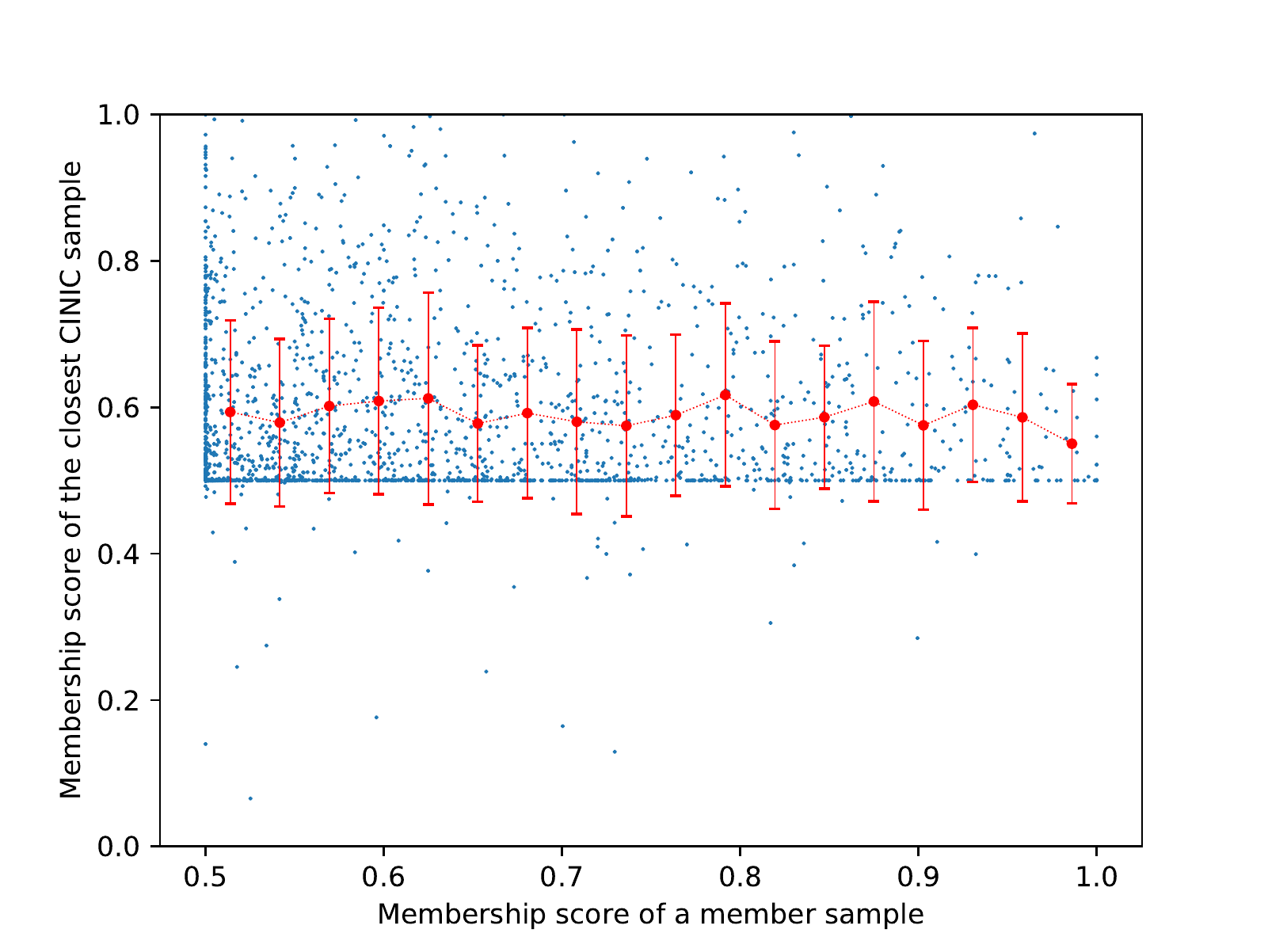}}
& \subfloat[Crafted samples using Algorithm \ref{algo-bigan}]{\includegraphics[width=0.23\linewidth]{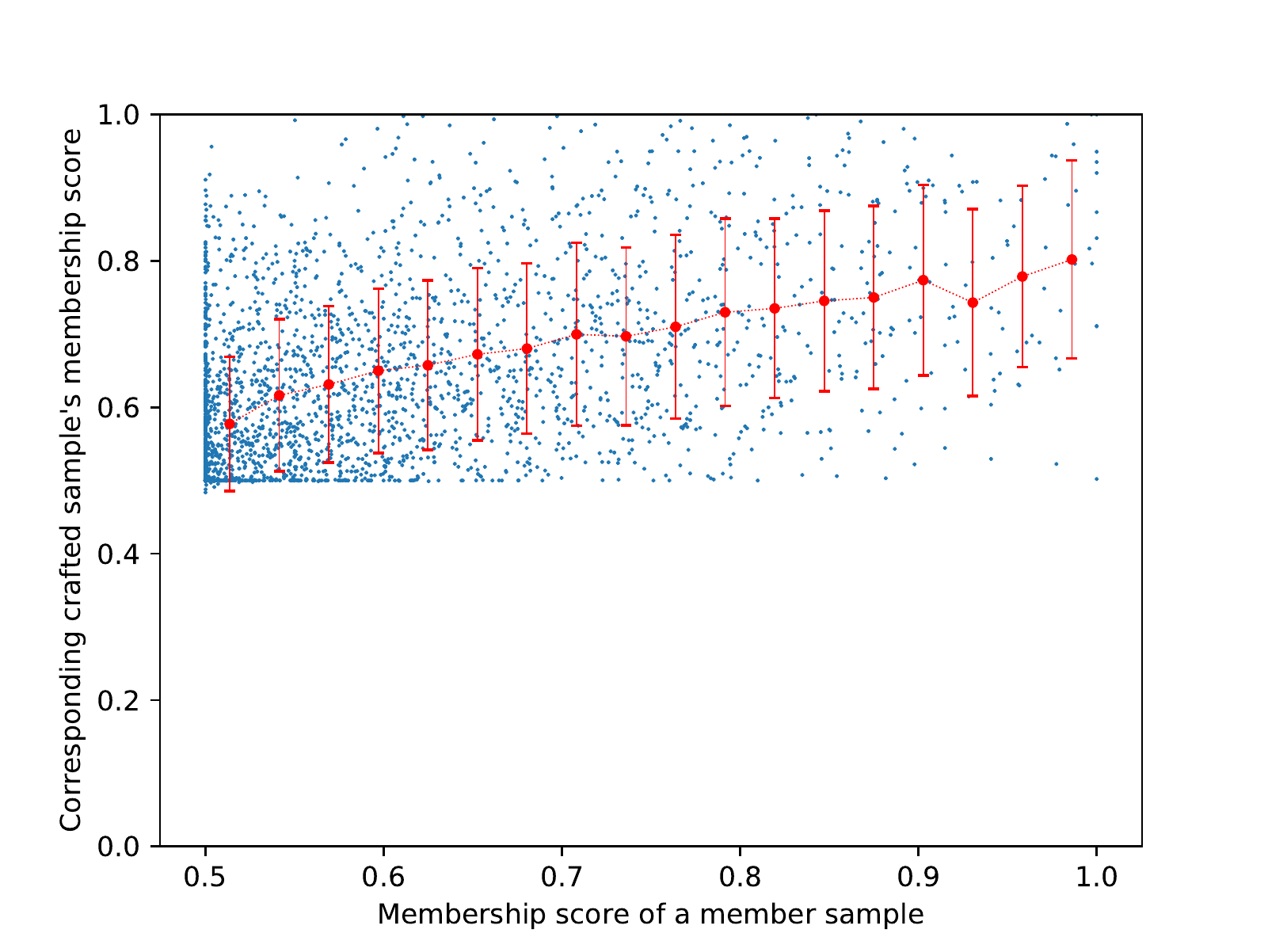}}
& \subfloat[Adversarial samples using Algorithm \ref{algo-adv}]{\includegraphics[width=0.23\linewidth]{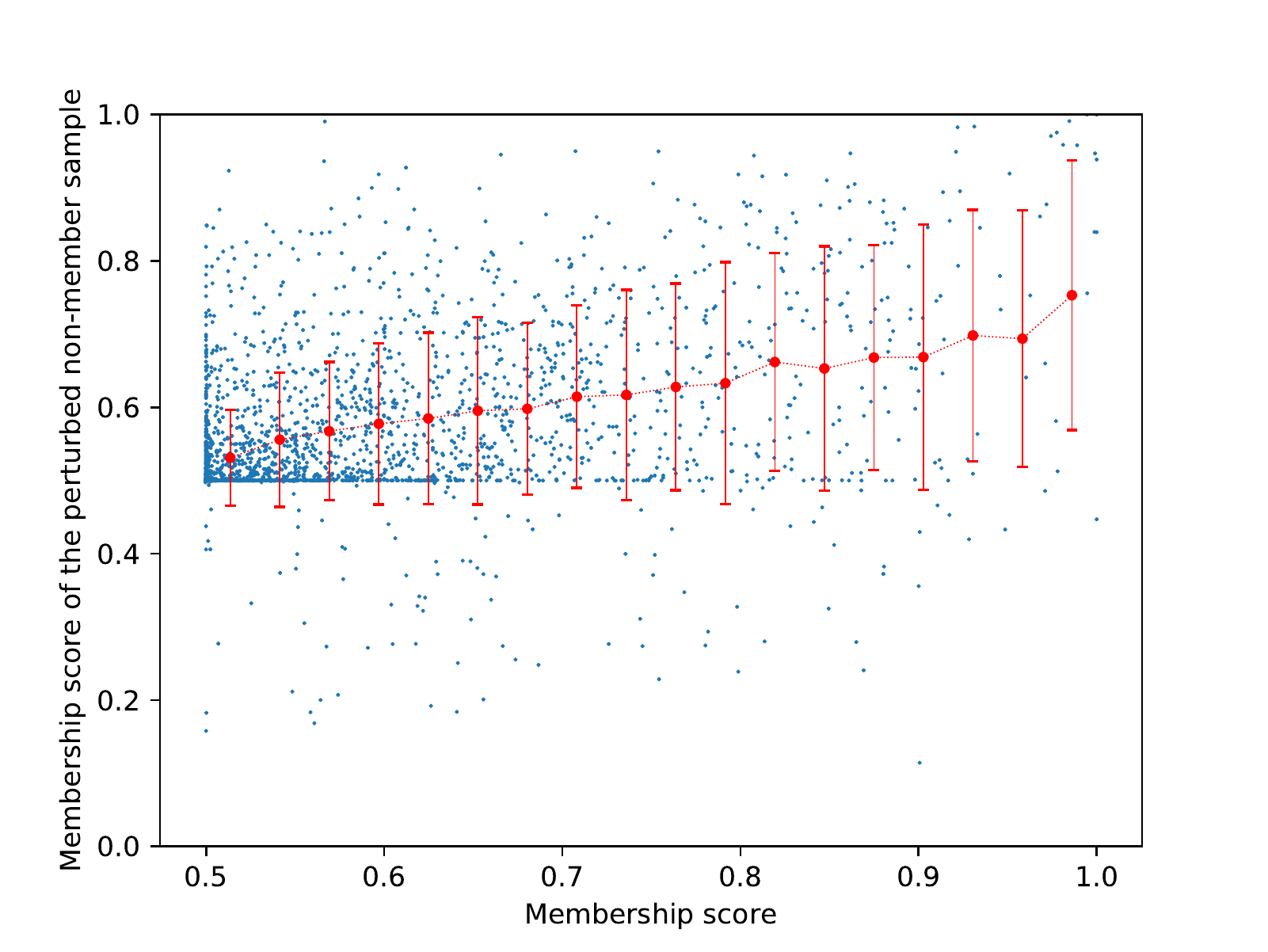}}\\
\end{tabular}
\caption{Correlation of (Watson attack) membership score between CIFAR-10 member samples (x-axis) and nonmember samples (y-axis). The criterion to select nonmember samples are specified by the title of each sub-figure.}
\label{fig-score-correlation}
\end{figure*}

Figure \ref{fig-fpr-fpr-c10-lenet-adv} demonstrates the false positive to false positive plot for a LeNet model trained CIFAR-10 dataset. In comparison with both Algorithm \ref{algo-search} (Figure \ref{fig-fpr-fpr-c10-lenet-natural}) and Algorithm \ref{algo-bigan} (Figure \ref{fig-fpr-fpr-c10-lenet-bigan}), using adversarial perturbation is a more effective on average. Even the perturbation of $\epsilon=0.01$ which does not produce perceptible artifacts is highly effective. It is worth emphasizing that the way the adversarial perturbation is used in this context is different from adversarial attack literature. In adversarial attack literature, the attacker has either white-box or black-box access to the model it tries to mislead, which would have been the MI attack in this case. However, in our scenario, the auditee who uses the adversarial attack does not even know the type of membership inference attack, let alone a query access or white-box access to it. The auditee, in this case, tries to perturb a sample so that it mimics the latent representation of another sample to which the MI attack has already assigned a high membership score.

Table \ref{tbl-adv} represents the results of the method on all datasets/models. Interestingly, in a few cases, the false positive is more than thousand times larger on discrediting samples. Given the simplicity of this approach in comparison with Algorithm \ref{algo-bigan} and the lack of the need for a large public dataset in comparison with \ref{algo-search}, the effectiveness of this approach as a discrediting tool is significant.

\section{Key Hypotheses and Validation}
\label{sec-analysis}

In this section, we investigate two hypotheses implicitly used as a cornerstone of the three discrediting algorithms. Here, the notion of closeness and neighborhood are all in the latent representation space, not the pixel space, unless specified otherwise. For more efficient visualization, we only show a small random set of samples in scatter plots. The average and standard error, however, is computed over all samples.

\begin{figure}
\centering
\includegraphics[width = 0.7\linewidth]{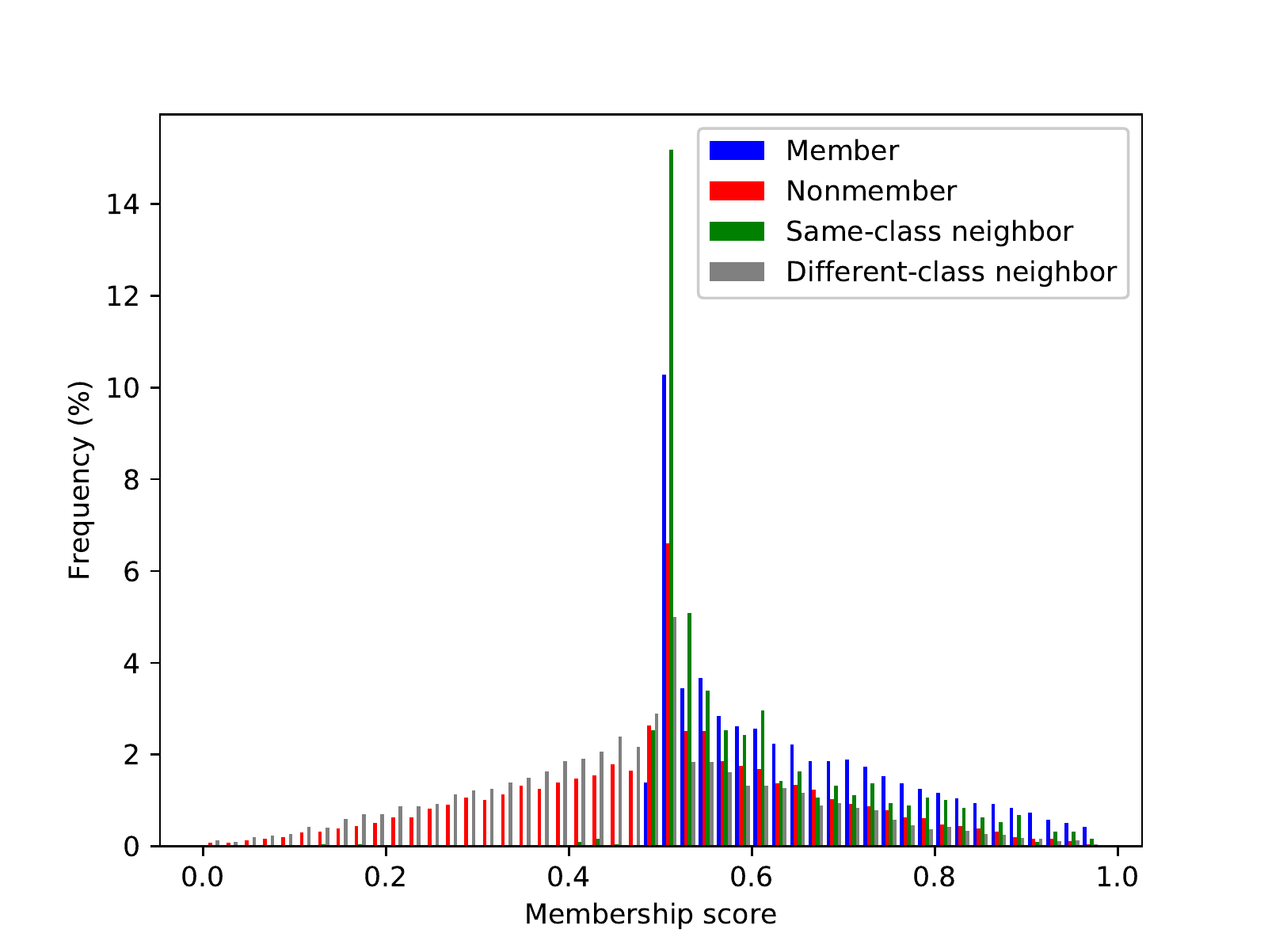}
\caption{Distribution of Watson attack's membership score for member samples, non-member samples, and same-class neighbors, and different-class neighbors from CINIC dataset}
\label{fig-distributions}
\end{figure}

\textbf{Hypothesis 1.} \textit{There is a correlation between the membership score of a member sample and its neighboring nonmember samples.}

This is the key assumptions used in all three discrediting algorithms. By sorting the $D_c$ dataset with respect to the membership score and finding/crafting samples based on them, we implicitly incorporating this assumption in all algorithms. To investigate this assumption, for each member sample in CIFAR-10 dataset, we use algorithm \ref{algo-search} (using CINIC dataset) and \ref{algo-bigan} to find/craft neighboring samples. Here, we use Watson attack \cite{watson2021importance} to compute the normalized membership score. Additionally, for each member sample, we randomly select a nonmember sample without any particular constraint to illustrate the case where no discrediting algorithm is used.

\begin{figure*}[h]
\centering
\centering
\begin{tabular}{ccc}
\subfloat[Closest sample from CINIC (Algorithm \ref{algo-search})]{\includegraphics[width=0.28\linewidth]{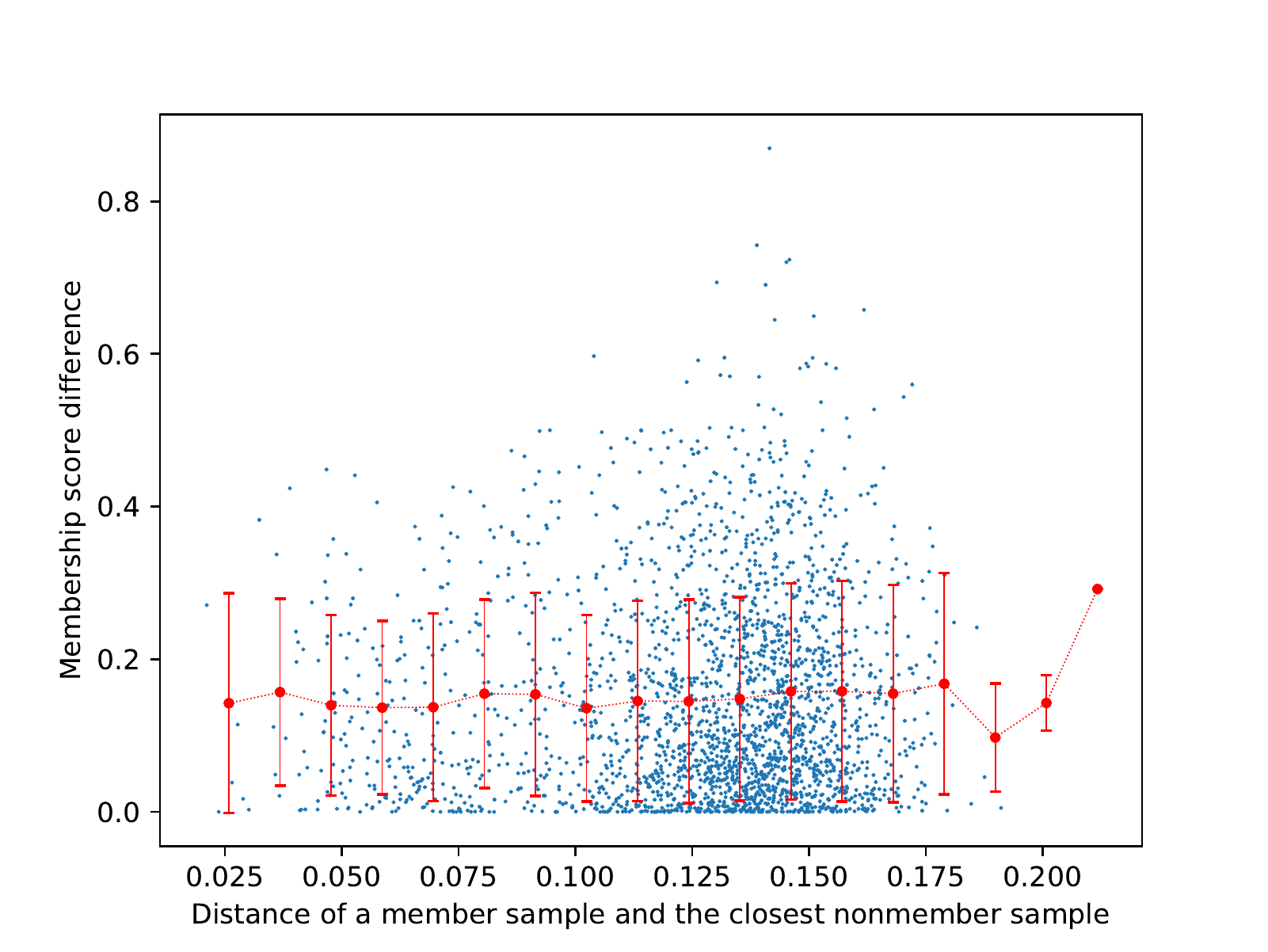}} 
& \subfloat[Crafted samples using Algorithm \ref{algo-bigan}]{\includegraphics[width=0.28\linewidth]{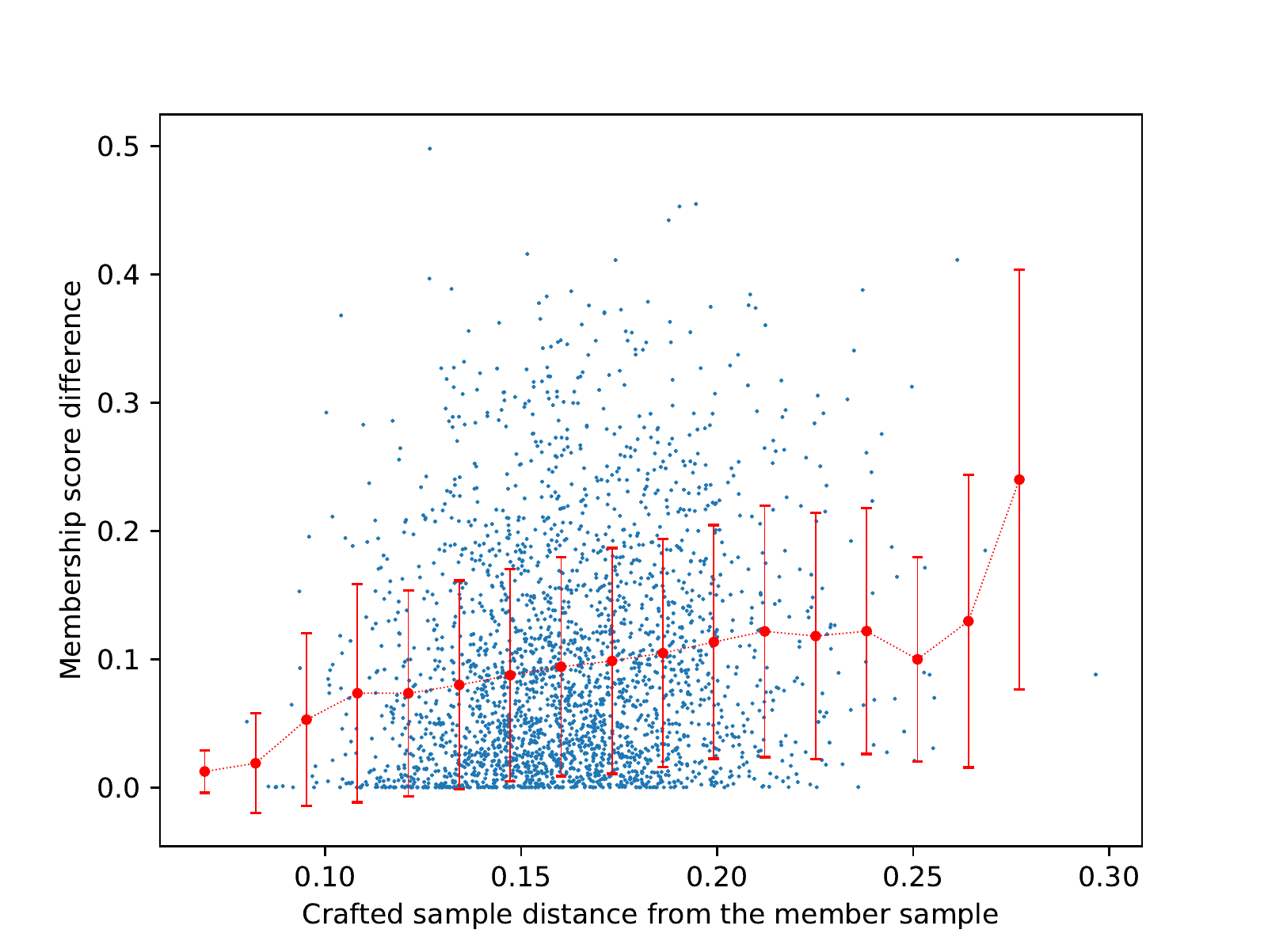}}
& \subfloat[Adversarial samples using Algorithm \ref{algo-adv}]{\includegraphics[width=0.28\linewidth]{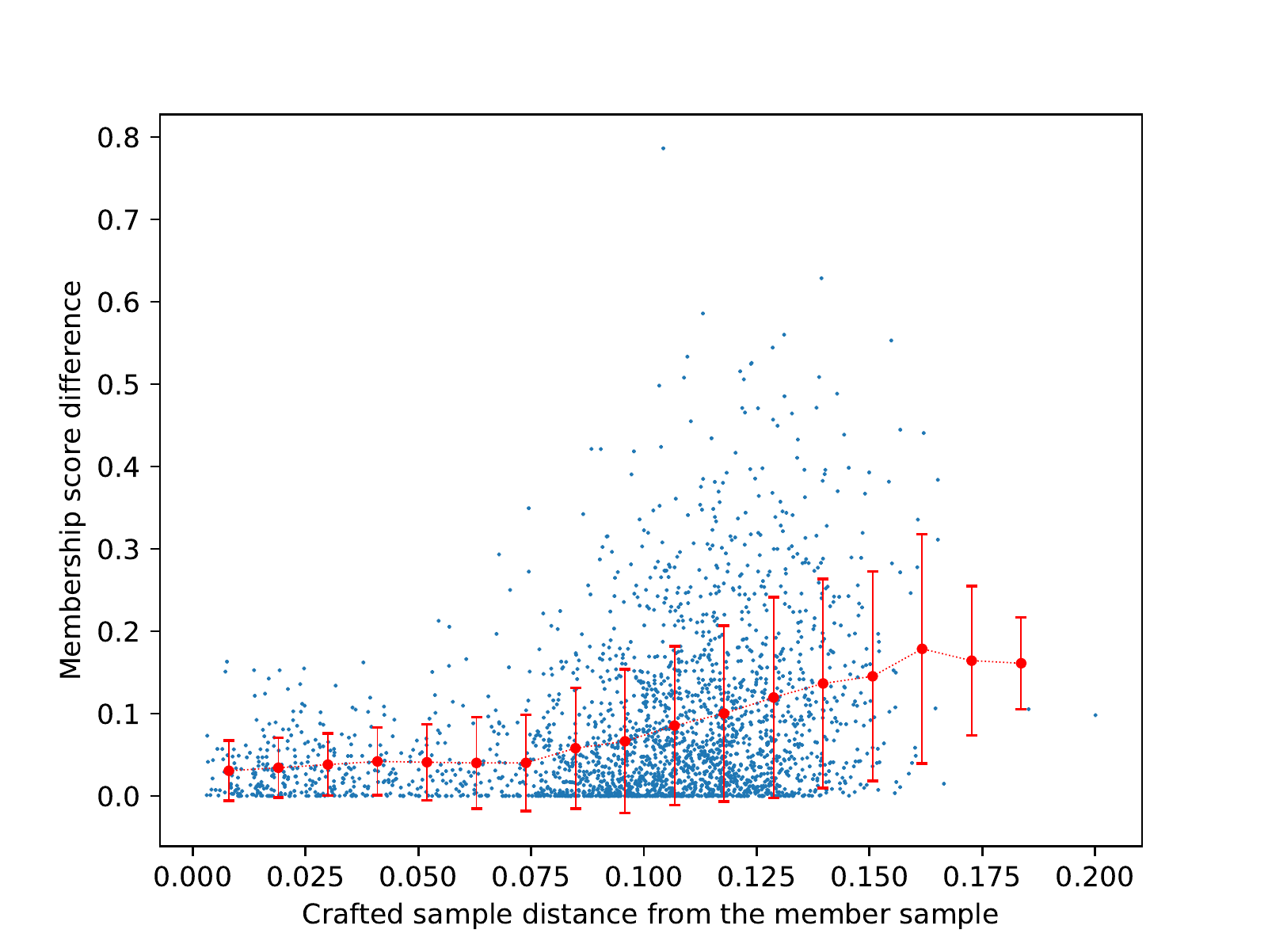}}\\
\end{tabular}
\caption{Correlation between the distance and the membership score difference of a member sample and its neighboring nonmember sample.}
\label{fig-dist-score-correlation}
\end{figure*}

\begin{figure}
\centering
\includegraphics[width = 0.7\linewidth]{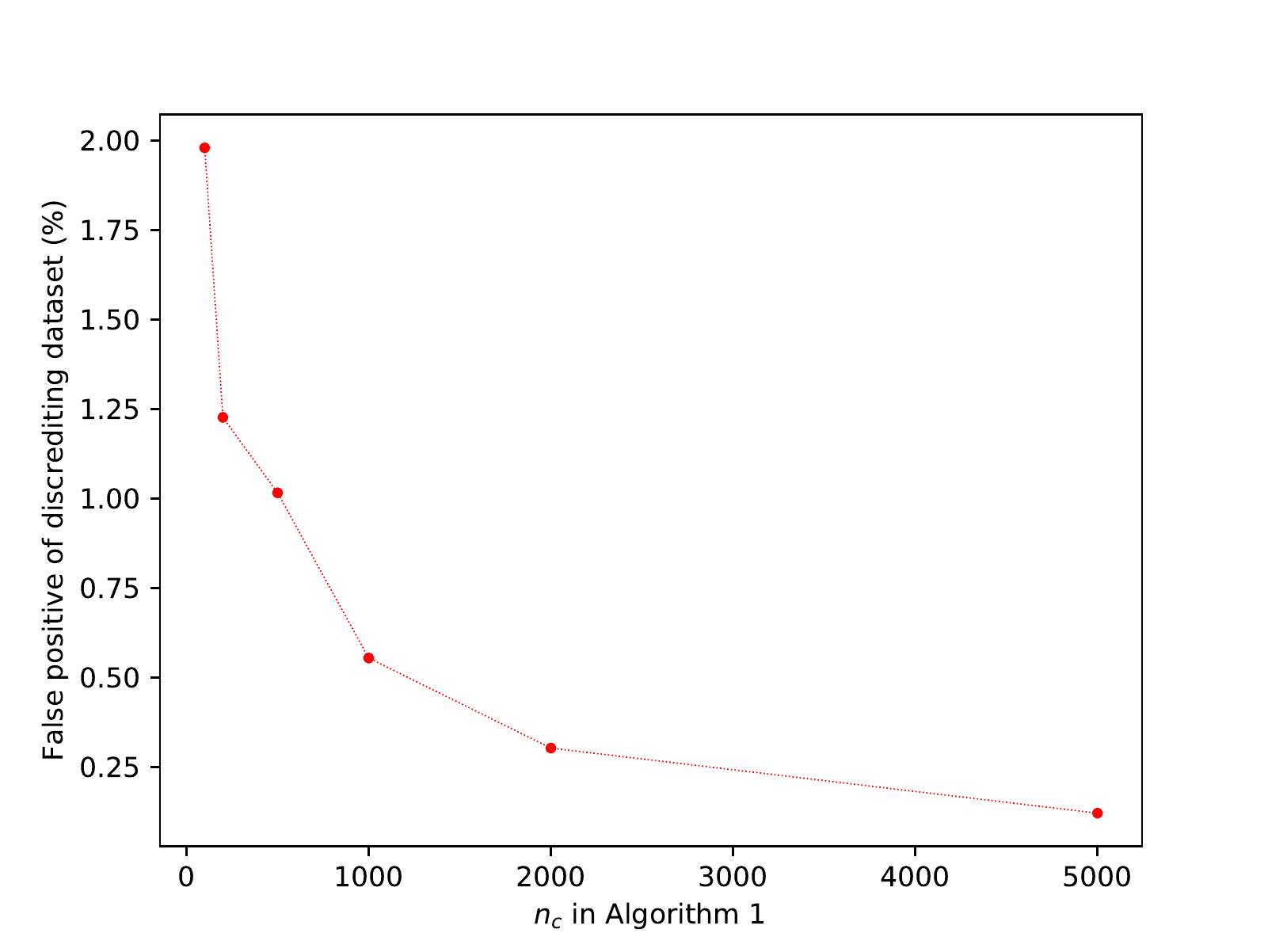}
\caption{$n_c$ in Algorithm \ref{algo-search} versus false positive rate of the discrediting dataset. 
%As more samples from $D_c$ set is involved in the process, the discrediting capability of the algorithm diminishes.
}
\label{fig-alg-1-analysis}
\end{figure}

Figure \ref{fig-score-correlation} (a) presents a case where no discrediting algorithm is used. X-axis shows the membership score of member samples, and the y-axis shows the score of a random sample from the nonmember set. As shown, the membership score of member samples are between $0.5$ and $1$. The membership score of nonmember samples, however, can be any value between $0$ and $1$. Figure \ref{fig-score-correlation} (b-d) demonstrates the case where our discrediting Algorithms are used. It is clear that the discrediting algorithms eliminate a majority of samples with low membership score. The output of discrediting algorithms are a set of nonmember samples whose membership score is between $0.5$ and $1$, similar to member samples.

Figure \ref{fig-score-correlation} (b-d) also illustrates the potential correlation between membership score of a member sample and its neighboring nonmember sample. It seems that there is no correlation when searching neighboring samples in CINIC dataset. The correlation analysis for this case is inconclusive and we speculate that if a much larger public dataset covering the entire portion of input space was available the results would have been different. The correlation can be better investigated with the generator model that allows us to generate arbitrary nonmember samples with different distance to the member samples. In this case, as shows in Figure \ref{fig-score-correlation} (c), there is a clear positive correlation between membership score of a member sample and its neighboring sample. The positive correlation is also clearly depicted in Figure \ref{fig-score-correlation} (d) for adversarially perturbed samples.

The effectiveness of using neighboring samples become more clear by looking at the distribution of membership scores of member, nonmember, and same-class neighbors from Algorithm \ref{algo-search}, as shown in Figure \ref{fig-distributions}. Here, same-class neighbors are closest samples whose class labels are the same as their neighbor member samples (corresponding to the if statement at line 6 in Algorithm \ref{algo-search}) but are not members themselves. Different-class neighbors are closest samples whose class labels are different from their member neighbors. We filter out different-class neighbors in Algorithm \ref{algo-search} and \ref{algo-bigan} for the following reason: The distance in latent space does not have a fixed scale and it is only meaningful locally. In other words, two samples $\epsilon$ away from each other in one region of the latent space might be semantically very similar and two other samples $\epsilon$ away from each other in another region of the latent space might be semantically very different. To filter out the samples that are likely to be semantically different, we match the class label as a rudimentary criterion. More research is needed to find a proper region-dependent scale for semantic similarity in latent space. As shown in Figure \ref{fig-distributions}, the distribution of same-class neighbors are much closer to the member samples and the distribution of different-class neighbors are closer to nonmember samples. That is the reason why MI attacks cannot avoid large false positive on discrediting samples.

Interestingly, the observation from Figure \ref{fig-score-correlation} (b)  that the membership scores of nonmember neighbors do not have clear positive correlation may lead to the perception that any member sample can be used as a part of $D_c$ to create discrediting dataset. There is a fundamental limitation in the experiment related to Figure \ref{fig-score-correlation} (b): When searching for the closest neighbor for each member sample in CINIC dataset, many duplicate samples are picked. In other words, many member samples share the same closest sample in CINIC dataset. Consequently, although member samples in x-axis of Figure \ref{fig-score-correlation} (b) are all unique, the corresponding neighbor samples in y-axis are not necessarily unique. This is important because the discrediting dataset provided to the judge should not have duplicate samples, otherwise the discrediting process was trivial. That is another reason why such experiment is inconclusive for algorithm \ref{algo-search} in Figure \ref{fig-score-correlation} (b). 

To investigate the correlation between the membership score of a member sample and the quality of corresponding discrediting dataset, we conduct an extra experiment. Instead of using all member samples, we use algorithm \ref{algo-search} with different $n_c$. The larger the $n_c$ is, the more samples with lower membership score are involved in the process. Here, we set the threshold such that the false positive rate is $0.01\%$ on the test dataset. Then, we use that threshold to compute the false positive on the discrediting dataset. As shown in Figure \ref{fig-alg-1-analysis}, it is clear that including samples with smaller membership score degrades the discredibility quality. Hence, it implies a positive correlation between the membership score of a member sample and the membership score of the corresponding nonmember neighbor.

% Interestingly, the observation from Figure \ref{fig-score-correlation} (b)  that the membership scores of nonmember neighbors do not have clear positive correlation with the membership score of their corresponding member samples has a useful byproduct. Instead of using $D_c$ in Algorithm \ref{algo-search}, we can use random samples from the training dataset. In other words, the auditee does not even need to know which member samples have higher membership scores w.r.t to the auditor's MI attack. Essentially, it eliminates the only information auditee needs from the auditor in Algorithm \ref{algo-search}. Hence, by replacing $D_c$ with auditee training data, the discrediting algorithm becomes completely independent of any MI attack.

% \begin{figure}
% \centering
% \includegraphics[width = 0.85\linewidth]{images/analysis_alg1/watson_neighbor_id_score_diff_correlation-eps-converted-to.pdf}
% \caption{Correlation between the neighbor index (Algorithm \ref{algo-search}) and membership score difference. X-axis shows the neighbor index, e.g., 1 is the closest neighbor, 2 is the second closest neighbor, and so on. The numbers on top of each index represent the number of samples in that x-index that have the same label as the target member sample.}
% \label{fig-alg-1-analysis}
% \end{figure}

\textbf{Hypothesis 2.} \textit{The closer the neighboring nonmember sample is to the member sample, the more similar their membership score would be.}

In Algorithm \ref{algo-search} we sort all neighbors with respect to their distance and explicitly prioritize the closest samples. The natural question is if there is a correlation between distance and the membership score. Figure \ref{fig-dist-score-correlation} demonstrates the correlation between the distance of a member sample to its nonmember neighbor and the absolute membership score difference. Similar to the previous experiment, there is a clear positive correlation in the case of crafted samples using Algorithm \ref{algo-bigan} and apparent lack of correlation in the case of natural samples. As discussed earlier, an experiment with a larger set of natural samples is needed to investigate the correlation for Algorithm \ref{algo-search} conclusively.

\begin{figure}
\centering
\includegraphics[width = 0.7\linewidth]{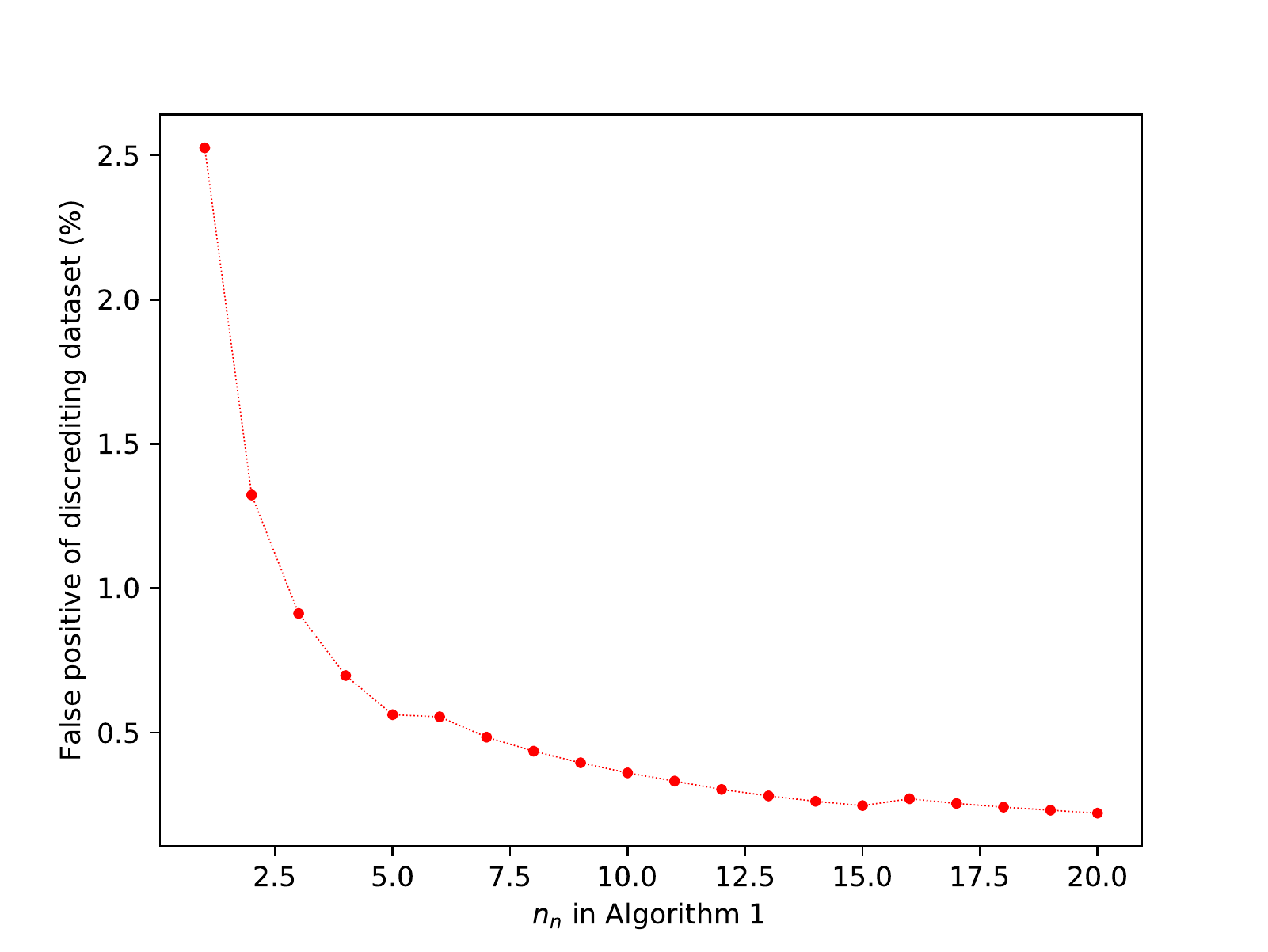}
\caption{$n_n$ in Algorithm \ref{algo-search} versus false positive rate of the discrediting dataset. 
%As more samples from $D_c$ set is involved in the process, the discrediting capability of the algorithm diminishes.
}
\label{fig-alg-1-analysis2}
\end{figure}

It is also interesting to see the correlation of the index of the neighbors and their membership score. In Figure \ref{fig-alg-1-analysis2}, as we include second, third, and n-th closest sample in the discrediting dataset, the false positive rate diminishes. It conveys that the further away from a sample we go, the membership score decreases. Although the previous experiment in Figure \ref{fig-dist-score-correlation} (a) is inconclusive about the correlation between the distance and the membership score, this experiment implies the correlation.

% we illustrate the correlation between the index of the closest neighbor and its membership score difference. The number on the top shows the number of x-th closest neighbors whose class label was similar to the target member sample. From CINIC dataset, we can conclude that the distribution of the first 10 (same-label) closest samples are similar with respect to the membership scores. The only notable observation is that the number of samples with same class label as the target member sample declines (the number on the top of the figure) as we consider further neighbors, which is intuitive.

% \begin{figure*}
% \centering
% \centering
% \begin{tabular}{ccc}
% \subfloat[]{\includegraphics[width=0.45\linewidth]{images/analysis_alg1/watson_neighbor_id_score_diff_correlation-eps-converted-to.pdf}} 
% & \subfloat[]{\includegraphics[width=0.45\linewidth]{images/analysis_alg1/watson_neighbor_id_dist_correlation-eps-converted-to.pdf}}\\
% \end{tabular}
% \caption{Correlation between the neighbor index (Algorithm \ref{algo-search}) and membership score difference/distance. x-axis shows the neighbor index, that is, 1 is the closest neighbor, 2 is the second closest neighbor, and so on. The numbers on top of each index represents the number of samples in that x-index that have the same label as the target member sample.}
% \label{fig-alg-1-analysis}
% \end{figure*}

\section{Discussion}

%\textbf{From Auditing Example to General MI:} 

\textbf{Auditing as an MI Application:} While many MI attacks have been proposed in the literature, not much discussion exists on how MI attacks can be used in real world scenarios. The auditing example we propose is a first attempt to address this limitation, by providing a potential real-world application of MI. Our work demonstrates the limitation of \textit{existing} MI attack techniques. It has two implications: 1) Strengths of MI attacks' output as an evidence of membership is weak when there is no prior to exclude semantically similar samples. 2) current MI attacks may better suited for user-level or subpopulation-level MI attacks, as discussed next.
% it does not imply that MI attacks cannot be useful/practical. Especially, considering the above-mentioned user-level MI attack, we hope that the potential usage of MI methodologies can inspire new research directions on MI.

\textbf{Membership Inference Application:} The ability to identify memorized subpopulations is useful in certain applications. For example, if a notion of subpopulation in latent space indicates individual users, it can be used for user-level membership inference, similar to \cite{li2022user}. A prominent example is face recognition where MI attacker (auditor) aims to know if a person's images have been unlawfully used or not. Interestingly, what we have shown in this paper suggests that the attacker does not need to know the exact training images to perform user-level membership inference. Hence, the MI attack in this case may be more practical than previously thought.

\textbf{Implication of Discredibility:} The implications of the discredibility is beyond the example of auditing we discuss in the paper. What we have shown is that the membership score distribution of member samples are similar to their nonmember neighbors. This issue exists in all MI attacks we studied. Using the loose definition of subpopulation, referring to samples close in the latent space, we argue that current membership inference attacks identify the \textit{memorized subpopulations}, not the \textit{memorized samples}. In other words, MI attacks can identify that a sample from a subpopulation is a member, but they cannot reliably identify which exact sample in that subpopulation is in the train set and which is not. Hence, it is different from what the existing MI attacks imply.
% The notion of memorized subpopulations might be interesting by itself in certain applications as discussed next. However, it certainly is not what membership inference attacks promise to deliver.

\textbf{Experimental vs Practical Setting: } As argued in \cite{carlini2022membership}, MI attack reports should include the true positive rate at low false positive rate like various areas of computer security \cite{ho2017detecting, kantchelian2015better, kolter2006learning, metsis2006spam}. Despite the similarities, there is an inherent difference between MI and other computer security applications. In membership inference, the ratio of positive samples are very small in comparison with all natural samples, similar to other computer security applications. However, the number of positive samples are \textit{fixed}, unlike other applications. Now, let's assume the common practice in MI literature where the entire fixed positive (member) samples are included in the performance evaluation, i.e., a closed-set experimental evaluation. Now, if we randomly collect billions of samples and add to the evaluation dataset, we only increase the number of negative samples because all positive samples had already been included. This means that the ratio of the number of true positive (TP) to the number of false positive (FP) depends on the size of the evaluating dataset because FP can infinitely grow in practice while TP is fixed. That is why the \textit{low false positive} ratio in the evaluation setting does not necessarily indicate \textit{small false positives} in practice. The existence of regions of high false positive rate, as shown in this paper, means that in practice when a large number of negative samples exists in the wild, the false positive samples dramatically outnumber the true positive samples. Therefore, further research is needed to introduce a more comprehensive and practical evaluation scheme.

% \textbf{Evaluation Criteria and Forensic Science: } As we discussed in the paper, it makes more sense to evaluate false positive on a subpopulation that the claimed member belongs to as it is a common practice in forensic science. Second, forensic science has been used primarily in two phases of the criminal-justice process \cite{lander2016forensic}: investigation and prosecution which has higher standard than investigation. We argue that although MI methodologies can be a useful tool during investigation phase, it should not be considered as a \textit{golden standard} for a prosecution in court at this stage.

\textbf{Limitations:} Our analysis lacks comprehensiveness in two areas. First, we do not have much larger dataset than CINIC to make sure that the majority of input space is covered. It might not be even possible. Although we indirectly shows the evidence of such a positive correlation in Section \ref{sec-analysis}, better experimental setting/dataset is needed for Algorithm \ref{algo-search}. Second, the BiGAN architecture proposed in \cite{rezaei2022efficient} to train a generator is far from perfect. Since we use the BiGAN in Algorithm \ref{algo-bigan}, and as a part of Rezaei's MI attack \cite{rezaei2022efficient}, it affects the performance of both. Hence, a better generator model may dramatically change the results of these two methods. It remains unclear whether a better generator helps the MI attack more or helps the discrediting algorithm more.

\section{Conclusion}
In this paper, we show that non-member samples from the subpopulation of a positively identified member sample often falsely identified as member by MI attack models. Consequently, the false positive of MI attacks are significantly higher on the exact samples that they identify as members. To demonstrate that this can be problematic, we showcase a real-world application scenario of MI attacks used as an investigative tool for auditing. High false positive rate of MI attacks on member samples allows an auditee to discredit the auditor's MI attacks.

% Then, we showcase a practical scenario where the membership inference attacks are used in a trail by an auditor (investigator or MI attacker) to prove to the judge that the auditee (MI victim) unlawfully used private data. Then, we show that the auditee can provide unlimited samples from the aforementioned subpopulation and seriously challenge the credibility of the auditor (MI attack). 

To achieve this goal, we propose three algorithms. The goal of all these algorithms is to search/craft samples whose latent representation is similar to a claimed member sample. We show that false positive rate of SOTA algorithms can jump from $0.01\%$ to hundreds or thousands time larger when evaluated on samples from the subpopulation of members. Therefore, we demonstrate that the discredibility issue is a serious concern when MI attacks are used in practice. In future, we investigate the possibility of new types of membership inference attacks immune to discredibility.

Finally, our findings suggest that the current membership inference attacks are not suitable for record-level membership inference. They may be better used for subpopulation-based MI attack, e.g., used-level membership inference. Moreover, we believe that a better experimental evaluation scenario needs to be designed that resembles open-set experimental design and also take the target sample's subpopulation into account.

% \begin{acks}
% \end{acks}

\bibliographystyle{ACM-Reference-Format}

% \bibliography{ref}

\begin{thebibliography}{42}

%%% the Web address.

\ifx \showCODEN    \undefined \def \showCODEN     #1{\unskip}     \fi
\ifx \showDOI      \undefined \def \showDOI       #1{#1}\fi
\ifx \showISBNx    \undefined \def \showISBNx     #1{\unskip}     \fi
\ifx \showISBNxiii \undefined \def \showISBNxiii  #1{\unskip}     \fi
\ifx \showISSN     \undefined \def \showISSN      #1{\unskip}     \fi
\ifx \showLCCN     \undefined \def \showLCCN      #1{\unskip}     \fi
\ifx \shownote     \undefined \def \shownote      #1{#1}          \fi
\ifx \showarticletitle \undefined \def \showarticletitle #1{#1}   \fi
\ifx \showURL      \undefined \def \showURL       {\relax}        \fi
% invisible to TeX
\providecommand\bibfield[2]{#2}
\providecommand\bibinfo[2]{#2}
\providecommand\natexlab[1]{#1}
\providecommand\showeprint[2][]{arXiv:#2}

\bibitem[Balding and Nichols(1994)]%
        {balding1994dna}
\bibfield{author}{\bibinfo{person}{David~J Balding} {and}
  \bibinfo{person}{Richard~A Nichols}.} \bibinfo{year}{1994}\natexlab{}.
\newblock \showarticletitle{DNA profile match probability calculation: how to
  allow for population stratification, relatedness, database selection and
  single bands}.
\newblock \bibinfo{journal}{\emph{Forensic science international}}
  \bibinfo{volume}{64}, \bibinfo{number}{2-3} (\bibinfo{year}{1994}),
  \bibinfo{pages}{125--140}.
\newblock


\bibitem[Baldwin et~al\mbox{.}(2014)]%
        {baldwin2014study}
\bibfield{author}{\bibinfo{person}{David~P Baldwin}, \bibinfo{person}{Stanley~J
  Bajic}, \bibinfo{person}{Max Morris}, {and} \bibinfo{person}{Daniel Zamzow}.}
  \bibinfo{year}{2014}\natexlab{}.
\newblock \bibinfo{booktitle}{\emph{A study of false-positive and
  false-negative error rates in cartridge case comparisons}}.
\newblock \bibinfo{type}{{T}echnical {R}eport}. \bibinfo{institution}{AMES LAB
  IA}.
\newblock


\bibitem[Carlini et~al\mbox{.}(2022)]%
        {carlini2022membership}
\bibfield{author}{\bibinfo{person}{Nicholas Carlini}, \bibinfo{person}{Steve
  Chien}, \bibinfo{person}{Milad Nasr}, \bibinfo{person}{Shuang Song},
  \bibinfo{person}{Andreas Terzis}, {and} \bibinfo{person}{Florian Tramer}.}
  \bibinfo{year}{2022}\natexlab{}.
\newblock \showarticletitle{Membership inference attacks from first
  principles}. In \bibinfo{booktitle}{\emph{2022 IEEE Symposium on Security and
  Privacy (SP)}}. IEEE, \bibinfo{pages}{1897--1914}.
\newblock


\bibitem[Chapnick et~al\mbox{.}(2021)]%
        {chapnick2021results}
\bibfield{author}{\bibinfo{person}{Chad Chapnick}, \bibinfo{person}{Todd~J
  Weller}, \bibinfo{person}{Pierre Duez}, \bibinfo{person}{Eric Meschke},
  \bibinfo{person}{John Marshall}, {and} \bibinfo{person}{Ryan Lilien}.}
  \bibinfo{year}{2021}\natexlab{}.
\newblock \showarticletitle{Results of the 3D virtual comparison microscopy
  error rate (VCMER) study for firearm forensics}.
\newblock \bibinfo{journal}{\emph{Journal of forensic sciences}}
  \bibinfo{volume}{66}, \bibinfo{number}{2} (\bibinfo{year}{2021}),
  \bibinfo{pages}{557--570}.
\newblock


\bibitem[Choo et~al\mbox{.}(2020)]%
        {choo2020label}
\bibfield{author}{\bibinfo{person}{Christopher A~Choquette Choo},
  \bibinfo{person}{Florian Tramer}, \bibinfo{person}{Nicholas Carlini}, {and}
  \bibinfo{person}{Nicolas Papernot}.} \bibinfo{year}{2020}\natexlab{}.
\newblock \showarticletitle{Label-only membership inference attacks}.
\newblock \bibinfo{journal}{\emph{arXiv preprint arXiv:2007.14321}}
  (\bibinfo{year}{2020}).
\newblock


\bibitem[Darlow et~al\mbox{.}(2018)]%
        {darlow2018cinic}
\bibfield{author}{\bibinfo{person}{Luke~N Darlow}, \bibinfo{person}{Elliot~J
  Crowley}, \bibinfo{person}{Antreas Antoniou}, {and} \bibinfo{person}{Amos~J
  Storkey}.} \bibinfo{year}{2018}\natexlab{}.
\newblock \showarticletitle{Cinic-10 is not imagenet or cifar-10}.
\newblock \bibinfo{journal}{\emph{arXiv preprint arXiv:1810.03505}}
  (\bibinfo{year}{2018}).
\newblock


\bibitem[Duez et~al\mbox{.}(2018)]%
        {duez2018development}
\bibfield{author}{\bibinfo{person}{Pierre Duez}, \bibinfo{person}{Todd Weller},
  \bibinfo{person}{Marcus Brubaker}, \bibinfo{person}{Richard~E Hockensmith},
  {and} \bibinfo{person}{Ryan Lilien}.} \bibinfo{year}{2018}\natexlab{}.
\newblock \showarticletitle{Development and validation of a virtual examination
  tool for firearm forensics}.
\newblock \bibinfo{journal}{\emph{Journal of forensic sciences}}
  \bibinfo{volume}{63}, \bibinfo{number}{4} (\bibinfo{year}{2018}),
  \bibinfo{pages}{1069--1084}.
\newblock


\bibitem[Ho et~al\mbox{.}(2017)]%
        {ho2017detecting}
\bibfield{author}{\bibinfo{person}{Grant Ho}, \bibinfo{person}{Aashish Sharma},
  \bibinfo{person}{Mobin Javed}, \bibinfo{person}{Vern Paxson}, {and}
  \bibinfo{person}{David Wagner}.} \bibinfo{year}{2017}\natexlab{}.
\newblock \showarticletitle{Detecting credential spearphishing in enterprise
  settings}. In \bibinfo{booktitle}{\emph{26th USENIX Security Symposium
  (USENIX Security 17)}}. \bibinfo{pages}{469--485}.
\newblock


\bibitem[Jayaraman et~al\mbox{.}(2021)]%
        {jayaraman2020revisiting}
\bibfield{author}{\bibinfo{person}{Bargav Jayaraman}, \bibinfo{person}{Lingxiao
  Wang}, \bibinfo{person}{Katherine Knipmeyer}, \bibinfo{person}{Quanquan Gu},
  {and} \bibinfo{person}{David Evans}.} \bibinfo{year}{2021}\natexlab{}.
\newblock \showarticletitle{Revisiting membership inference under realistic
  assumptions}.
\newblock \bibinfo{journal}{\emph{In Proceedings on Privacy Enhancing
  Technologies (PoPETs)}} (\bibinfo{year}{2021}).
\newblock


\bibitem[Kantchelian et~al\mbox{.}(2015)]%
        {kantchelian2015better}
\bibfield{author}{\bibinfo{person}{Alex Kantchelian},
  \bibinfo{person}{Michael~Carl Tschantz}, \bibinfo{person}{Sadia Afroz},
  \bibinfo{person}{Brad Miller}, \bibinfo{person}{Vaishaal Shankar},
  \bibinfo{person}{Rekha Bachwani}, \bibinfo{person}{Anthony~D Joseph}, {and}
  \bibinfo{person}{J~Doug Tygar}.} \bibinfo{year}{2015}\natexlab{}.
\newblock \showarticletitle{Better malware ground truth: Techniques for
  weighting anti-virus vendor labels}. In \bibinfo{booktitle}{\emph{Proceedings
  of the 8th ACM Workshop on Artificial Intelligence and Security}}.
  \bibinfo{pages}{45--56}.
\newblock


\bibitem[Kolter and Maloof(2006)]%
        {kolter2006learning}
\bibfield{author}{\bibinfo{person}{J~Zico Kolter} {and}
  \bibinfo{person}{Marcus~A Maloof}.} \bibinfo{year}{2006}\natexlab{}.
\newblock \showarticletitle{Learning to detect and classify malicious
  executables in the wild.}
\newblock \bibinfo{journal}{\emph{Journal of Machine Learning Research}}
  \bibinfo{volume}{7}, \bibinfo{number}{12} (\bibinfo{year}{2006}).
\newblock


\bibitem[Krizhevsky et~al\mbox{.}(2009)]%
        {krizhevsky2009learning}
\bibfield{author}{\bibinfo{person}{Alex Krizhevsky}, \bibinfo{person}{Geoffrey
  Hinton}, {et~al\mbox{.}}} \bibinfo{year}{2009}\natexlab{}.
\newblock \bibinfo{title}{Learning multiple layers of features from tiny
  images}.
\newblock
\newblock


\bibitem[LeCun et~al\mbox{.}(1998)]%
        {lecun1998gradient}
\bibfield{author}{\bibinfo{person}{Yann LeCun}, \bibinfo{person}{L{\'e}on
  Bottou}, \bibinfo{person}{Yoshua Bengio}, {and} \bibinfo{person}{Patrick
  Haffner}.} \bibinfo{year}{1998}\natexlab{}.
\newblock \showarticletitle{Gradient-based learning applied to document
  recognition}.
\newblock \bibinfo{journal}{\emph{Proc. IEEE}} \bibinfo{volume}{86},
  \bibinfo{number}{11} (\bibinfo{year}{1998}), \bibinfo{pages}{2278--2324}.
\newblock


\bibitem[Leino and Fredrikson(2020)]%
        {leino2020stolen}
\bibfield{author}{\bibinfo{person}{Klas Leino} {and} \bibinfo{person}{Matt
  Fredrikson}.} \bibinfo{year}{2020}\natexlab{}.
\newblock \showarticletitle{Stolen memories: Leveraging model memorization for
  calibrated white-box membership inference}. In \bibinfo{booktitle}{\emph{29th
  $\{$USENIX$\}$ Security Symposium ($\{$USENIX$\}$ Security 20)}}.
  \bibinfo{pages}{1605--1622}.
\newblock


\bibitem[Li et~al\mbox{.}(2022)]%
        {li2022user}
\bibfield{author}{\bibinfo{person}{Guoyao Li}, \bibinfo{person}{Shahbaz
  Rezaei}, {and} \bibinfo{person}{Xin Liu}.} \bibinfo{year}{2022}\natexlab{}.
\newblock \showarticletitle{User-Level Membership Inference Attack against
  Metric Embedding Learning}.
\newblock \bibinfo{journal}{\emph{arXiv preprint arXiv:2203.02077}}
  (\bibinfo{year}{2022}).
\newblock


\bibitem[Li and Zhang(2020)]%
        {li2020label}
\bibfield{author}{\bibinfo{person}{Zheng Li} {and} \bibinfo{person}{Yang
  Zhang}.} \bibinfo{year}{2020}\natexlab{}.
\newblock \showarticletitle{Label-Leaks: Membership Inference Attack with
  Label}.
\newblock \bibinfo{journal}{\emph{arXiv preprint arXiv:2007.15528}}
  (\bibinfo{year}{2020}).
\newblock


\bibitem[Liu et~al\mbox{.}(2019)]%
        {liu2019socinf}
\bibfield{author}{\bibinfo{person}{Gaoyang Liu}, \bibinfo{person}{Chen Wang},
  \bibinfo{person}{Kai Peng}, \bibinfo{person}{Haojun Huang},
  \bibinfo{person}{Yutong Li}, {and} \bibinfo{person}{Wenqing Cheng}.}
  \bibinfo{year}{2019}\natexlab{}.
\newblock \showarticletitle{Socinf: Membership inference attacks on social
  media health data with machine learning}.
\newblock \bibinfo{journal}{\emph{IEEE Transactions on Computational Social
  Systems}} \bibinfo{volume}{6}, \bibinfo{number}{5} (\bibinfo{year}{2019}),
  \bibinfo{pages}{907--921}.
\newblock


\bibitem[Long et~al\mbox{.}(2017)]%
        {long2017towards}
\bibfield{author}{\bibinfo{person}{Yunhui Long}, \bibinfo{person}{Vincent
  Bindschaedler}, {and} \bibinfo{person}{Carl~A Gunter}.}
  \bibinfo{year}{2017}\natexlab{}.
\newblock \showarticletitle{Towards measuring membership privacy}.
\newblock \bibinfo{journal}{\emph{arXiv preprint arXiv:1712.09136}}
  (\bibinfo{year}{2017}).
\newblock


\bibitem[Long et~al\mbox{.}(2018)]%
        {long2018understanding}
\bibfield{author}{\bibinfo{person}{Yunhui Long}, \bibinfo{person}{Vincent
  Bindschaedler}, \bibinfo{person}{Lei Wang}, \bibinfo{person}{Diyue Bu},
  \bibinfo{person}{Xiaofeng Wang}, \bibinfo{person}{Haixu Tang},
  \bibinfo{person}{Carl~A Gunter}, {and} \bibinfo{person}{Kai Chen}.}
  \bibinfo{year}{2018}\natexlab{}.
\newblock \showarticletitle{Understanding membership inferences on
  well-generalized learning models}.
\newblock \bibinfo{journal}{\emph{arXiv preprint arXiv:1802.04889}}
  (\bibinfo{year}{2018}).
\newblock


\bibitem[Long et~al\mbox{.}(2020)]%
        {long2020pragmatic}
\bibfield{author}{\bibinfo{person}{Yunhui Long}, \bibinfo{person}{Lei Wang},
  \bibinfo{person}{Diyue Bu}, \bibinfo{person}{Vincent Bindschaedler},
  \bibinfo{person}{Xiaofeng Wang}, \bibinfo{person}{Haixu Tang},
  \bibinfo{person}{Carl~A Gunter}, {and} \bibinfo{person}{Kai Chen}.}
  \bibinfo{year}{2020}\natexlab{}.
\newblock \showarticletitle{A pragmatic approach to membership inferences on
  machine learning models}. In \bibinfo{booktitle}{\emph{2020 IEEE European
  Symposium on Security and Privacy (EuroS\&P)}}. IEEE,
  \bibinfo{pages}{521--534}.
\newblock


\bibitem[Mattijssen et~al\mbox{.}(2020)]%
        {mattijssen2020validity}
\bibfield{author}{\bibinfo{person}{Erwin~JAT Mattijssen},
  \bibinfo{person}{Cilia~LM Witteman}, \bibinfo{person}{Charles~EH Berger},
  \bibinfo{person}{Nicolaas~W Brand}, {and} \bibinfo{person}{Reinoud~D Stoel}.}
  \bibinfo{year}{2020}\natexlab{}.
\newblock \showarticletitle{Validity and reliability of forensic firearm
  examiners}.
\newblock \bibinfo{journal}{\emph{Forensic science international}}
  \bibinfo{volume}{307} (\bibinfo{year}{2020}), \bibinfo{pages}{110112}.
\newblock


\bibitem[Metsis et~al\mbox{.}(2006)]%
        {metsis2006spam}
\bibfield{author}{\bibinfo{person}{Vangelis Metsis}, \bibinfo{person}{Ion
  Androutsopoulos}, {and} \bibinfo{person}{Georgios Paliouras}.}
  \bibinfo{year}{2006}\natexlab{}.
\newblock \showarticletitle{Spam filtering with naive bayes-which naive
  bayes?}. In \bibinfo{booktitle}{\emph{CEAS}}, Vol.~\bibinfo{volume}{17}.
  Mountain View, CA, \bibinfo{pages}{28--69}.
\newblock


\bibitem[Monson et~al\mbox{.}(2022)]%
        {monson2022planning}
\bibfield{author}{\bibinfo{person}{Keith~L Monson}, \bibinfo{person}{Erich~D
  Smith}, {and} \bibinfo{person}{Stanley~J Bajic}.}
  \bibinfo{year}{2022}\natexlab{}.
\newblock \showarticletitle{Planning, design and logistics of a decision
  analysis study: The FBI/Ames study involving forensic firearms examiners}.
\newblock \bibinfo{journal}{\emph{Forensic Science International: Synergy}}
  \bibinfo{volume}{4} (\bibinfo{year}{2022}), \bibinfo{pages}{100221}.
\newblock


\bibitem[Murakonda and Shokri(2020)]%
        {murakonda2020ml}
\bibfield{author}{\bibinfo{person}{Sasi~Kumar Murakonda} {and}
  \bibinfo{person}{Reza Shokri}.} \bibinfo{year}{2020}\natexlab{}.
\newblock \showarticletitle{Ml privacy meter: Aiding regulatory compliance by
  quantifying the privacy risks of machine learning}.
\newblock \bibinfo{journal}{\emph{arXiv preprint arXiv:2007.09339}}
  (\bibinfo{year}{2020}).
\newblock


\bibitem[Netzer et~al\mbox{.}(2011)]%
        {netzer2011reading}
\bibfield{author}{\bibinfo{person}{Yuval Netzer}, \bibinfo{person}{Tao Wang},
  \bibinfo{person}{Adam Coates}, \bibinfo{person}{Alessandro Bissacco},
  \bibinfo{person}{Bo Wu}, {and} \bibinfo{person}{Andrew~Y Ng}.}
  \bibinfo{year}{2011}\natexlab{}.
\newblock \showarticletitle{Reading digits in natural images with unsupervised
  feature learning}. In \bibinfo{booktitle}{\emph{NIPS Workshop}}.
\newblock


\bibitem[Pascanu et~al\mbox{.}(2013)]%
        {pascanu2013number}
\bibfield{author}{\bibinfo{person}{Razvan Pascanu}, \bibinfo{person}{Guido
  Montufar}, {and} \bibinfo{person}{Yoshua Bengio}.}
  \bibinfo{year}{2013}\natexlab{}.
\newblock \showarticletitle{On the number of response regions of deep feed
  forward networks with piece-wise linear activations}.
\newblock \bibinfo{journal}{\emph{arXiv preprint arXiv:1312.6098}}
  (\bibinfo{year}{2013}).
\newblock


\bibitem[Pauw-Vugts et~al\mbox{.}(2013)]%
        {pauw2013faid2009}
\bibfield{author}{\bibinfo{person}{P Pauw-Vugts}, \bibinfo{person}{A Walters},
  \bibinfo{person}{L {\O}ren}, {and} \bibinfo{person}{L Pfoser}.}
  \bibinfo{year}{2013}\natexlab{}.
\newblock \showarticletitle{FAID2009: proficiency test and workshop}.
\newblock \bibinfo{journal}{\emph{AFTE Journal}} \bibinfo{volume}{45},
  \bibinfo{number}{2} (\bibinfo{year}{2013}), \bibinfo{pages}{115--127}.
\newblock


\bibitem[Rahimian et~al\mbox{.}(2020)]%
        {rahimian2020sampling}
\bibfield{author}{\bibinfo{person}{Shadi Rahimian},
  \bibinfo{person}{Tribhuvanesh Orekondy}, {and} \bibinfo{person}{Mario
  Fritz}.} \bibinfo{year}{2020}\natexlab{}.
\newblock \showarticletitle{Sampling Attacks: Amplification of Membership
  Inference Attacks by Repeated Queries}.
\newblock \bibinfo{journal}{\emph{arXiv preprint arXiv:2009.00395}}
  (\bibinfo{year}{2020}).
\newblock


\bibitem[Rezaei et~al\mbox{.}(2022)]%
        {rezaei2022efficient}
\bibfield{author}{\bibinfo{person}{Shahbaz Rezaei}, \bibinfo{person}{}, {and}
  \bibinfo{person}{Xin Liu}.} \bibinfo{year}{2022}\natexlab{}.
\newblock \showarticletitle{An Efficient Subpopulation-based Membership
  Inference Attack}.
\newblock \bibinfo{journal}{\emph{arXiv preprint arXiv:2203.02080}}
  (\bibinfo{year}{2022}).
\newblock


\bibitem[Rezaei and Liu(2021)]%
        {rezaei2020towards}
\bibfield{author}{\bibinfo{person}{Shahbaz Rezaei} {and} \bibinfo{person}{Xin
  Liu}.} \bibinfo{year}{2021}\natexlab{}.
\newblock \showarticletitle{On the Difficulty of Membership Inference Attacks}.
  In \bibinfo{booktitle}{\emph{Proceedings of the IEEE/CVF Conference on
  Computer Vision and Pattern Recognition}}.
\newblock


\bibitem[Rezaei et~al\mbox{.}(2021)]%
        {rezaei2021accuracy}
\bibfield{author}{\bibinfo{person}{Shahbaz Rezaei}, \bibinfo{person}{Zubair
  Shafiq}, {and} \bibinfo{person}{Xin Liu}.} \bibinfo{year}{2021}\natexlab{}.
\newblock \showarticletitle{Accuracy-Privacy Trade-off in Deep Ensemble: A
  Membership Inference Perspective}.
\newblock \bibinfo{journal}{\emph{arXiv preprint arXiv:2105.05381}}
  (\bibinfo{year}{2021}).
\newblock


\bibitem[Sablayrolles et~al\mbox{.}(2019)]%
        {sablayrolles2019white}
\bibfield{author}{\bibinfo{person}{Alexandre Sablayrolles},
  \bibinfo{person}{Matthijs Douze}, \bibinfo{person}{Cordelia Schmid},
  \bibinfo{person}{Yann Ollivier}, {and} \bibinfo{person}{Herv{\'e}
  J{\'e}gou}.} \bibinfo{year}{2019}\natexlab{}.
\newblock \showarticletitle{White-box vs black-box: Bayes optimal strategies
  for membership inference}. In \bibinfo{booktitle}{\emph{International
  Conference on Machine Learning}}. PMLR, \bibinfo{pages}{5558--5567}.
\newblock


\bibitem[Salem et~al\mbox{.}(2018)]%
        {salem2018ml}
\bibfield{author}{\bibinfo{person}{Ahmed Salem}, \bibinfo{person}{Yang Zhang},
  \bibinfo{person}{Mathias Humbert}, \bibinfo{person}{Pascal Berrang},
  \bibinfo{person}{Mario Fritz}, {and} \bibinfo{person}{Michael Backes}.}
  \bibinfo{year}{2018}\natexlab{}.
\newblock \showarticletitle{Ml-leaks: Model and data independent membership
  inference attacks and defenses on machine learning models}.
\newblock \bibinfo{journal}{\emph{arXiv preprint arXiv:1806.01246}}
  (\bibinfo{year}{2018}).
\newblock


\bibitem[Shokri et~al\mbox{.}(2017)]%
        {shokri2017membership}
\bibfield{author}{\bibinfo{person}{Reza Shokri}, \bibinfo{person}{Marco
  Stronati}, \bibinfo{person}{Congzheng Song}, {and} \bibinfo{person}{Vitaly
  Shmatikov}.} \bibinfo{year}{2017}\natexlab{}.
\newblock \showarticletitle{Membership inference attacks against machine
  learning models}. In \bibinfo{booktitle}{\emph{2017 IEEE Symposium on
  Security and Privacy (SP)}}. IEEE, \bibinfo{pages}{3--18}.
\newblock


\bibitem[Song and Mittal(2021)]%
        {song2021systematic}
\bibfield{author}{\bibinfo{person}{Liwei Song} {and} \bibinfo{person}{Prateek
  Mittal}.} \bibinfo{year}{2021}\natexlab{}.
\newblock \showarticletitle{Systematic evaluation of privacy risks of machine
  learning models}. In \bibinfo{booktitle}{\emph{30th USENIX Security Symposium
  (USENIX Security 21)}}. \bibinfo{pages}{2615--2632}.
\newblock


\bibitem[Song et~al\mbox{.}(2019)]%
        {song2019privacy}
\bibfield{author}{\bibinfo{person}{Liwei Song}, \bibinfo{person}{Reza Shokri},
  {and} \bibinfo{person}{Prateek Mittal}.} \bibinfo{year}{2019}\natexlab{}.
\newblock \showarticletitle{Privacy risks of securing machine learning models
  against adversarial examples}. In \bibinfo{booktitle}{\emph{Proceedings of
  the 2019 ACM SIGSAC Conference on Computer and Communications Security}}.
  \bibinfo{pages}{241--257}.
\newblock


\bibitem[Truex et~al\mbox{.}(2019)]%
        {truex2019demystifying}
\bibfield{author}{\bibinfo{person}{Stacey Truex}, \bibinfo{person}{Ling Liu},
  \bibinfo{person}{Mehmet~Emre Gursoy}, \bibinfo{person}{Lei Yu}, {and}
  \bibinfo{person}{Wenqi Wei}.} \bibinfo{year}{2019}\natexlab{}.
\newblock \showarticletitle{Demystifying membership inference attacks in
  machine learning as a service}.
\newblock \bibinfo{journal}{\emph{IEEE Transactions on Services Computing}}
  (\bibinfo{year}{2019}).
\newblock


\bibitem[Watson et~al\mbox{.}(2021)]%
        {watson2021importance}
\bibfield{author}{\bibinfo{person}{Lauren Watson}, \bibinfo{person}{Chuan Guo},
  \bibinfo{person}{Graham Cormode}, {and} \bibinfo{person}{Alex Sablayrolles}.}
  \bibinfo{year}{2021}\natexlab{}.
\newblock \showarticletitle{On the Importance of Difficulty Calibration in
  Membership Inference Attacks}.
\newblock \bibinfo{journal}{\emph{arXiv preprint arXiv:2111.08440}}
  (\bibinfo{year}{2021}).
\newblock


\bibitem[Xiao et~al\mbox{.}(2017)]%
        {fmnist}
\bibfield{author}{\bibinfo{person}{Han Xiao}, \bibinfo{person}{Kashif Rasul},
  {and} \bibinfo{person}{Roland Vollgraf}.} \bibinfo{year}{2017}\natexlab{}.
\newblock \bibinfo{title}{Fashion-MNIST: a Novel Image Dataset for Benchmarking
  Machine Learning Algorithms}.
\newblock
\newblock
\showeprint[arXiv]{cs.LG/1708.07747}~[cs.LG]


\bibitem[Ye et~al\mbox{.}(2021)]%
        {ye2021enhanced}
\bibfield{author}{\bibinfo{person}{Jiayuan Ye}, \bibinfo{person}{Aadyaa Maddi},
  \bibinfo{person}{Sasi~Kumar Murakonda}, \bibinfo{person}{Vincent
  Bindschaedler}, {and} \bibinfo{person}{Reza Shokri}.}
  \bibinfo{year}{2021}\natexlab{}.
\newblock \showarticletitle{Enhanced membership inference attacks against
  machine learning models}.
\newblock \bibinfo{journal}{\emph{arXiv preprint arXiv:2111.09679}}
  (\bibinfo{year}{2021}).
\newblock


\bibitem[Yeom et~al\mbox{.}(2018)]%
        {yeom2018privacy}
\bibfield{author}{\bibinfo{person}{Samuel Yeom}, \bibinfo{person}{Irene
  Giacomelli}, \bibinfo{person}{Matt Fredrikson}, {and} \bibinfo{person}{Somesh
  Jha}.} \bibinfo{year}{2018}\natexlab{}.
\newblock \showarticletitle{Privacy risk in machine learning: Analyzing the
  connection to overfitting}. In \bibinfo{booktitle}{\emph{2018 IEEE 31st
  Computer Security Foundations Symposium (CSF)}}. IEEE,
  \bibinfo{pages}{268--282}.
\newblock


\bibitem[Zou et~al\mbox{.}(2020)]%
        {zou2020privacy}
\bibfield{author}{\bibinfo{person}{Yang Zou}, \bibinfo{person}{Zhikun Zhang},
  \bibinfo{person}{Michael Backes}, {and} \bibinfo{person}{Yang Zhang}.}
  \bibinfo{year}{2020}\natexlab{}.
\newblock \showarticletitle{Privacy Analysis of Deep Learning in the Wild:
  Membership Inference Attacks against Transfer Learning}.
\newblock \bibinfo{journal}{\emph{arXiv preprint arXiv:2009.04872}}
  (\bibinfo{year}{2020}).
\newblock


\end{thebibliography}


\appendix

\section{Appendix}

\subsection{Natural Subpopulation}
\label{appendix-natural}
\begin{figure*}
\centering
\centering
\begin{tabular}{cc}
\subfloat[FPR/FPR plot]{\includegraphics[width=0.3\linewidth]{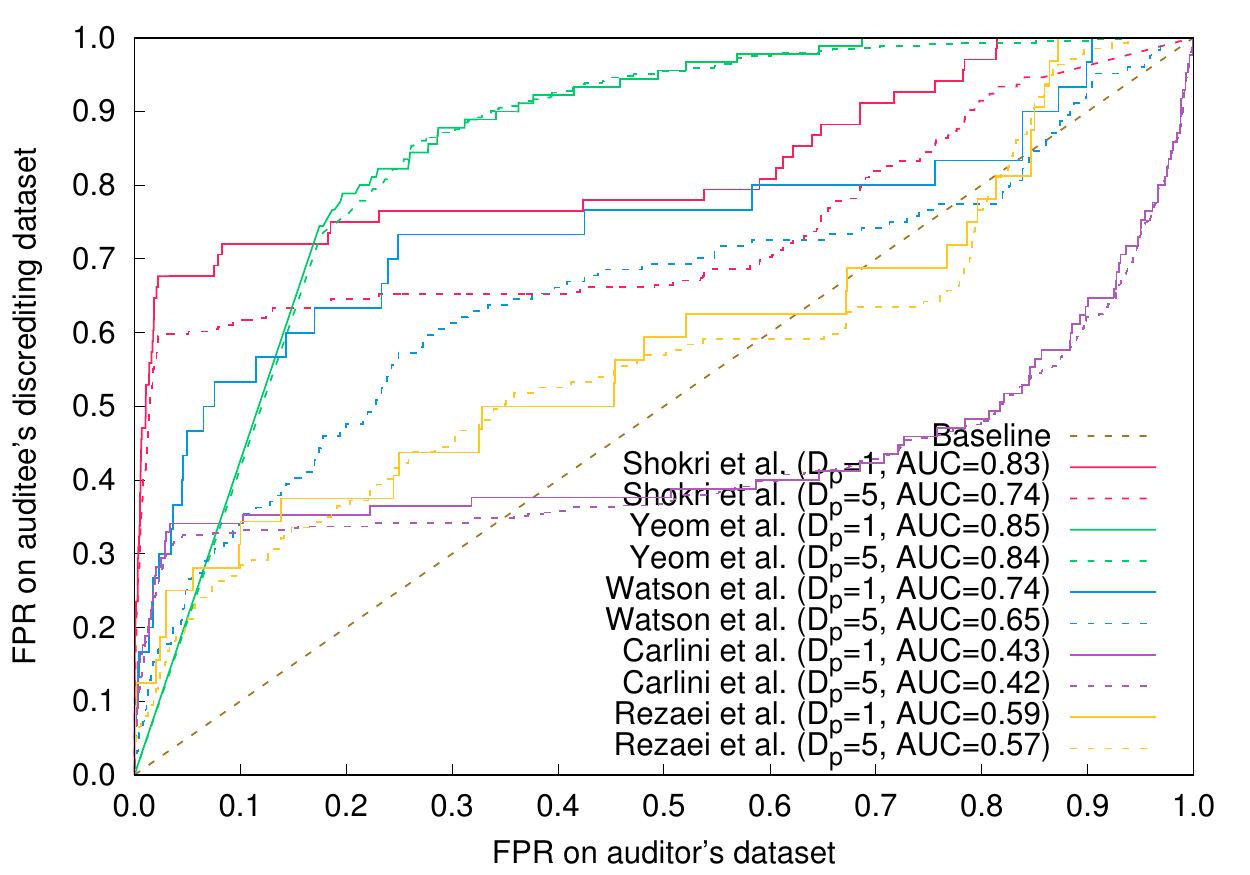}} 
& \subfloat[FPR/FPR logscale plot]{\includegraphics[width=0.3\linewidth]{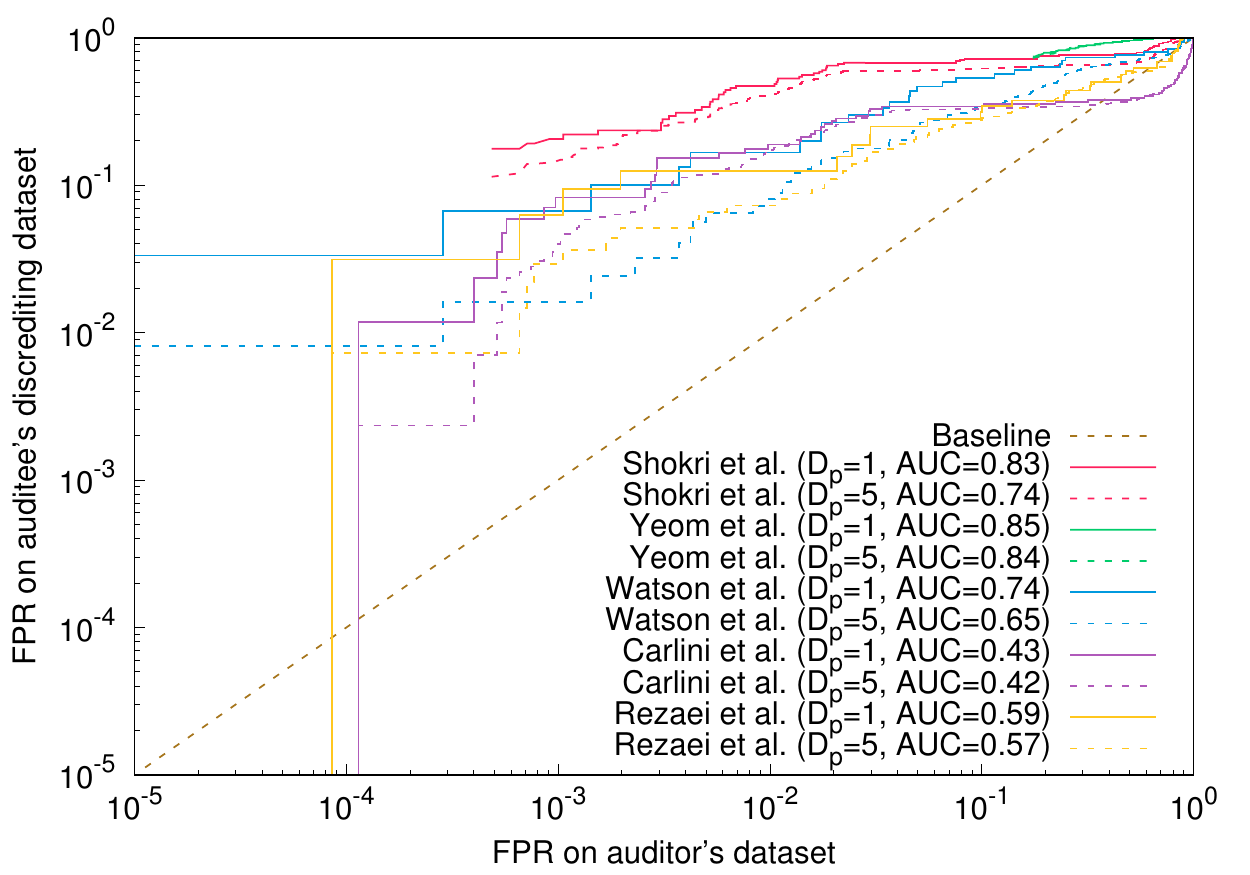}}\\
\end{tabular}
\caption{Cifar10/ResNet20 model. Searched on natural samples from CINIC.}
\label{fig-fpr-fpr-c10-resnet-natural}
\end{figure*}

\begin{figure*}
\centering
\centering
\begin{tabular}{cc}
\subfloat[FPR/FPR plot]{\includegraphics[width=0.3\linewidth]{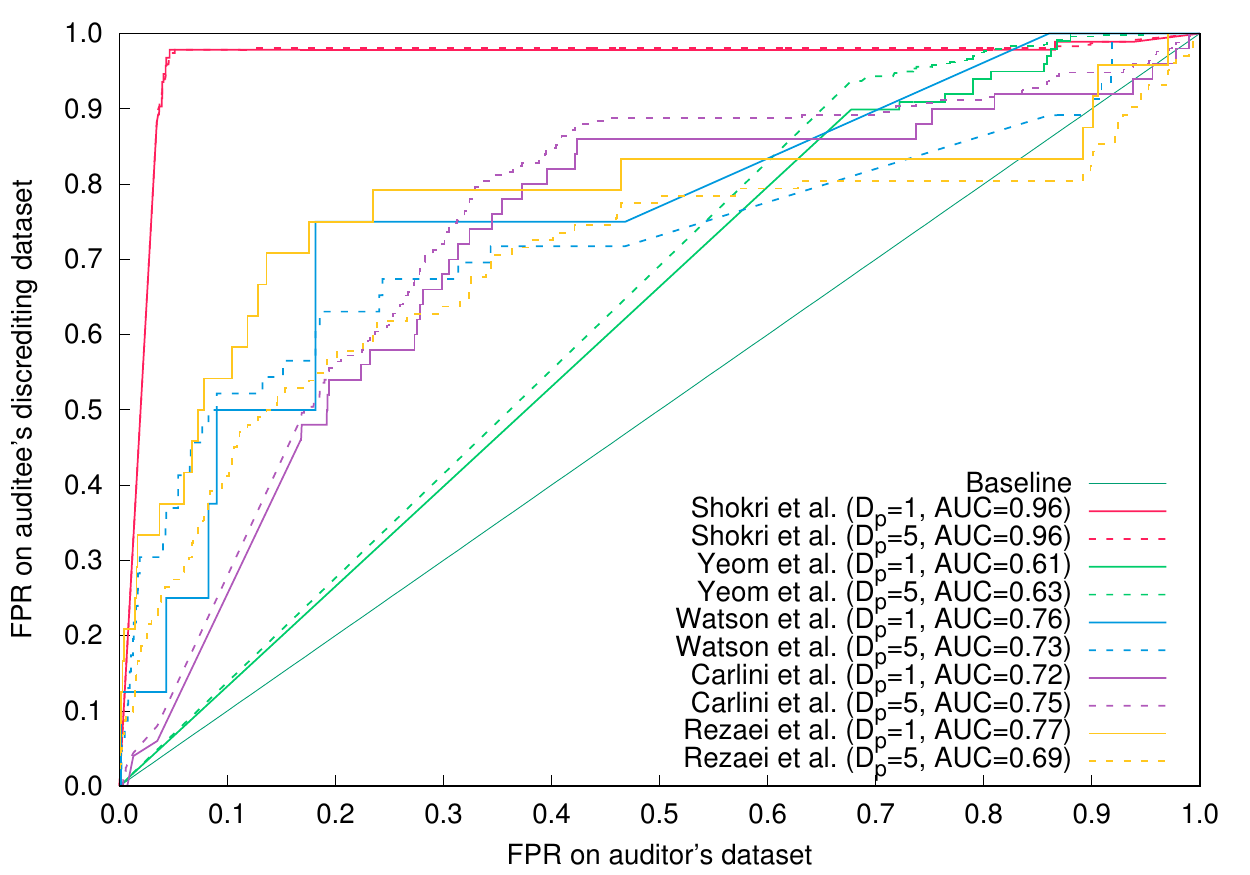}}
& \subfloat[FPR/FPR logscale plot]{\includegraphics[width=0.3\linewidth]{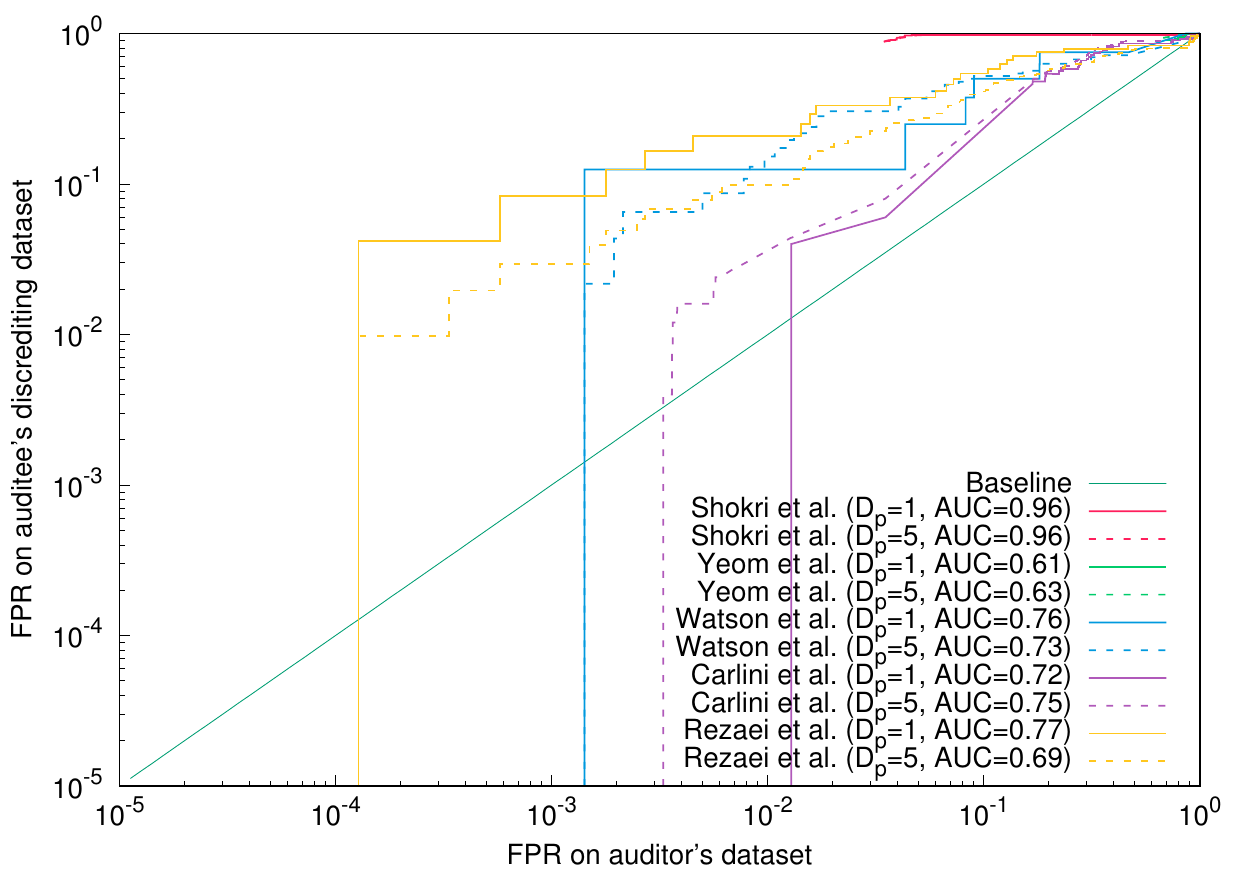}}\\
\end{tabular}
\caption{SVHN/LeNet model. Searched on natural samples from SVHN (extra).}
\label{fig-fpr-fpr-vhn-lenet-natural}
\end{figure*}

Figure \ref{fig-fpr-fpr-c10-resnet-natural} and \ref{fig-fpr-fpr-vhn-lenet-natural} shows the false positive ratio to false positive ratio of Algorithm \ref{algo-search} on ResNet and LeNet, trained on CIFAR10 and SVHN, respectively.

\subsection{Crafted Subpopulation}
\label{appendix-bigan}

\begin{figure*}
\centering
\centering
\begin{tabular}{cc}
\subfloat[FPR/FPR plot]{\includegraphics[width=0.3\linewidth]{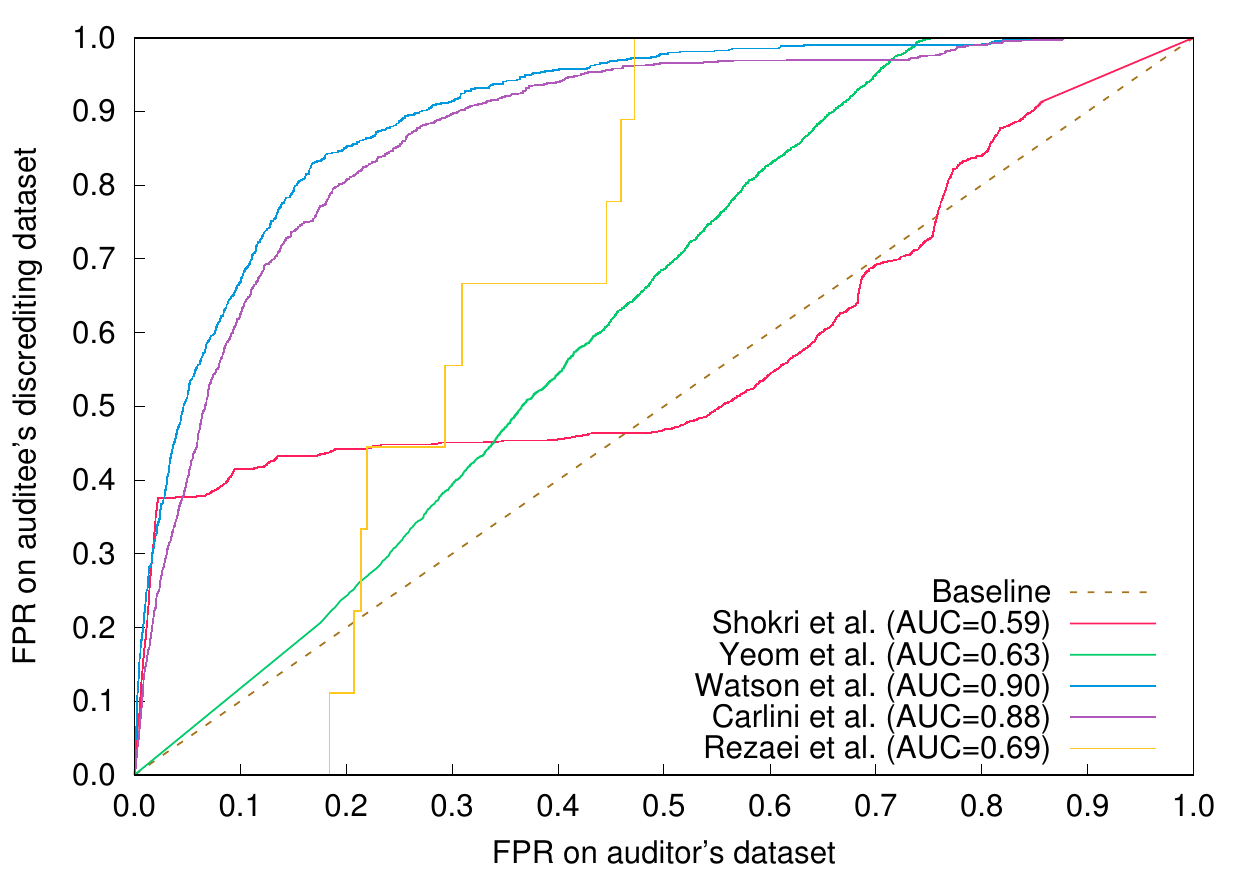}} 
& \subfloat[FPR/FPR logscale plot]{\includegraphics[width=0.3\linewidth]{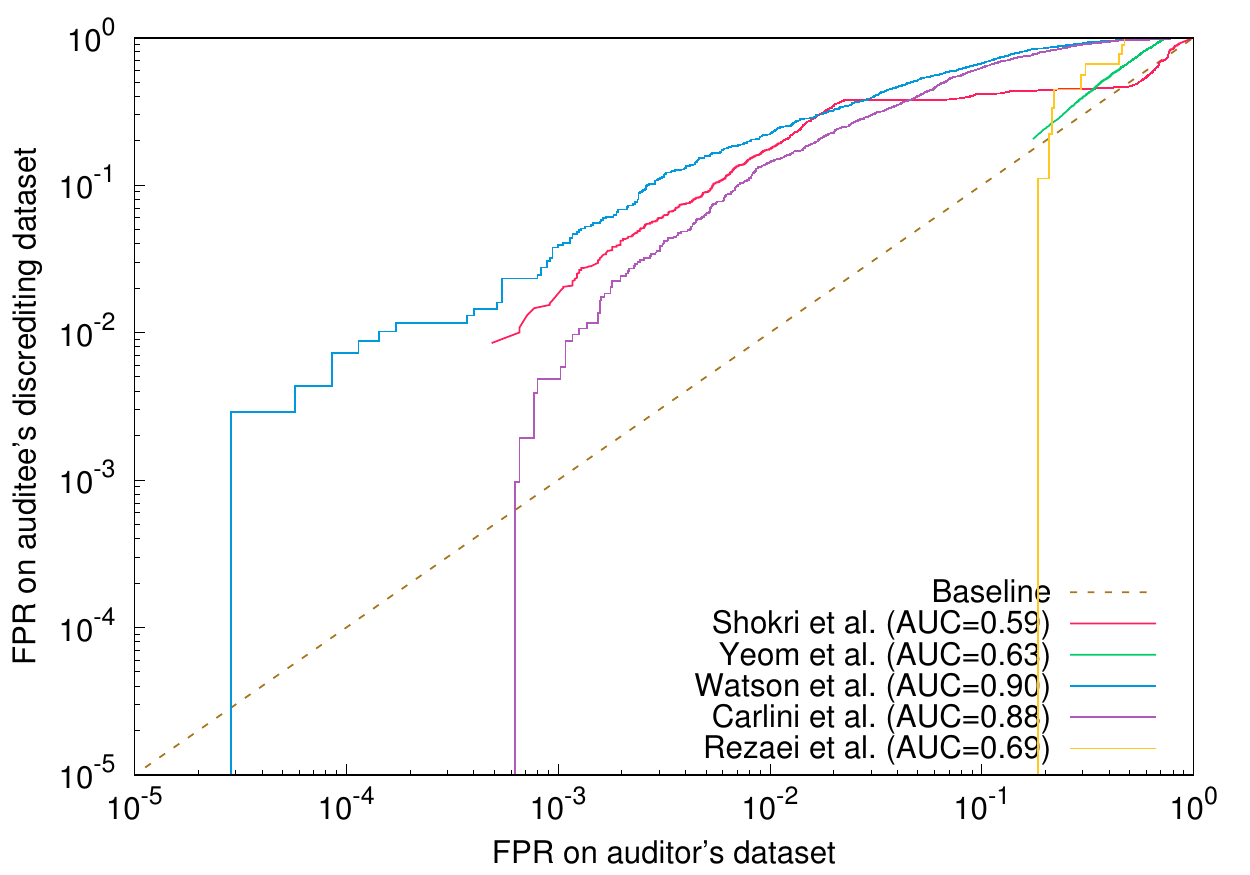}}\\
\end{tabular}
\caption{Cifar10/ResNet20 model. On crafted samples using BiGAN}
\label{fig-fpr-fpr-c10-resnet-bigan}
\end{figure*}

\begin{figure*}
\centering
\centering
\begin{tabular}{cc}
\subfloat[FPR/FPR plot]{\includegraphics[width=0.3\linewidth]{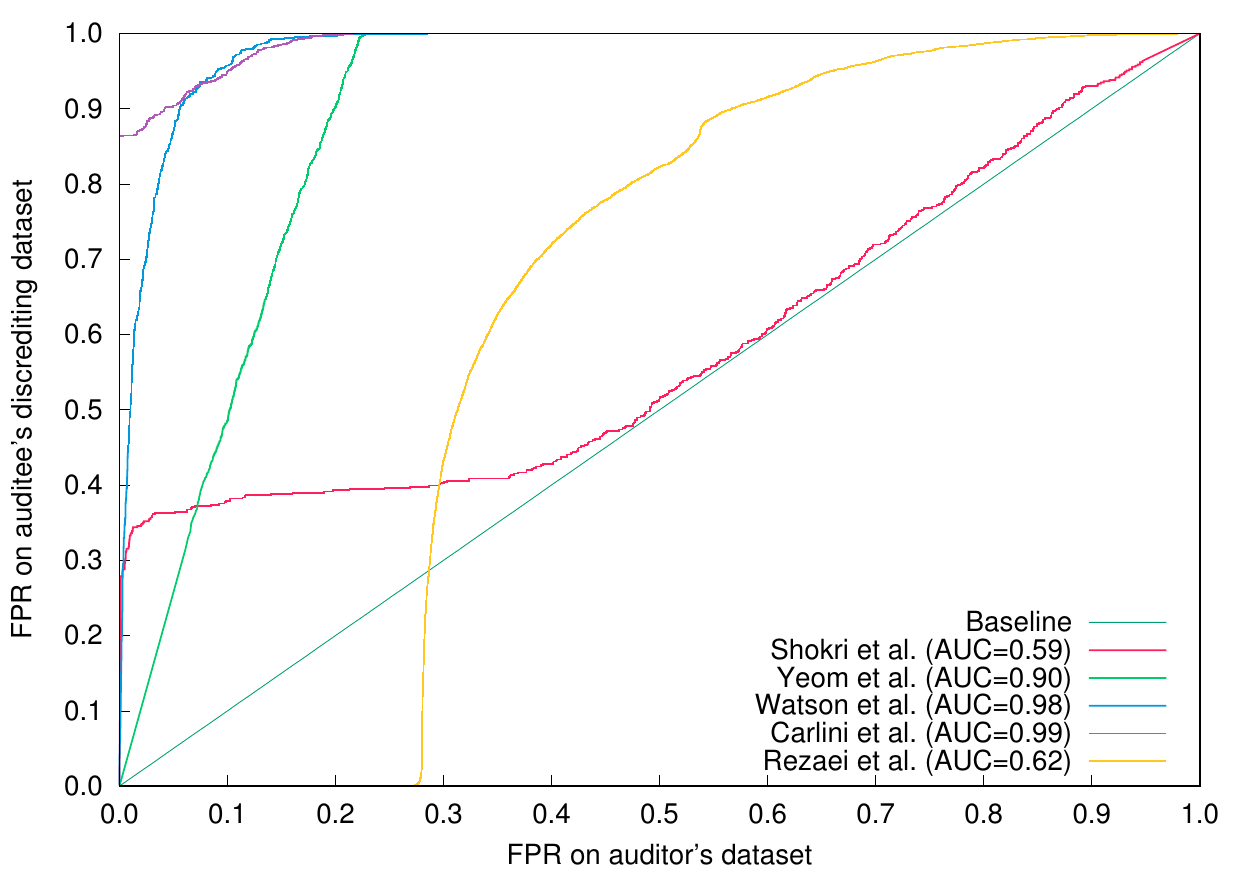}} 
& \subfloat[FPR/FPR logscale plot]{\includegraphics[width=0.3\linewidth]{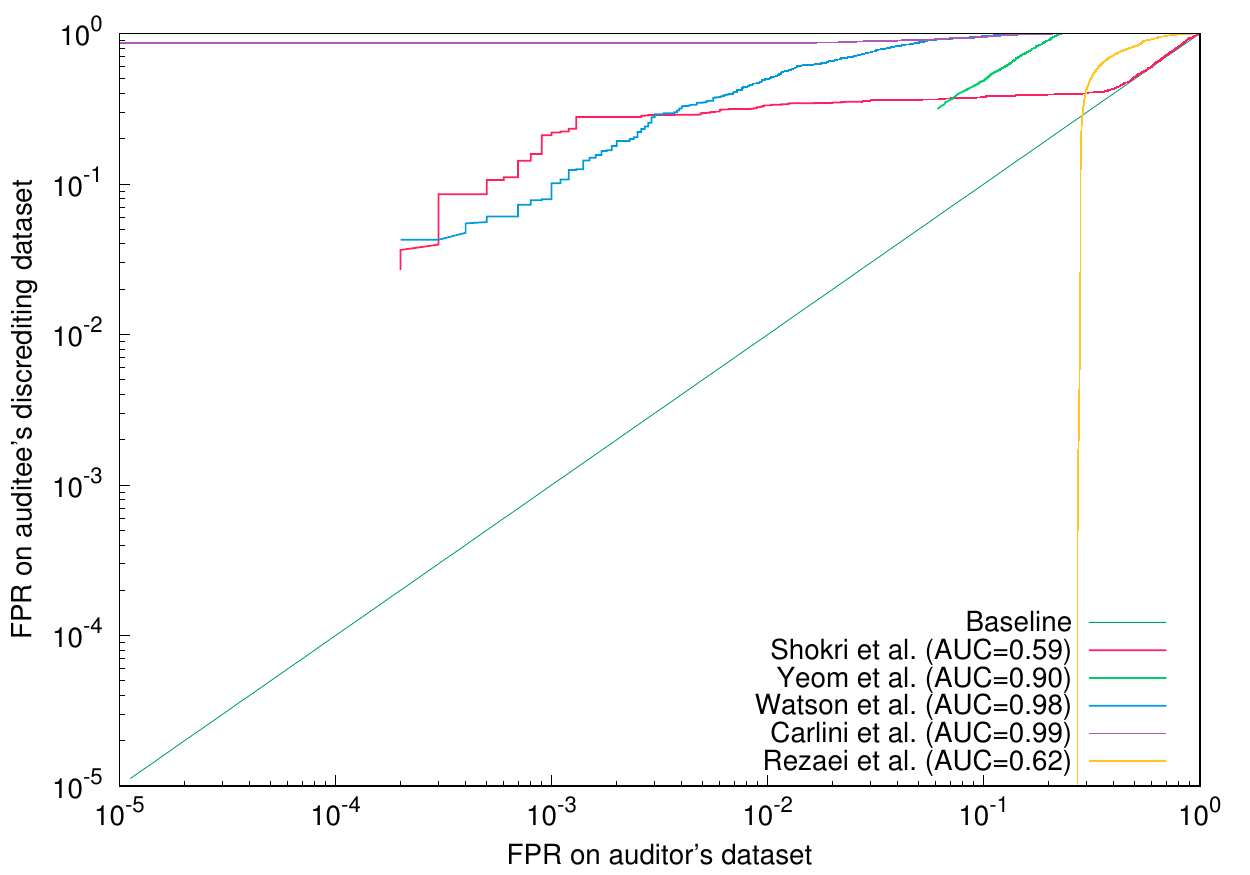}}\\
\end{tabular}
\caption{Cifar100/LeNet model. On crafted samples using BiGAN}
\label{fig-fpr-fpr-c100-lenet-bigan}
\end{figure*}

\begin{figure*}
\centering
\centering
\begin{tabular}{cc}
\subfloat[FPR/FPR plot]{\includegraphics[width=0.3\linewidth]{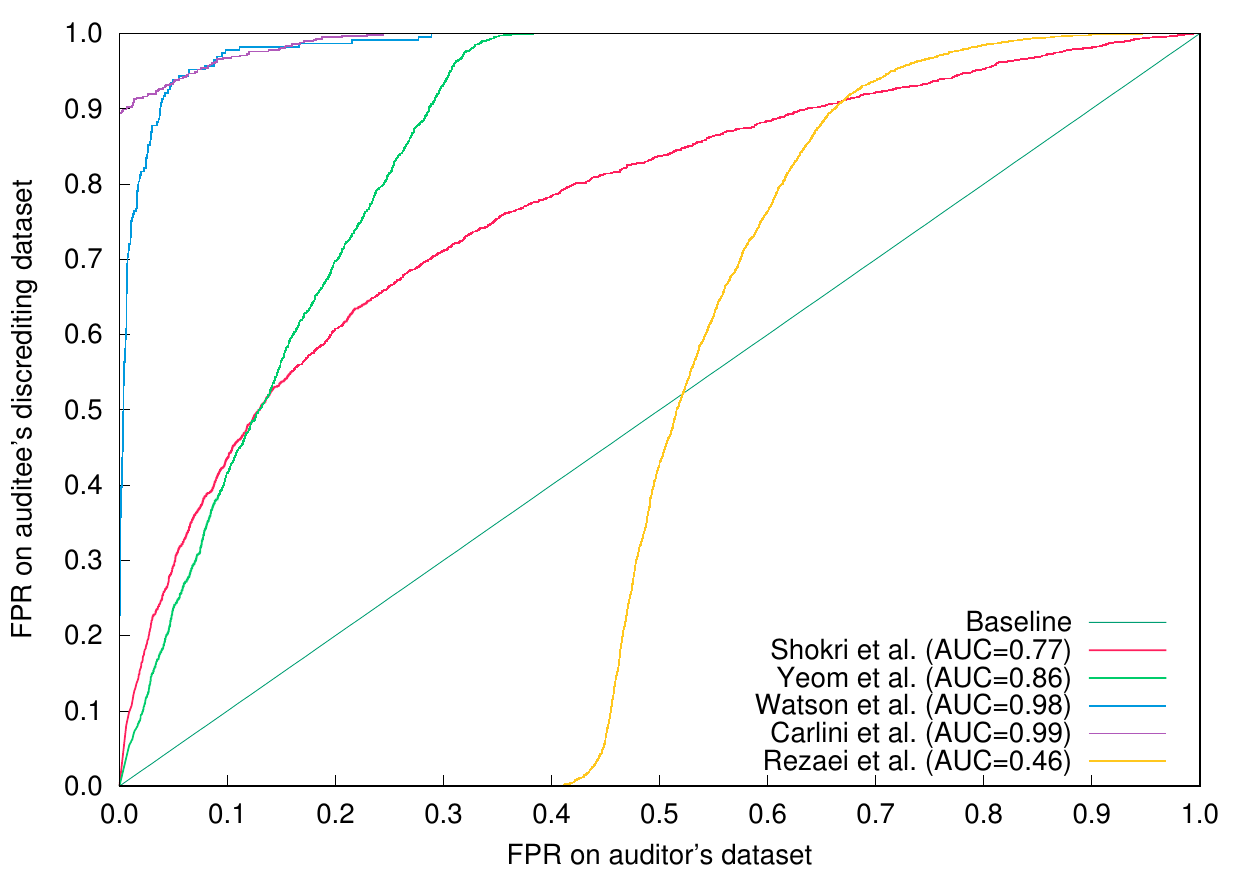}} 
& \subfloat[FPR/FPR logscale plot]{\includegraphics[width=0.3\linewidth]{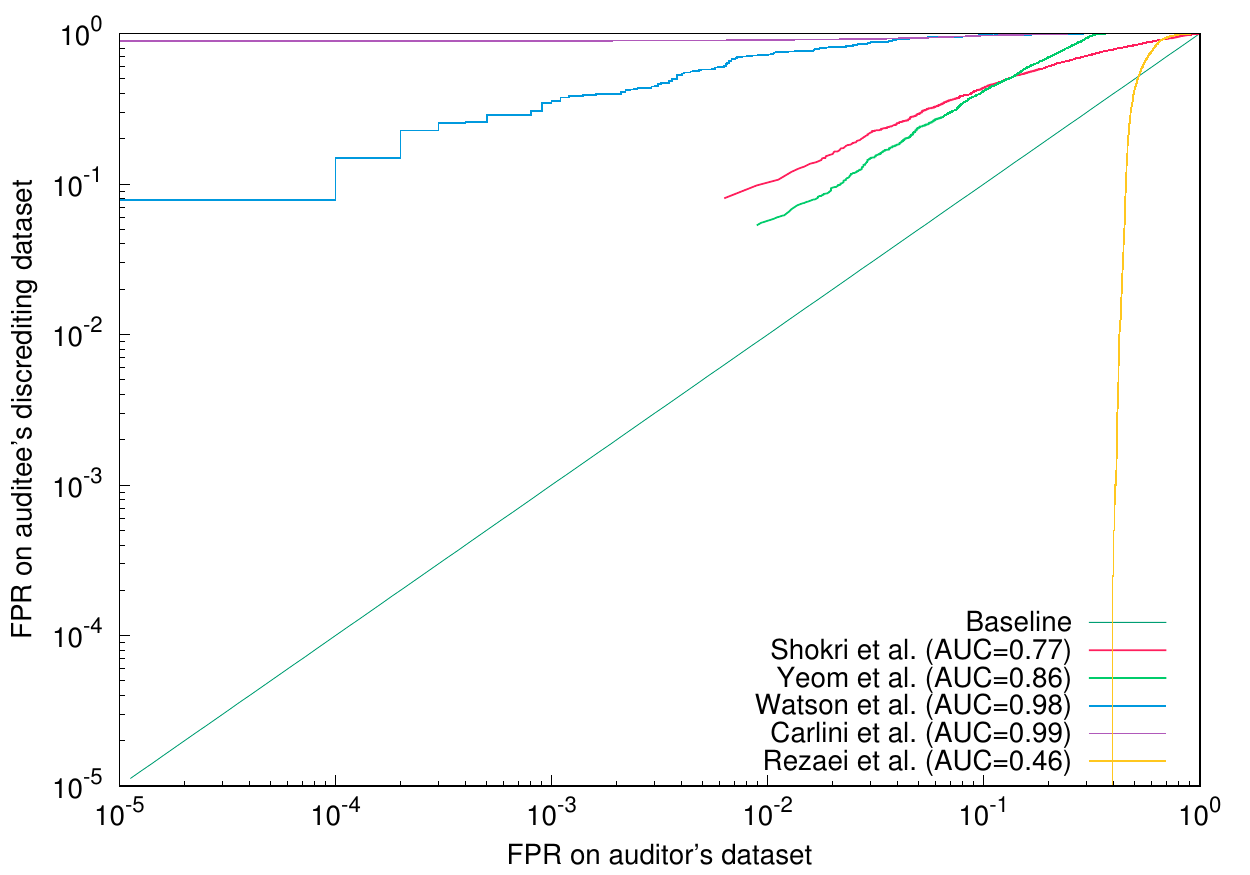}}\\
\end{tabular}
\caption{Cifar100/ResNet20 model. On crafted samples using BiGAN}
\label{fig-fpr-fpr-c100-resnet-bigan}
\end{figure*}

\begin{figure*}
\centering
\centering
\begin{tabular}{cc}
\subfloat[FPR/FPR plot]{\includegraphics[width=0.3\linewidth]{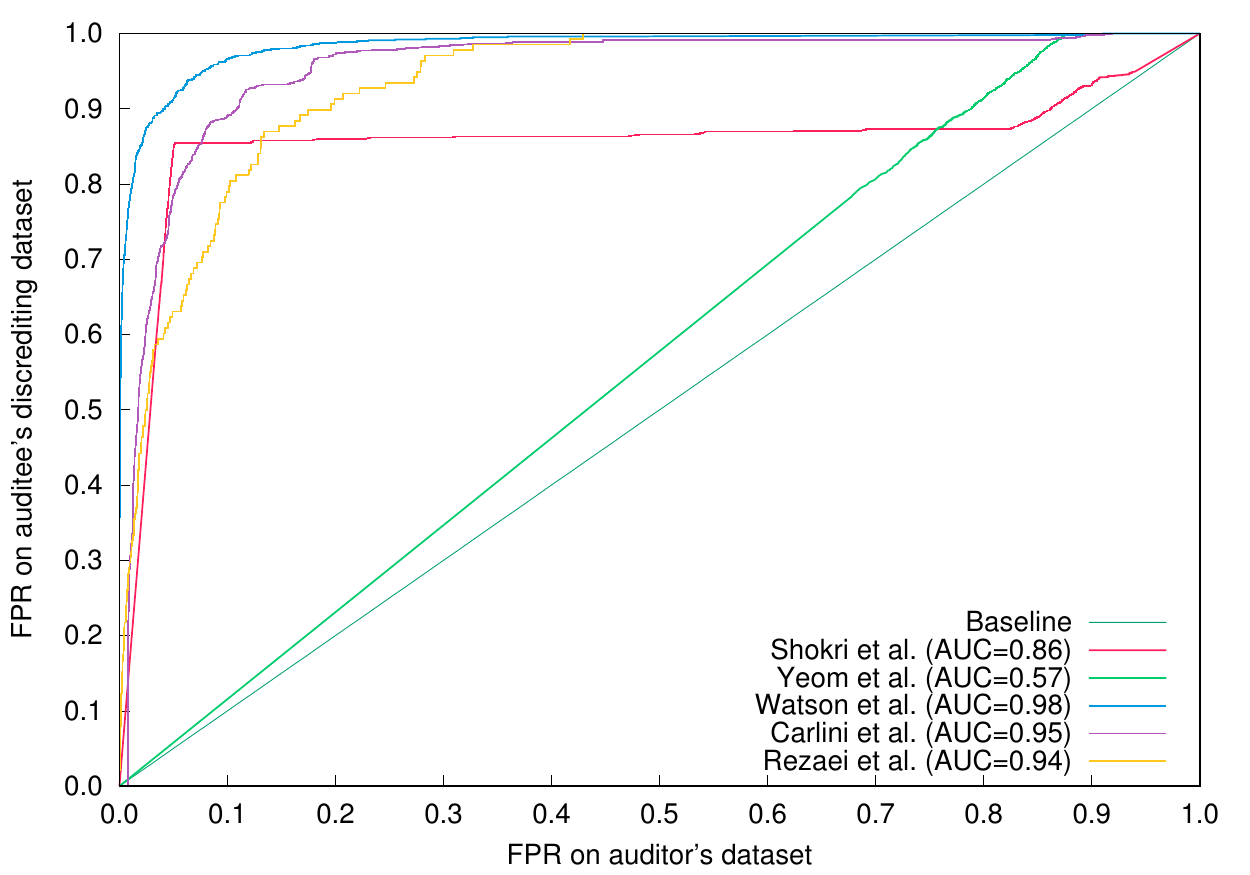}} 
& \subfloat[FPR/FPR logscale plot]{\includegraphics[width=0.3\linewidth]{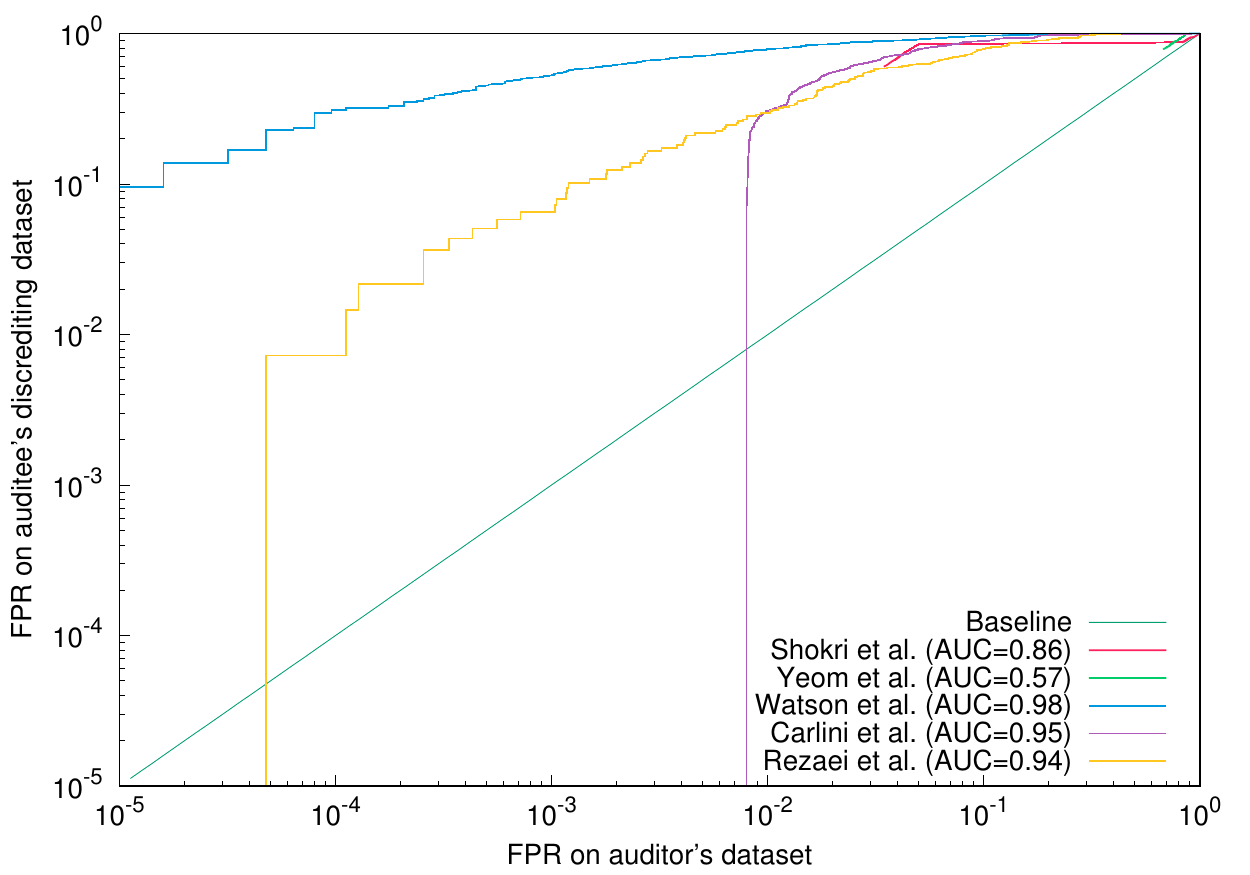}}\\
\end{tabular}
\caption{SVHN/LeNet model. On crafted samples using BiGAN}
\label{fig-fpr-fpr-svhn-lenet-bigan}
\end{figure*}

\begin{figure*}
\centering
\centering
\begin{tabular}{cc}
\subfloat[FPR/FPR plot]{\includegraphics[width=0.3\linewidth]{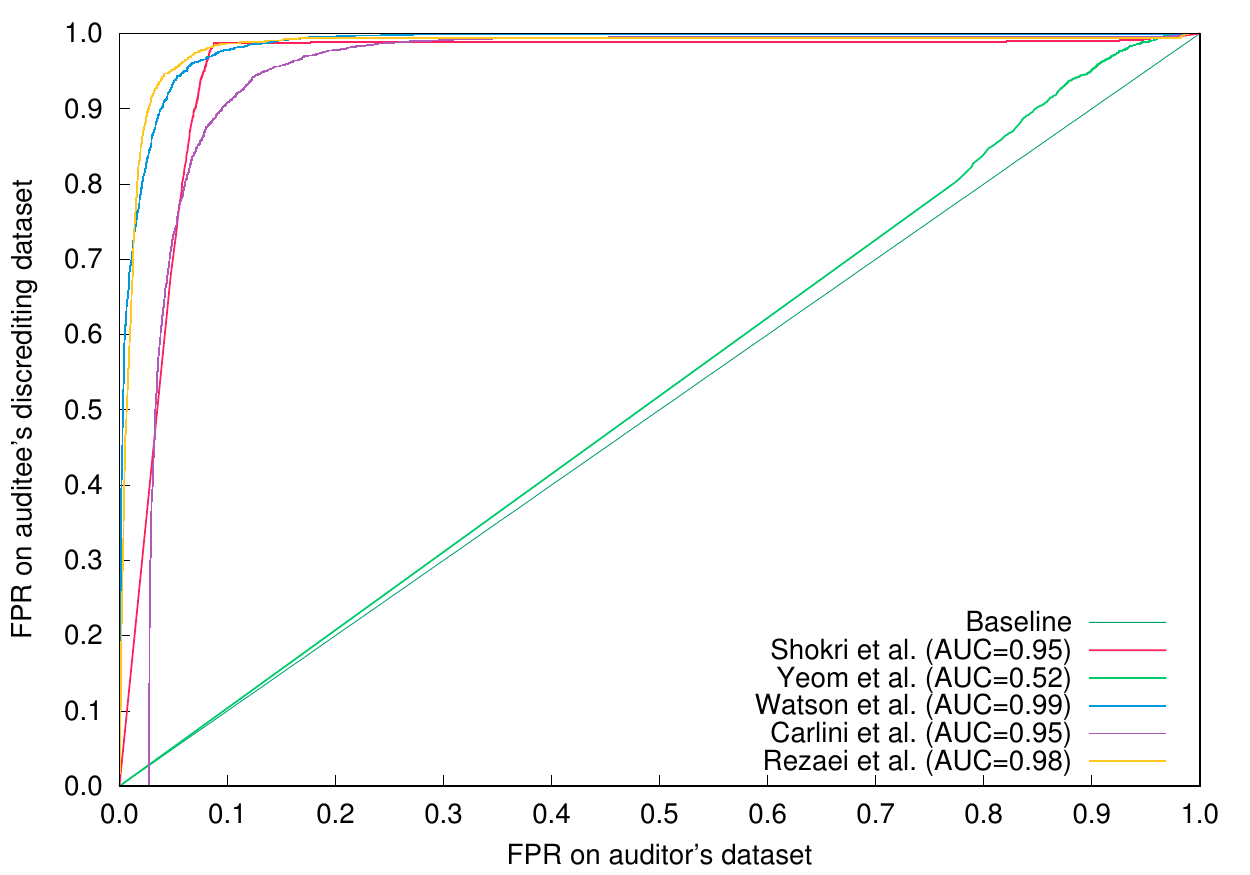}} 
& \subfloat[FPR/FPR logscale plot]{\includegraphics[width=0.3\linewidth]{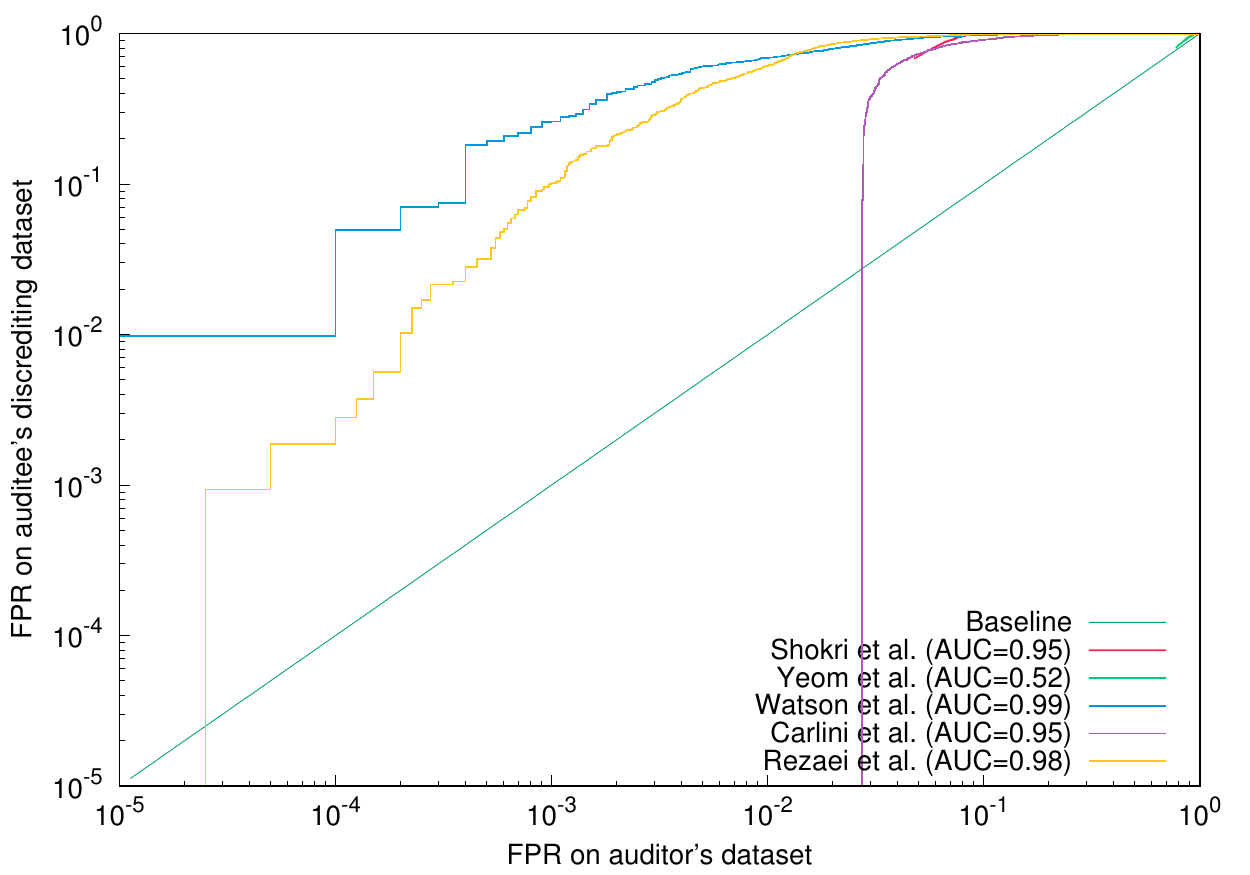}}\\
\end{tabular}
\caption{MNIST/MLP model. On crafted samples using BiGAN}
\label{fig-fpr-fpr-mnist-mlp-bigan}
\end{figure*}

\begin{figure*}
\centering
\centering
\begin{tabular}{cc}
\subfloat[FPR/FPR plot]{\includegraphics[width=0.3\linewidth]{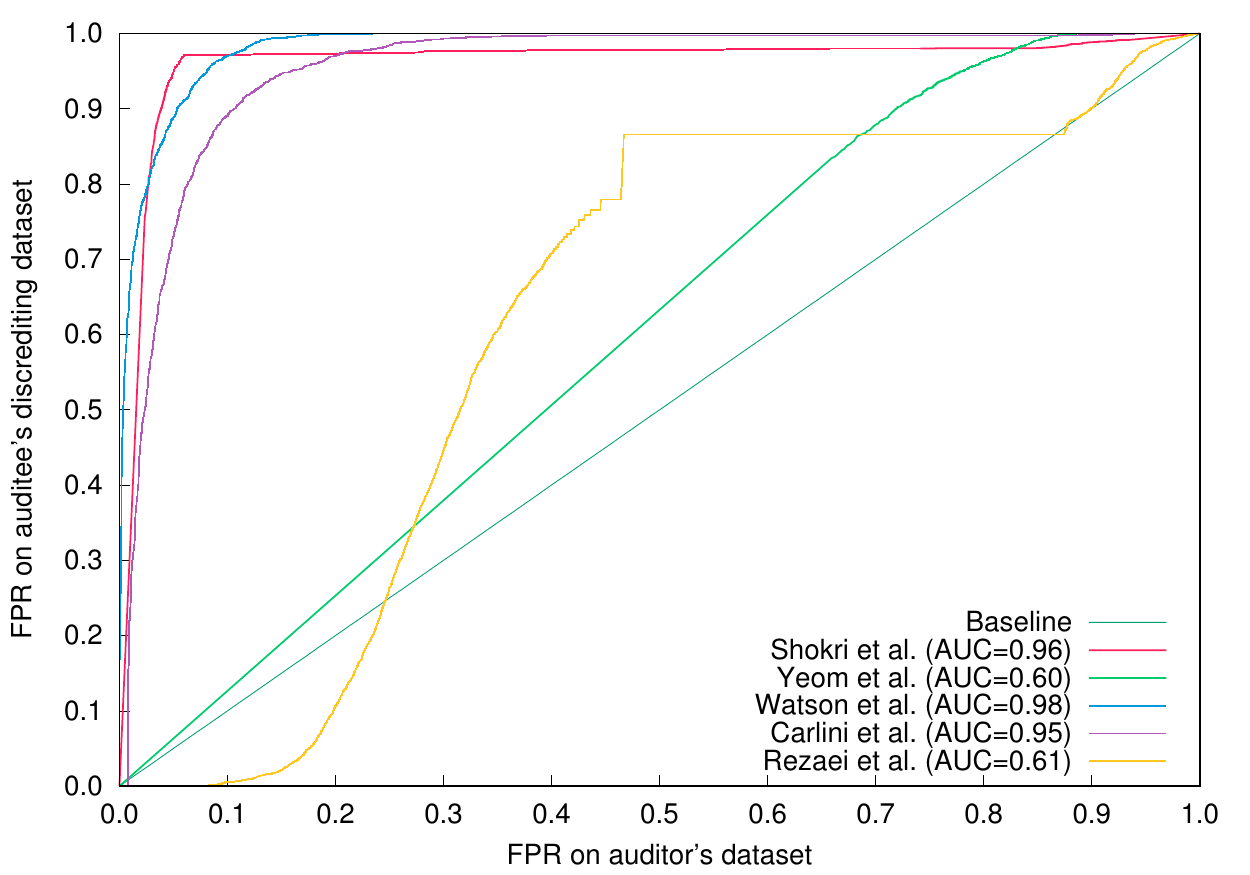}} 
& \subfloat[FPR/FPR logscale plot]{\includegraphics[width=0.3\linewidth]{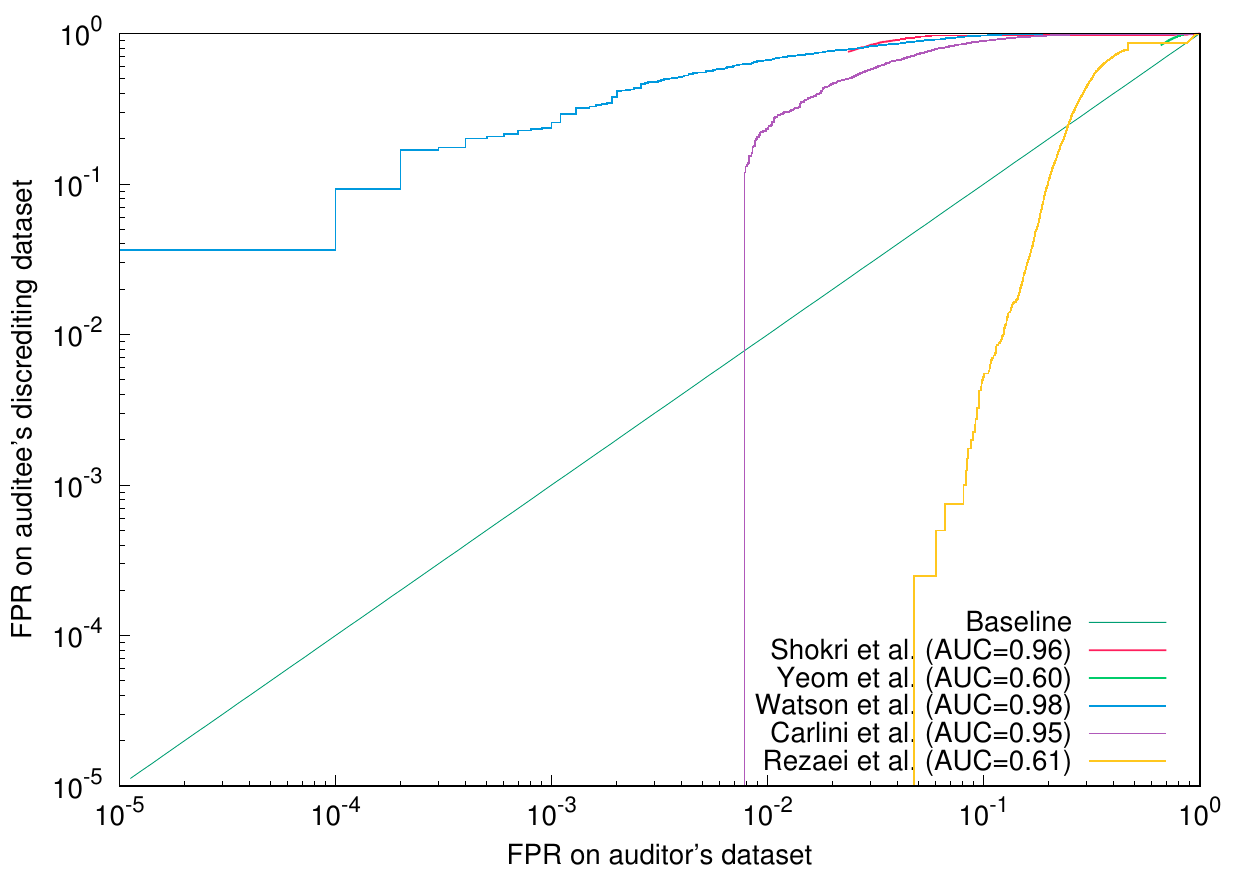}}\\
\end{tabular}
\caption{FMNIST/MLP model. On crafted samples using BiGAN}
\label{fig-fpr-fpr-fmnist-mlp-bigan}
\end{figure*}

Figure \ref{fig-fpr-fpr-c10-resnet-bigan} through \ref{fig-fpr-fpr-fmnist-mlp-bigan} shows the false positive ratio to false positive ratio of Algorithm \ref{algo-bigan} on several models/datasets.

\subsection{Adversarially Tuned Subpopulation}
\label{appendix-adv}
Figure \ref{fig-fpr-fpr-c10-renset-adv} through \ref{fig-fpr-fpr-fmnist-mlp-adv} shows the false positive ratio to false positive ratio of Algorithm \ref{algo-adv} on several models/datasets.

\begin{figure*}
\centering
\begin{tabular}{cc}
\subfloat[FPR/FPR plot]{\includegraphics[width=0.3\linewidth]{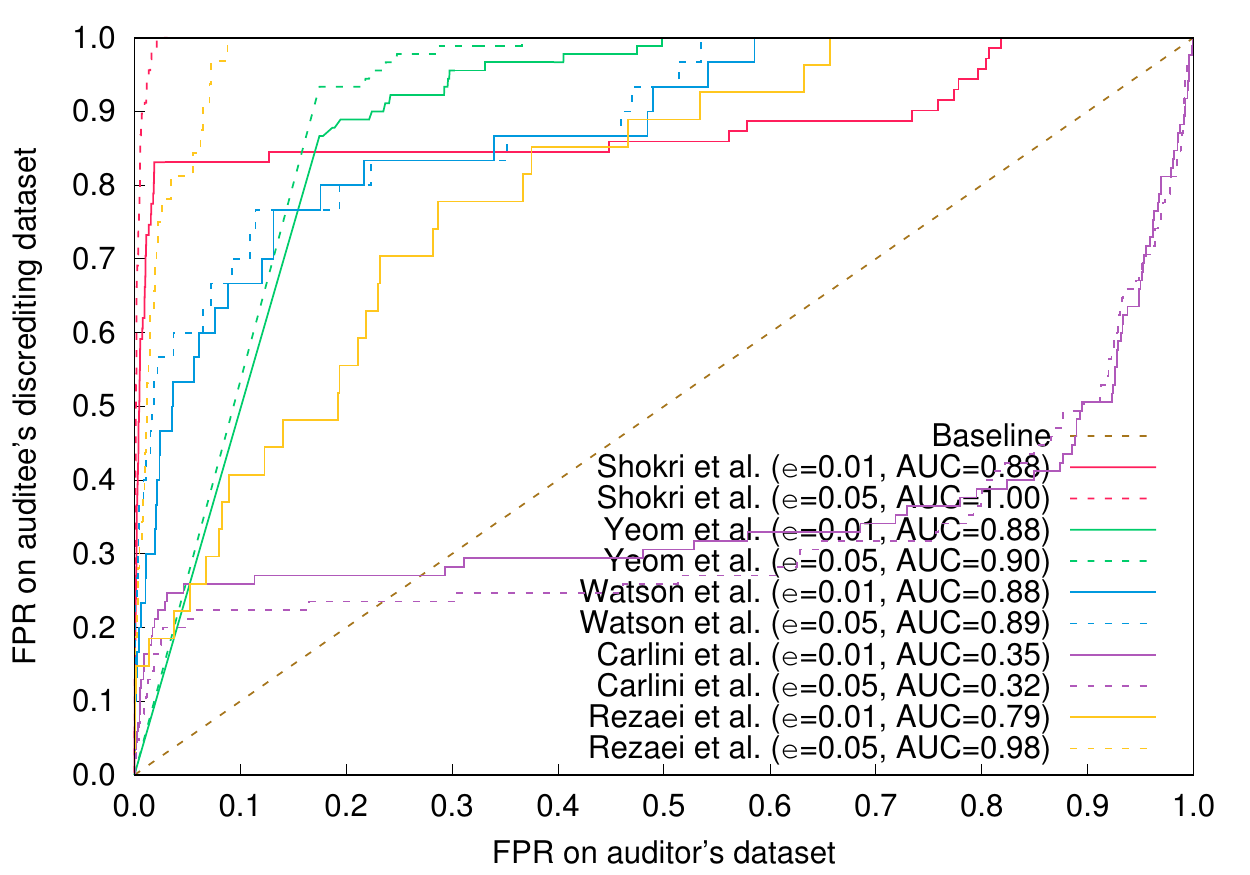}} 
& \subfloat[FPR/FPR logscale plot]{\includegraphics[width=0.3\linewidth]{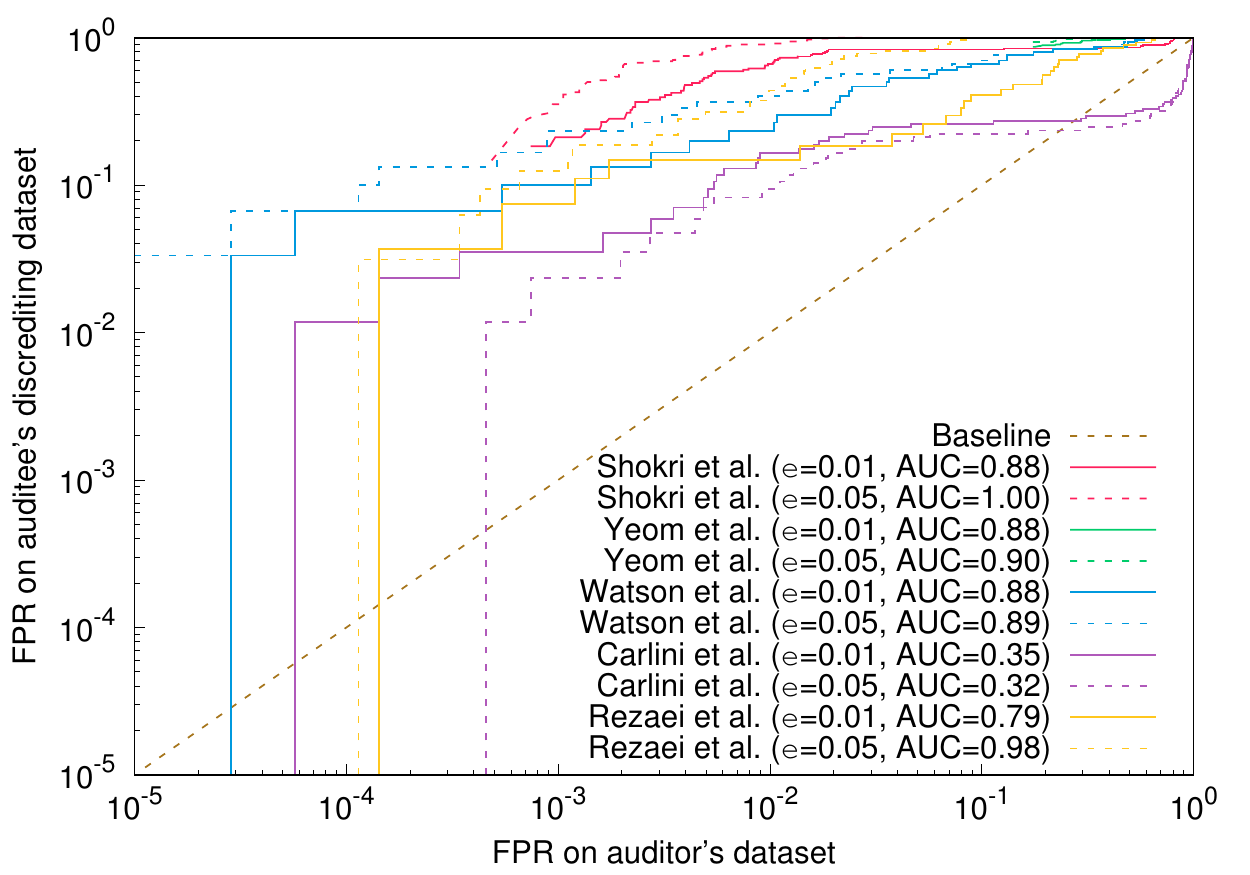}}\\
\end{tabular}
\caption{Cifar10/ResNet20 model. On adversarially tuned samples.}
\label{fig-fpr-fpr-c10-renset-adv}
\end{figure*}

\begin{figure*}
\centering
\centering
\begin{tabular}{cc}
\subfloat[FPR/FPR plot]{\includegraphics[width=0.3\linewidth]{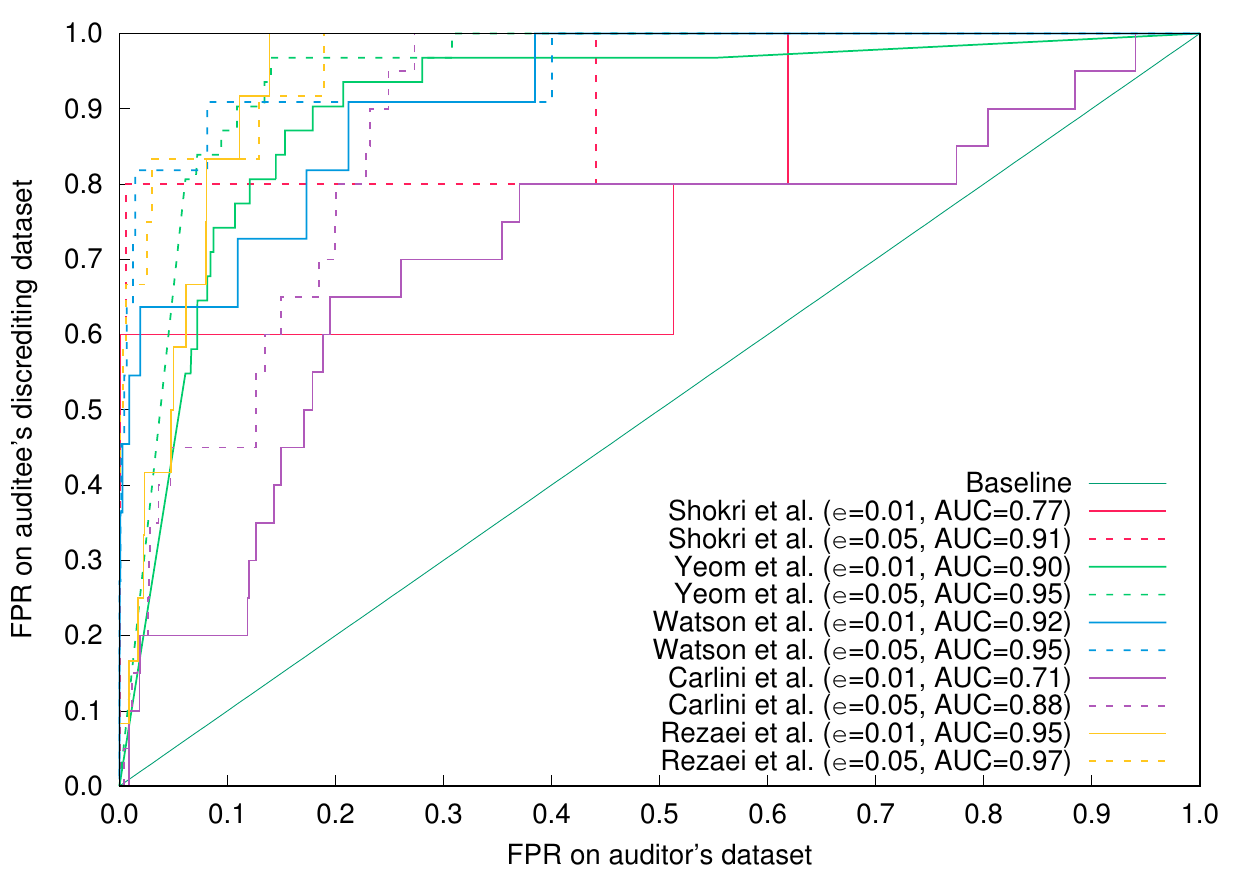}} 
& \subfloat[FPR/FPR logscale plot]{\includegraphics[width=0.3\linewidth]{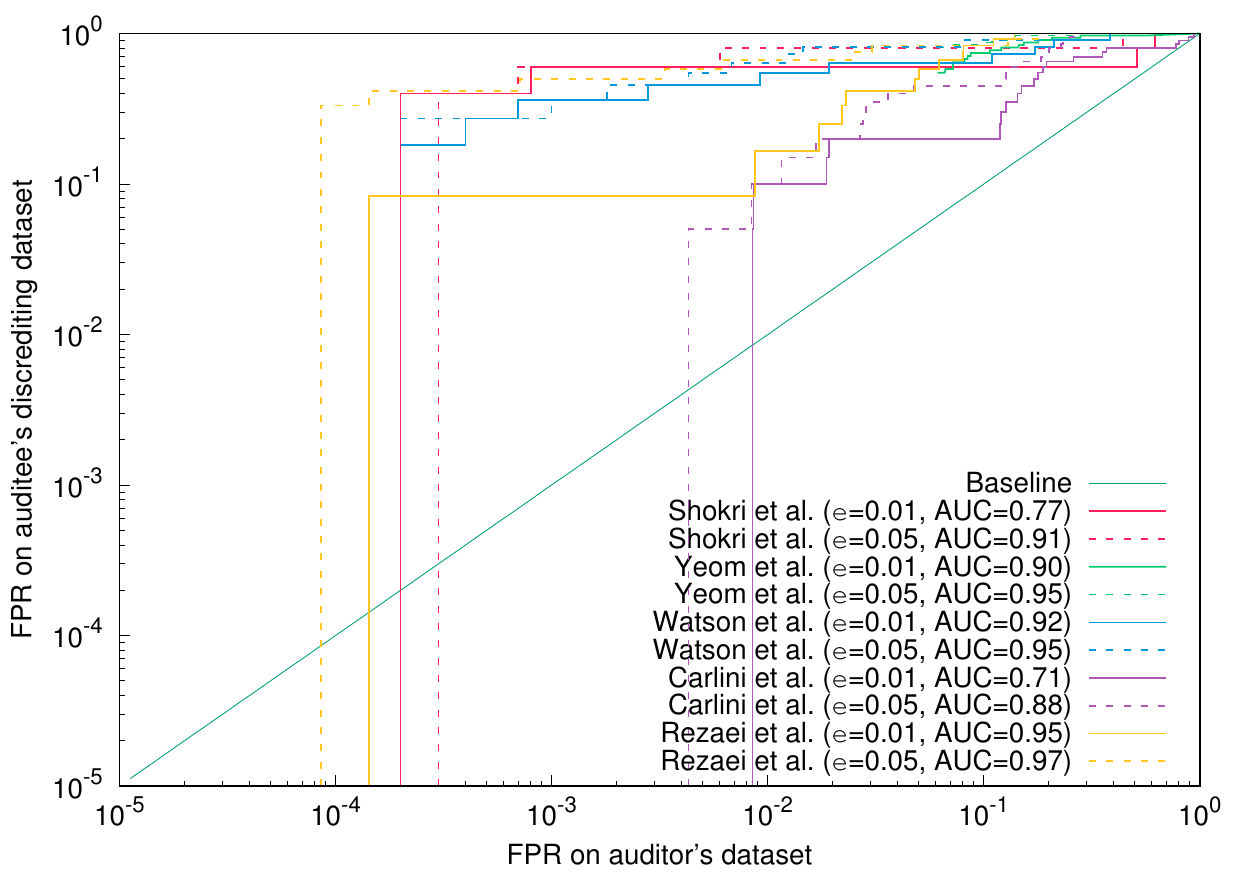}}\\
\end{tabular}
\caption{Cifar100/LeNet model. On adversarially tuned samples.}
\label{fig-fpr-fpr-c100-lenet-adv}
\end{figure*}

\begin{figure*}
\centering
\centering
\begin{tabular}{cc}
\subfloat[FPR/FPR plot]{\includegraphics[width=0.3\linewidth]{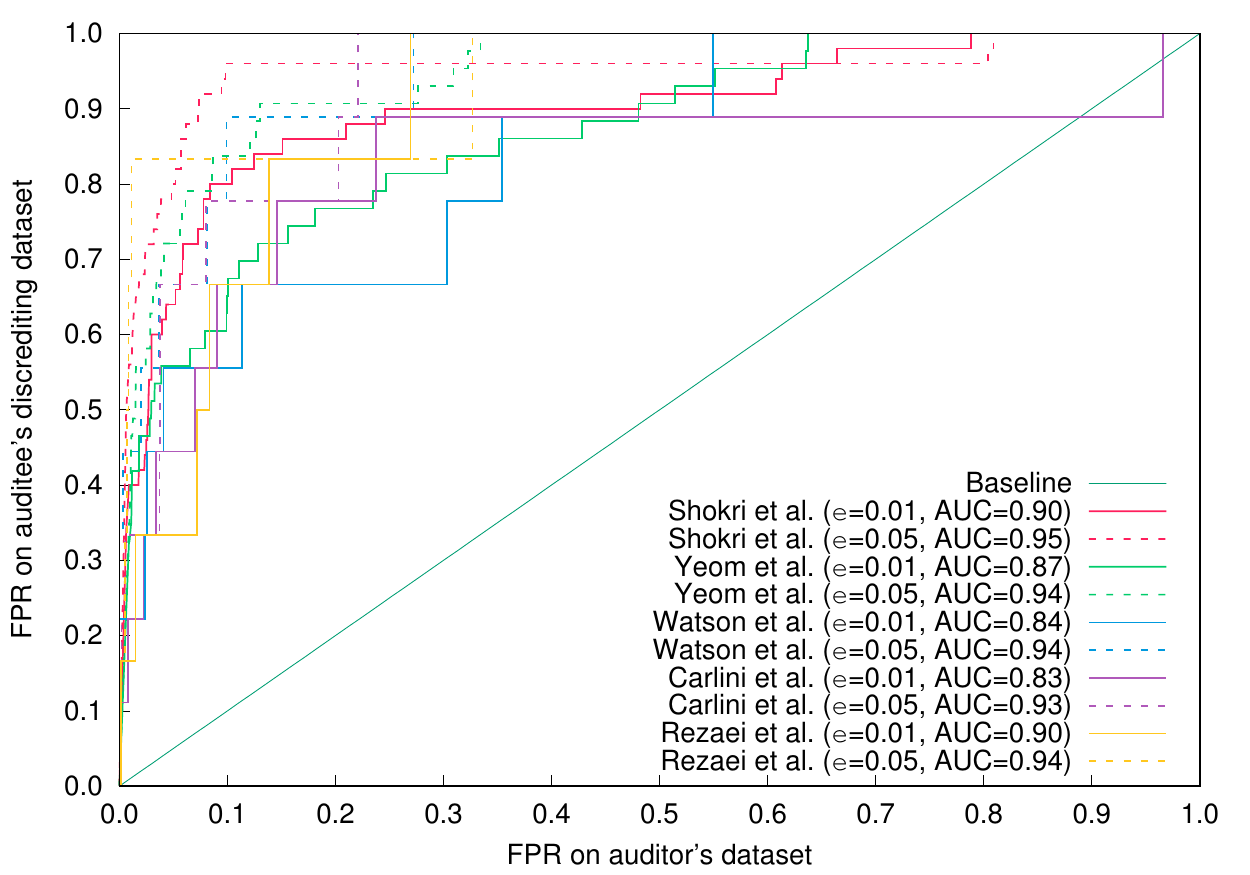}} 
& \subfloat[FPR/FPR logscale plot]{\includegraphics[width=0.3\linewidth]{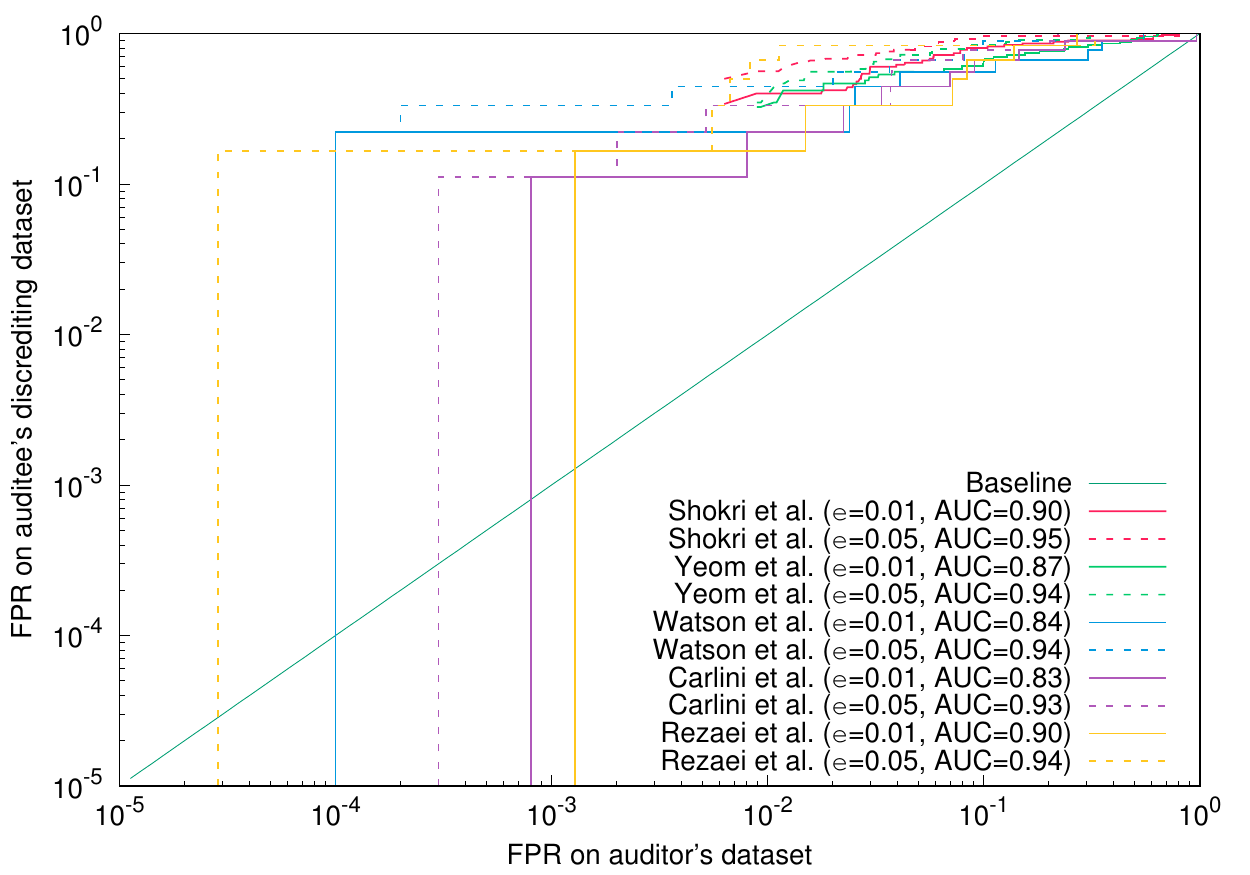}}\\
\end{tabular}
\caption{Cifar100/ResNet20 model. On adversarially tuned samples.}
\label{fig-fpr-fpr-c100-renset-adv}
\end{figure*}

\begin{figure*}
\centering
\centering
\begin{tabular}{cc}
\subfloat[FPR/FPR plot]{\includegraphics[width=0.3\linewidth]{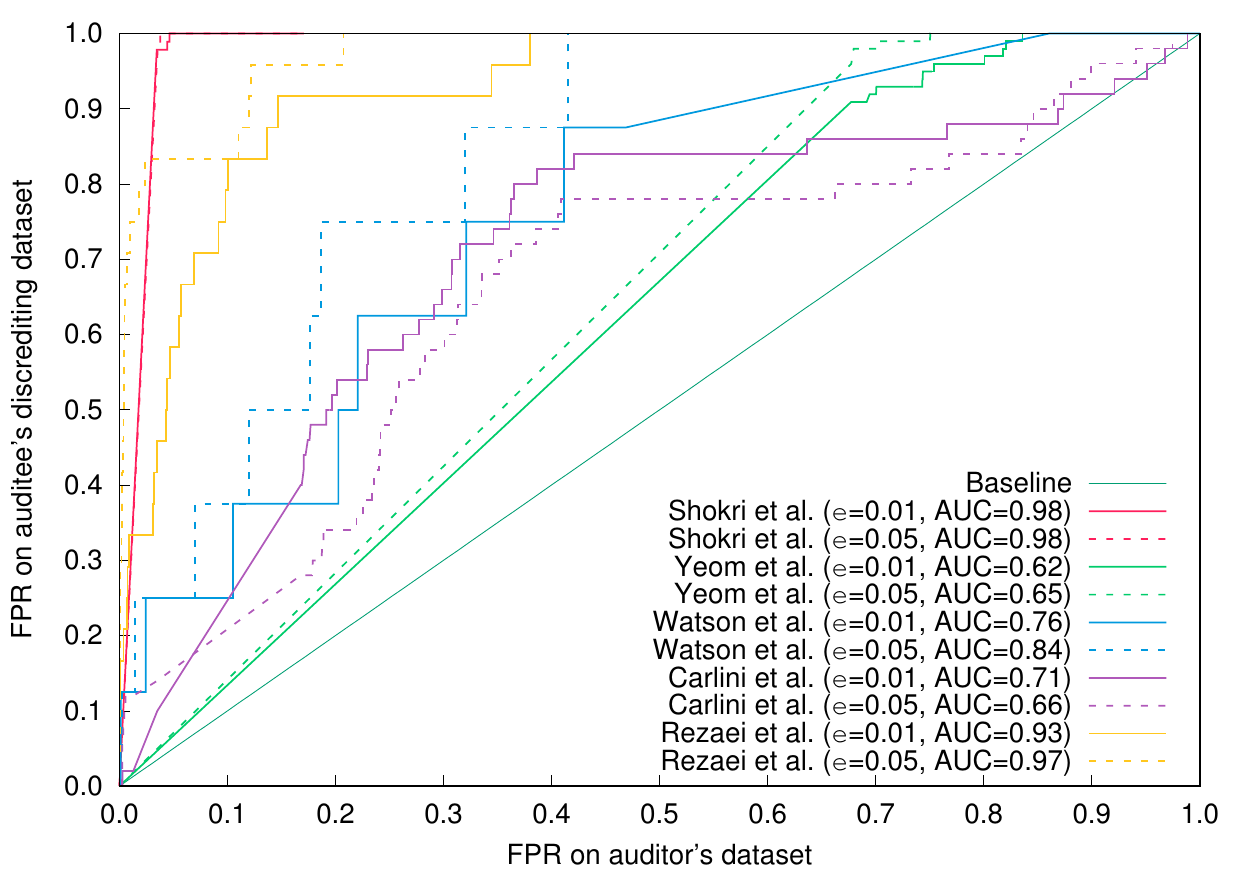}} 
& \subfloat[FPR/FPR logscale plot]{\includegraphics[width=0.3\linewidth]{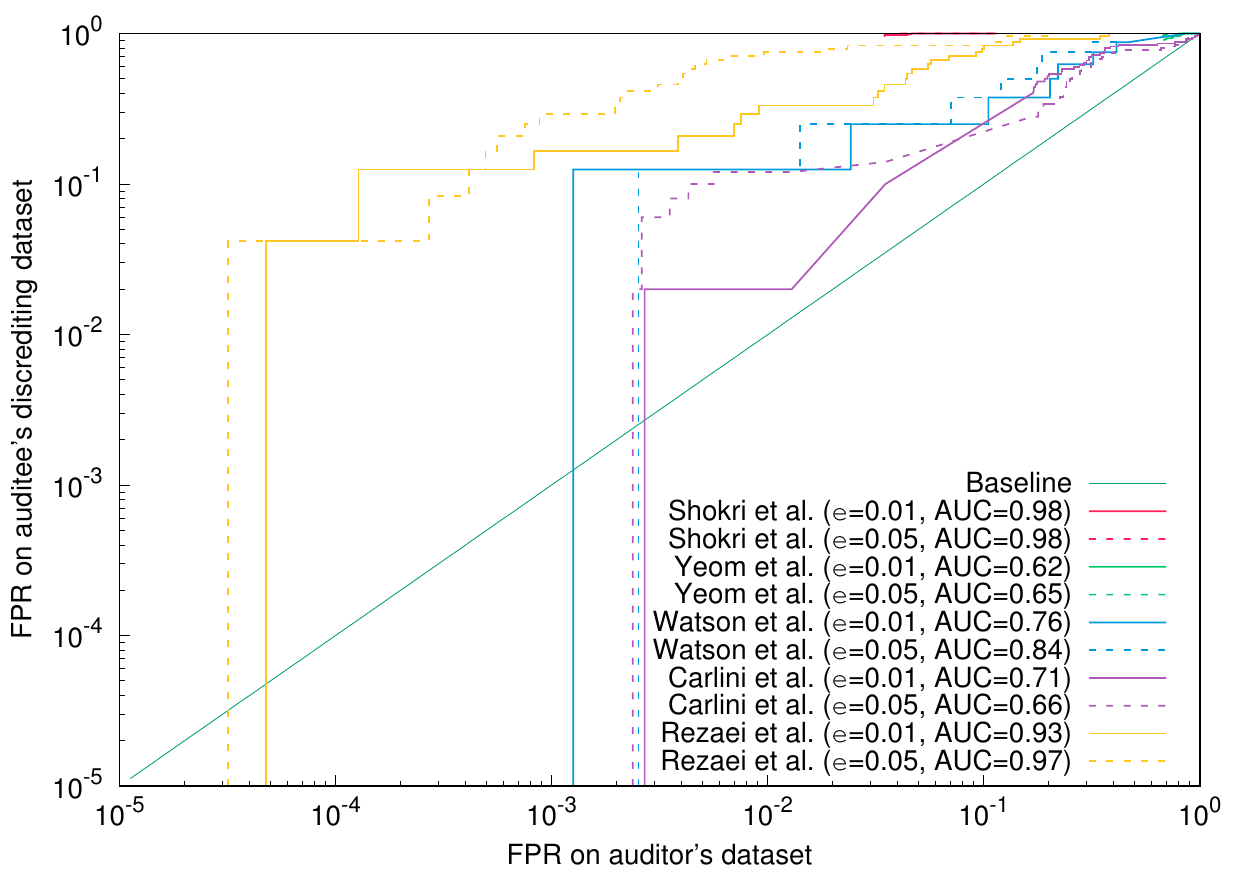}}\\
\end{tabular}
\caption{SVHN/LeNet model. On adversarially tuned samples.}
\label{fig-fpr-fpr-svhn-lenet-adv}
\end{figure*}

\begin{figure*}
\centering
\centering
\begin{tabular}{cc}
\subfloat[FPR/FPR plot]{\includegraphics[width=0.3\linewidth]{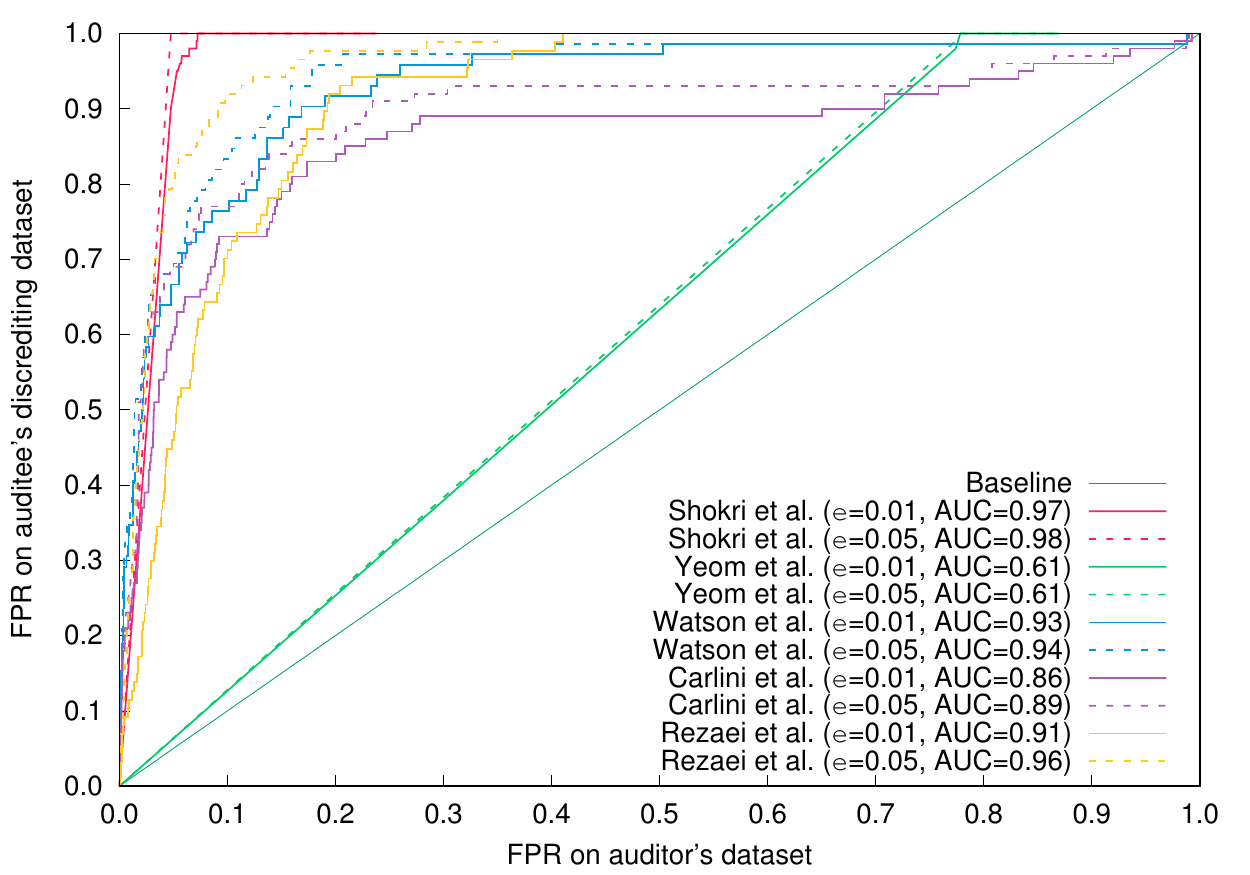}} 
& \subfloat[FPR/FPR logscale plot]{\includegraphics[width=0.3\linewidth]{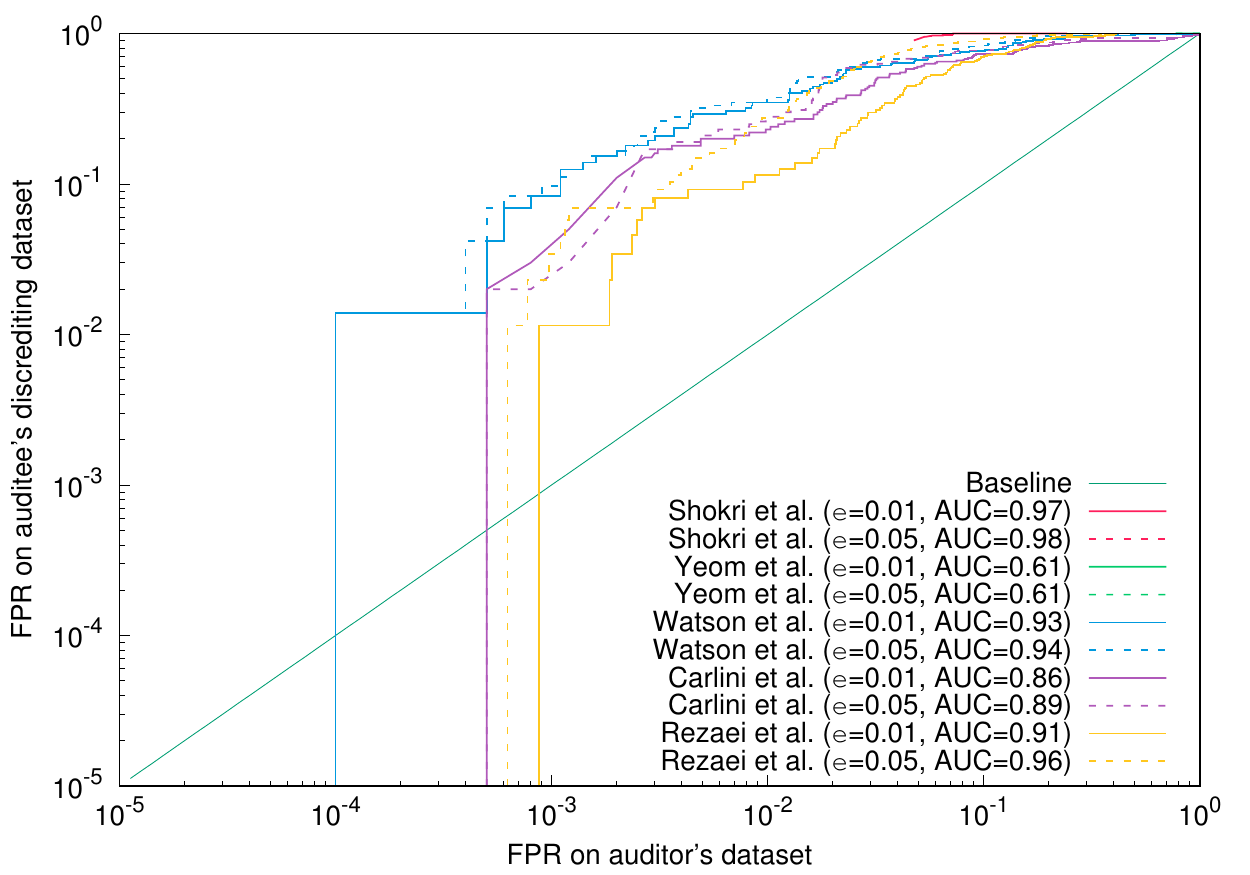}}\\
\end{tabular}
\caption{MNIST/MLP model. On adversarially tuned samples.}
\label{fig-fpr-fpr-mnist-mlp-adv}
\end{figure*}

\begin{figure*}
\centering
\centering
\begin{tabular}{cc}
\subfloat[FPR/FPR plot]{\includegraphics[width=0.3\linewidth]{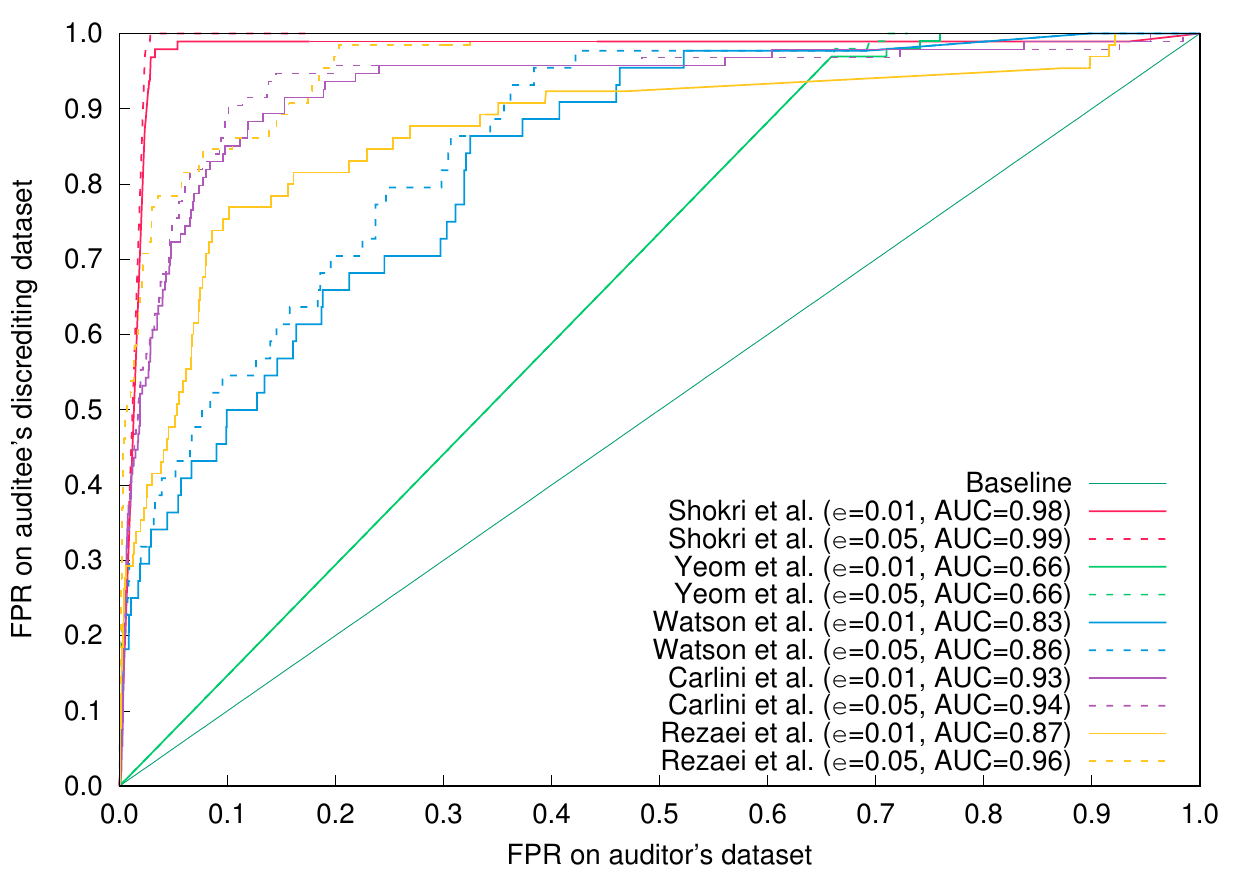}} 
& \subfloat[FPR/FPR logscale plot]{\includegraphics[width=0.3\linewidth]{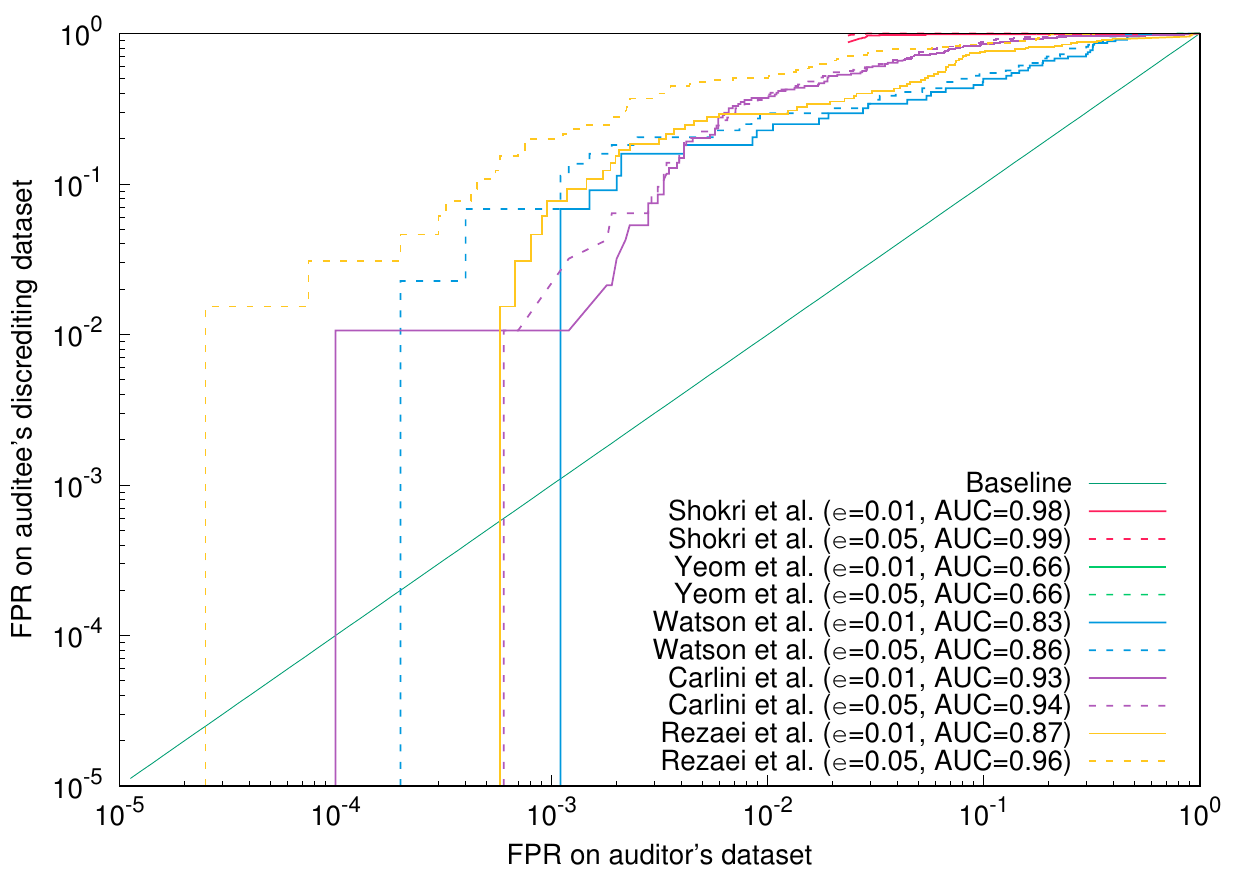}}\\
\end{tabular}
\caption{fMNIST/MLP model. On adversarially tuned samples.}
\label{fig-fpr-fpr-fmnist-mlp-adv}
\end{figure*}

\begin{figure}
\centering
\begin{tabular}{c}
\subfloat[FPR/FPR logscale plot]{\includegraphics[width=0.6\linewidth]{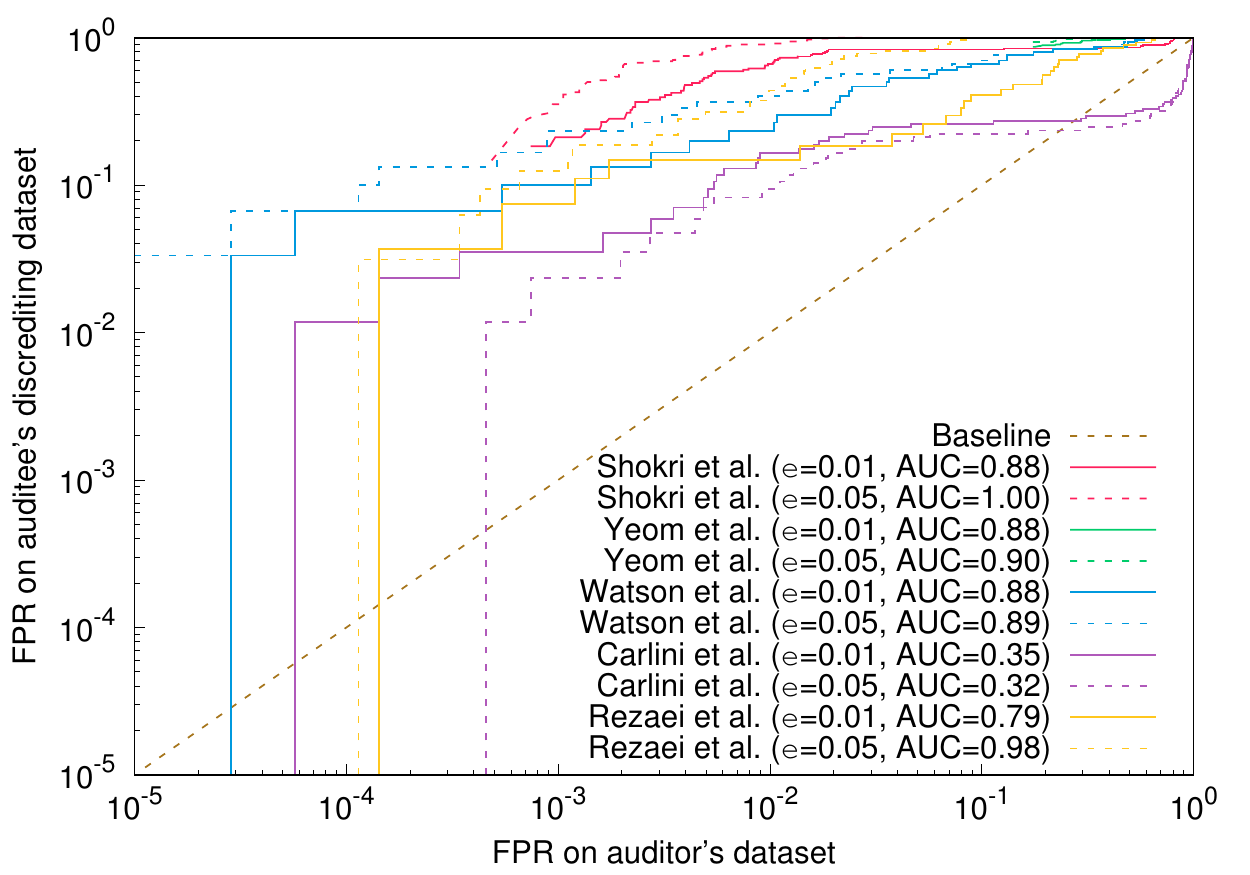}}\\
\end{tabular}
\caption{Cifar10/ResNet20 model. On adversarially tuned samples.}
\label{fig-fpr-fpr-c10-renset-adv}
\end{figure}

\begin{figure}
\centering
\centering
\begin{tabular}{c}
\subfloat[FPR/FPR logscale plot]{\includegraphics[width=0.6\linewidth]{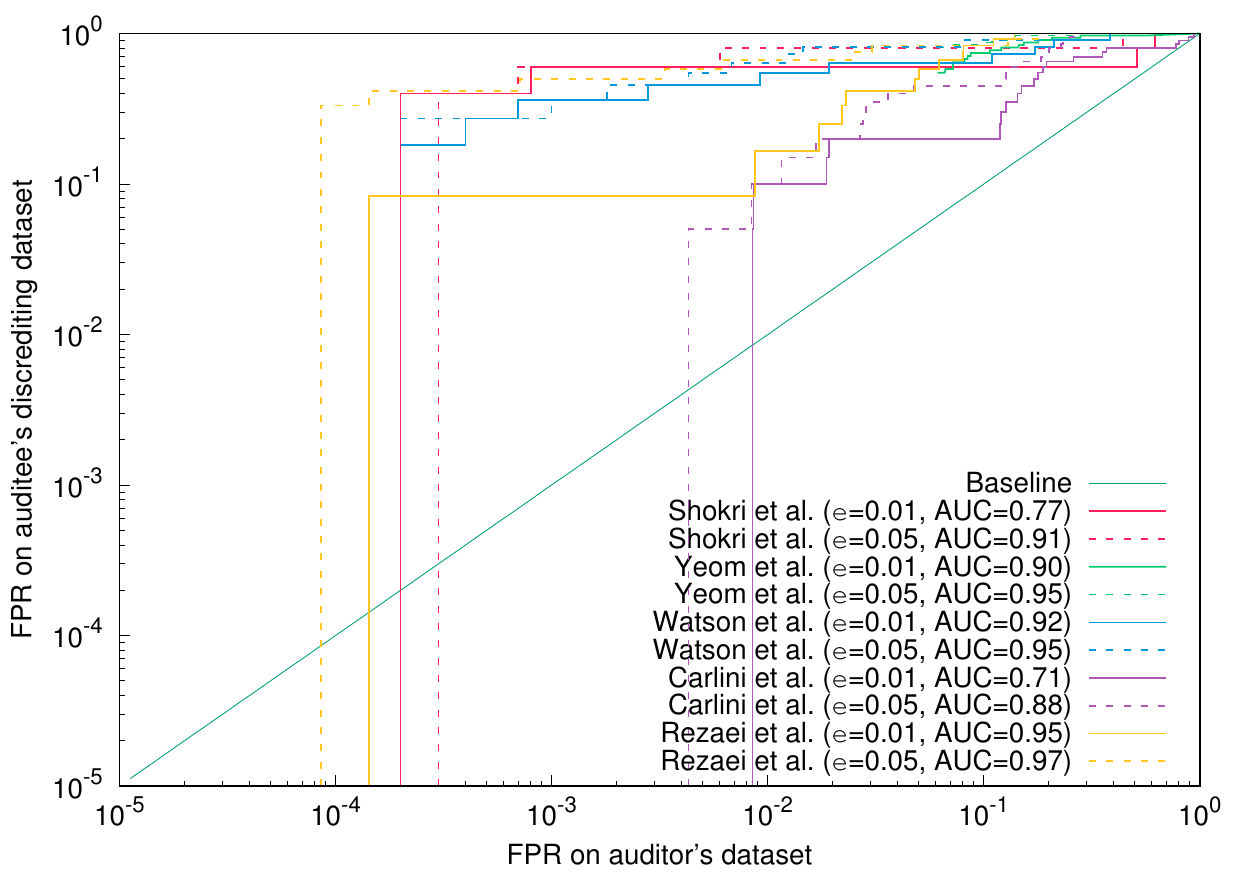}}\\
\end{tabular}
\caption{Cifar100/LeNet model. On adversarially tuned samples.}
\label{fig-fpr-fpr-c100-lenet-adv}
\end{figure}

\begin{figure}
\centering
\centering
\begin{tabular}{c}
\subfloat[FPR/FPR logscale plot]{\includegraphics[width=0.6\linewidth]{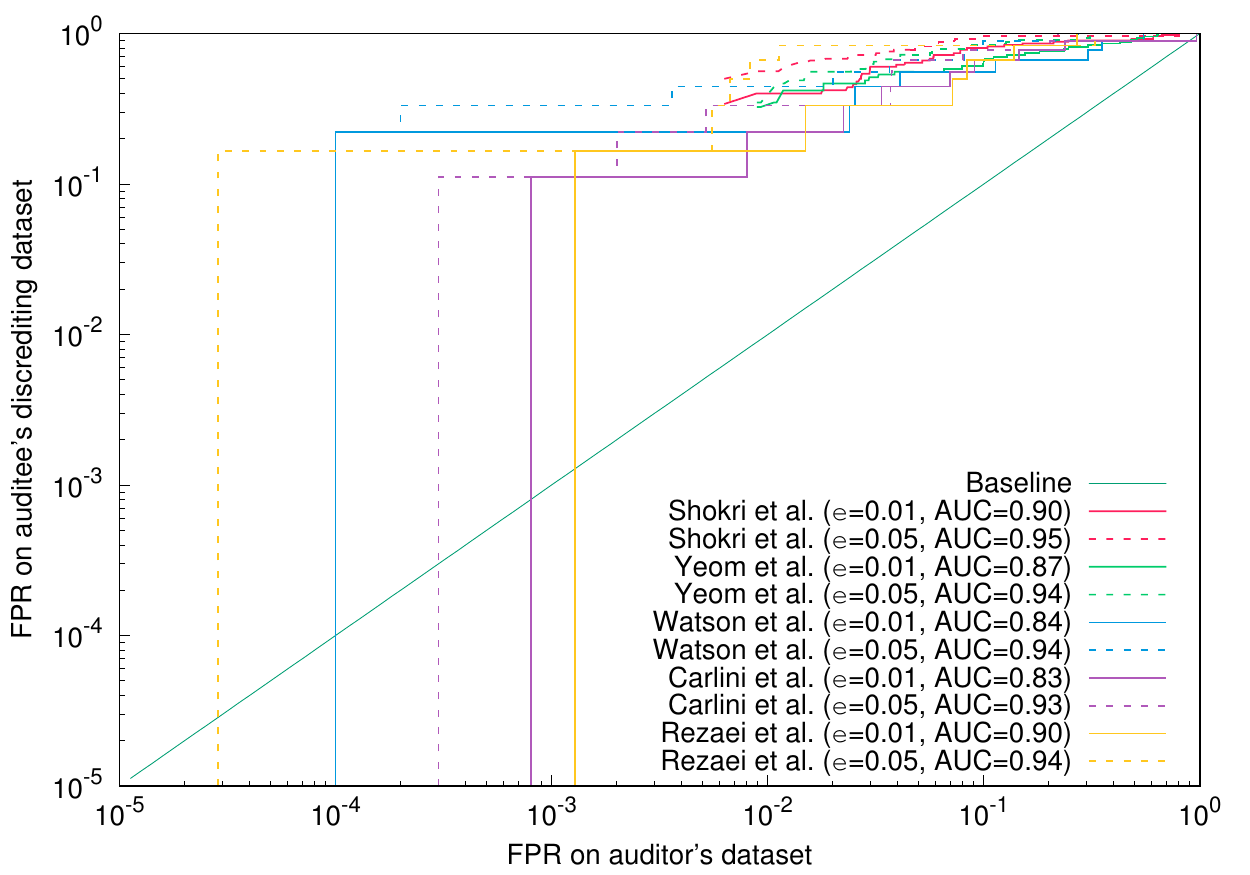}}\\
\end{tabular}
\caption{Cifar100/ResNet20 model. On adversarially tuned samples.}
\label{fig-fpr-fpr-c100-renset-adv}
\end{figure}

\begin{figure}
\centering
\centering
\begin{tabular}{c}
\subfloat[FPR/FPR logscale plot]{\includegraphics[width=0.6\linewidth]{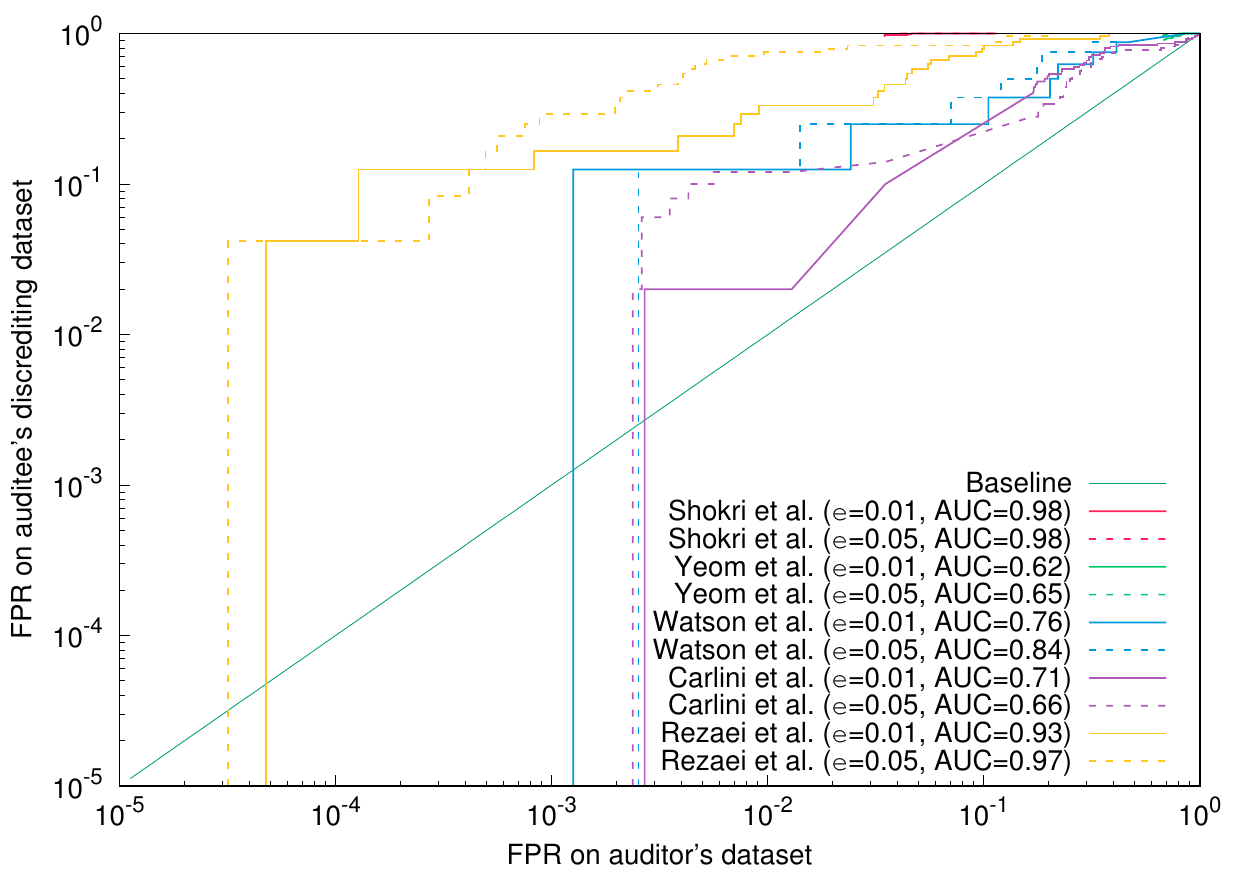}}\\
\end{tabular}
\caption{SVHN/LeNet model. On adversarially tuned samples.}
\label{fig-fpr-fpr-svhn-lenet-adv}
\end{figure}

\begin{figure}
\centering
\centering
\begin{tabular}{c}
\subfloat[FPR/FPR logscale plot]{\includegraphics[width=0.6\linewidth]{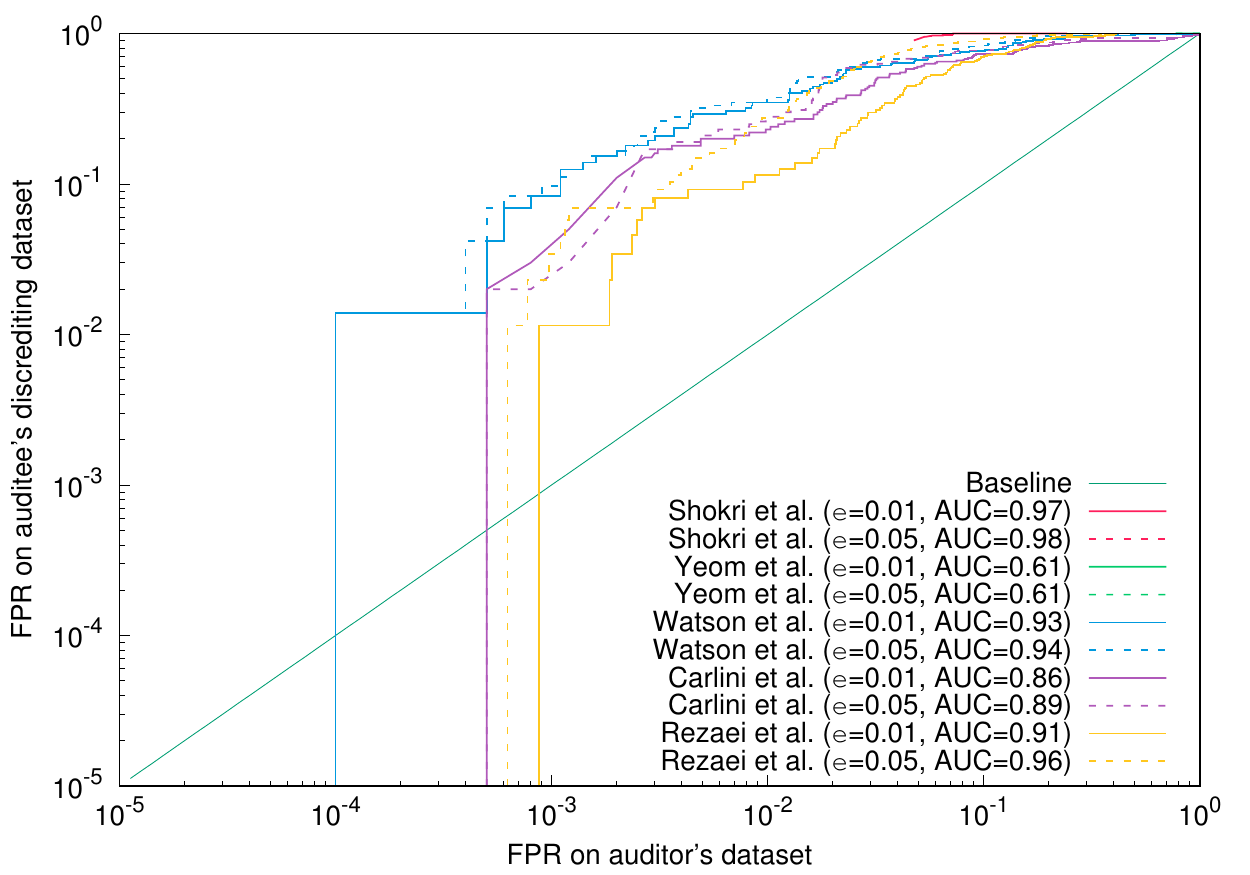}}\\
\end{tabular}
\caption{MNIST/MLP model. On adversarially tuned samples.}
\label{fig-fpr-fpr-mnist-mlp-adv}
\end{figure}

\begin{figure}
\centering
\begin{tabular}{c}
\subfloat[FPR/FPR logscale plot]{\includegraphics[width=0.6\linewidth]{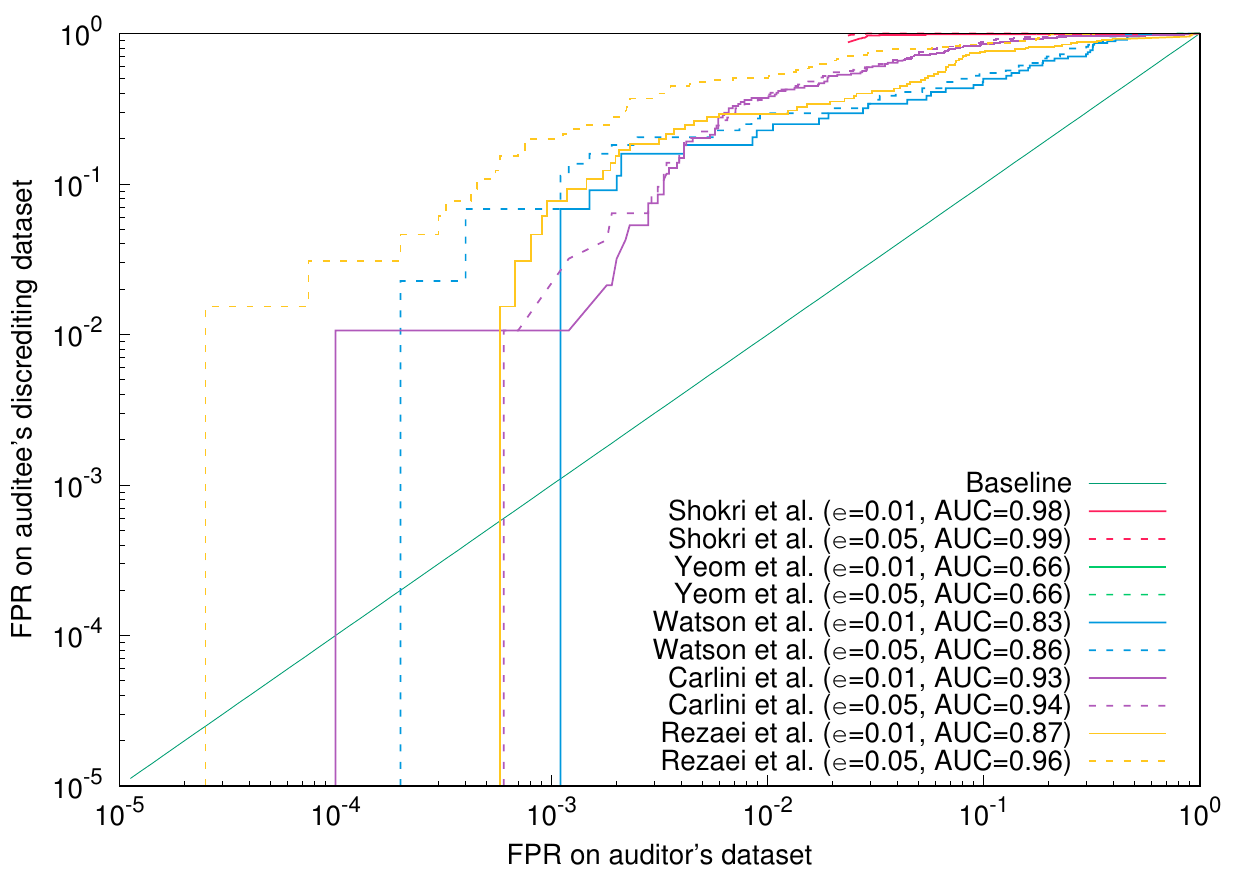}}\\
\end{tabular}
\caption{fMNIST/MLP model. On adversarially tuned samples.}
\label{fig-fpr-fpr-fmnist-mlp-adv}
\end{figure}

\subsection{Could Increase in FP be a Repercussion of Domain Shift?}

\begin{figure*}[h]
\centering
\centering
\begin{tabular}{ccc}
\subfloat[FPR/FPR plot]{\includegraphics[width=0.4\linewidth]{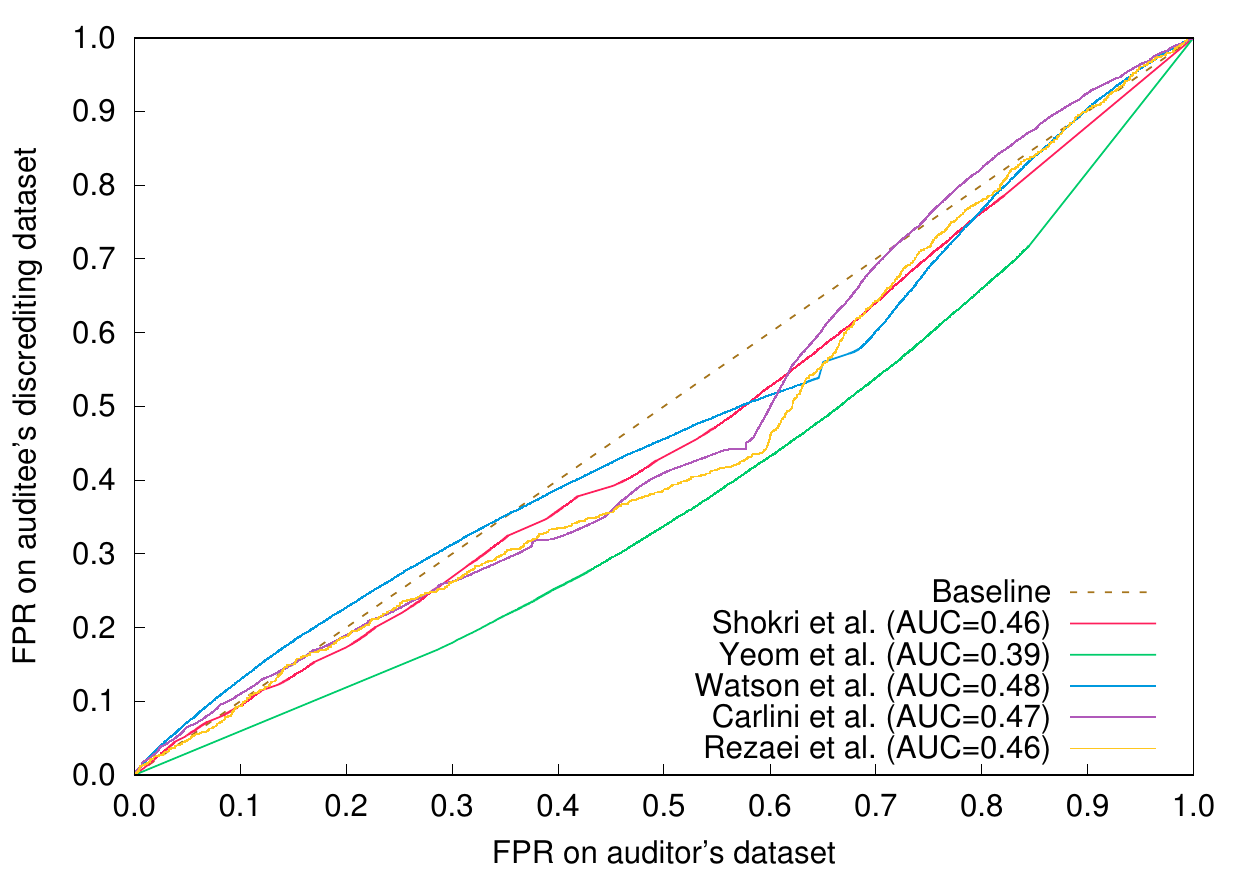}} 
& \subfloat[FPR/FPR logscale plot]{\includegraphics[width=0.4\linewidth]{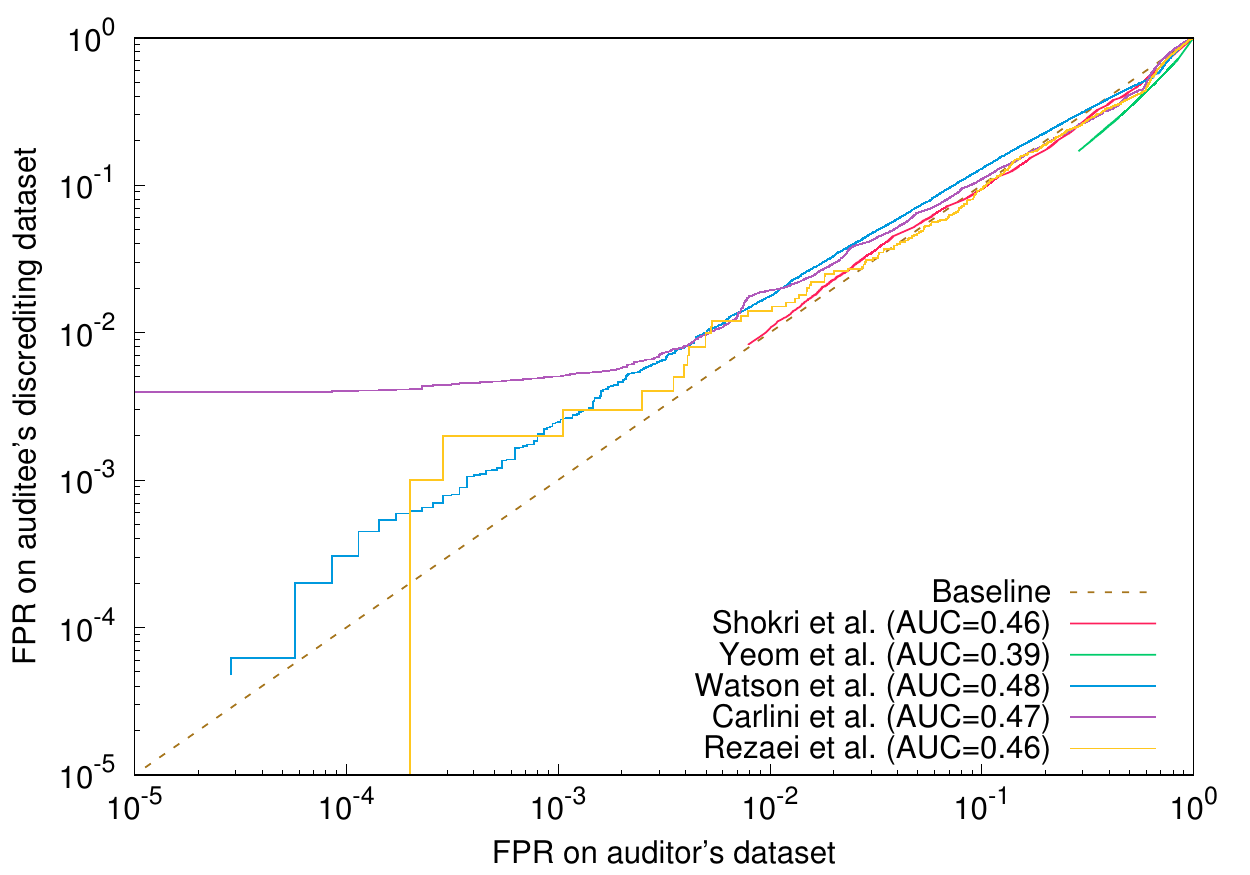}}\\
\end{tabular}
\caption{CIFAR-10/LeNet model. The effect of distribution shift. Here, instead of using discrediting algorithms, we use entire CINIC dataset as the discrediting dataset without filtering out any sample.}
\label{fig-fpr-fpr-c10-lenet-dist-shift}
\end{figure*}

A natural question upon the success of the three algorithms to significantly increase the false positive rate is if a domain shift across datasets are the real culprit. In other words, one may suspect that using the entire CINIC dataset as a discrediting dataset may achieve the same goal as the proposed algorithms because the MI attacks are vulnerable to domain shift. 

To refute the hypothesis, we illustrate the false positive to false positive plot in Figure \ref{fig-fpr-fpr-c10-lenet-dist-shift} for a LeNet model trained on CIFAR-10. Here, the auditor dataset is the test portion of the CIFAR-10 dataset. The MI attack models that require dataset for training use the unused portion of the training set of the CIFAR-10 dataset. The auditee's discrediting dataset is the entire CINIC dataset. Due to the huge computational complexity of training individual models containing each sample in CINIC dataset separately, here, we use the offline version of the Carlini attack \cite{carlini2022membership}.

Interestingly, it is clear that the domain shift works in favor of the auditor by slightly decreasing the false positive. The reason is that the auditee's model trained on CIFAR-10 is naturally less confident on samples from another distribution. Unless carefully picked by an algorithm, such as Algorithm \ref{algo-search}, the confidence output of the model is lower on average and, hence, less likely to be incorrectly labeled as member (positive). Therefore, discrediting process cannot be simply reduced to finding a dataset with different distributions.

\begin{figure*}
\centering
\centering
\begin{tabular}{cc}
\subfloat[FPR/FPR plot]{\includegraphics[width=0.3\linewidth]{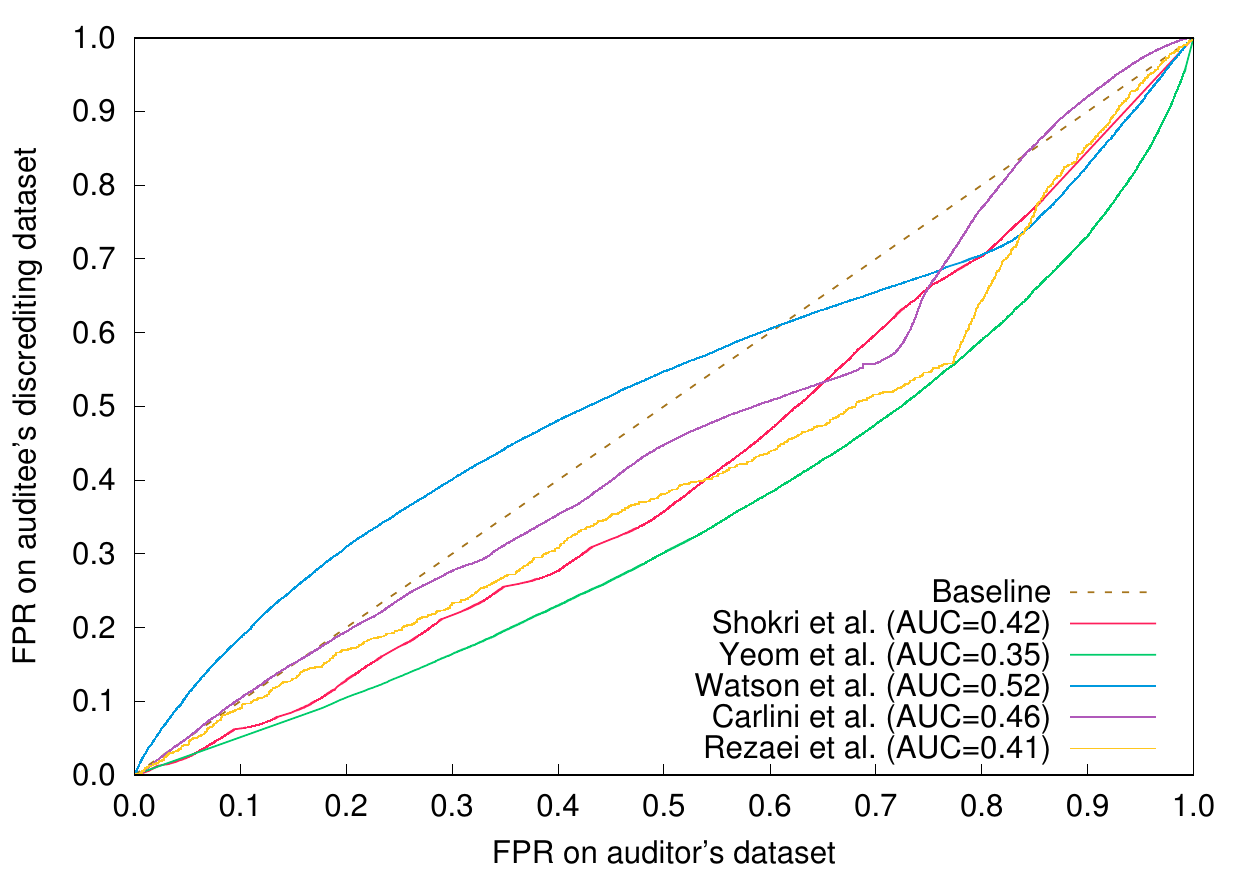}} 
& \subfloat[FPR/FPR logscale plot]{\includegraphics[width=0.3\linewidth]{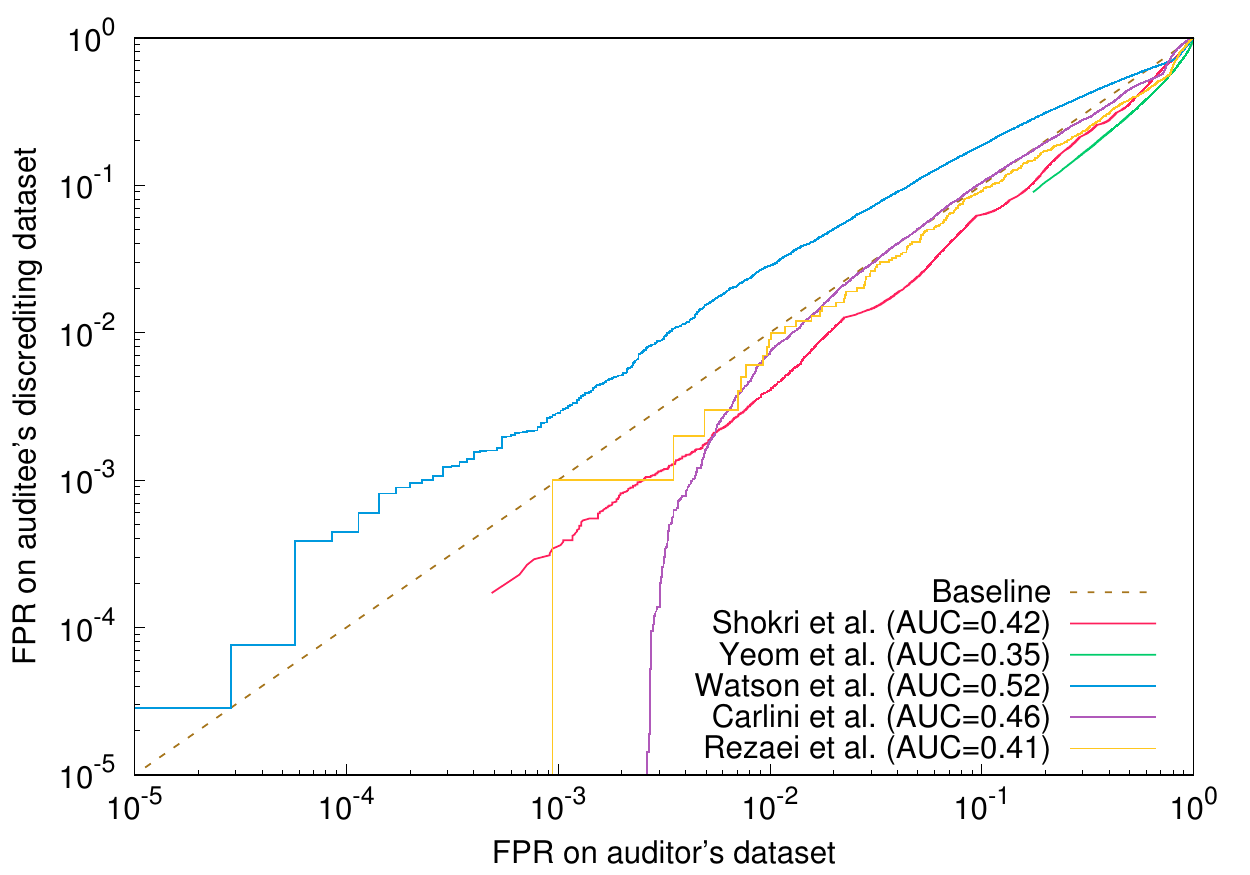}}\\
\end{tabular}
\caption{CIFAR-10/ResNet model. Here, the discrediting dataset is the entire CINIC dataset.}
\label{fig-fpr-fpr-c10-resnet-dist-shift}
\end{figure*}

\begin{figure*}
\centering
\centering
\begin{tabular}{cc}
\subfloat[FPR/FPR plot]{\includegraphics[width=0.3\linewidth]{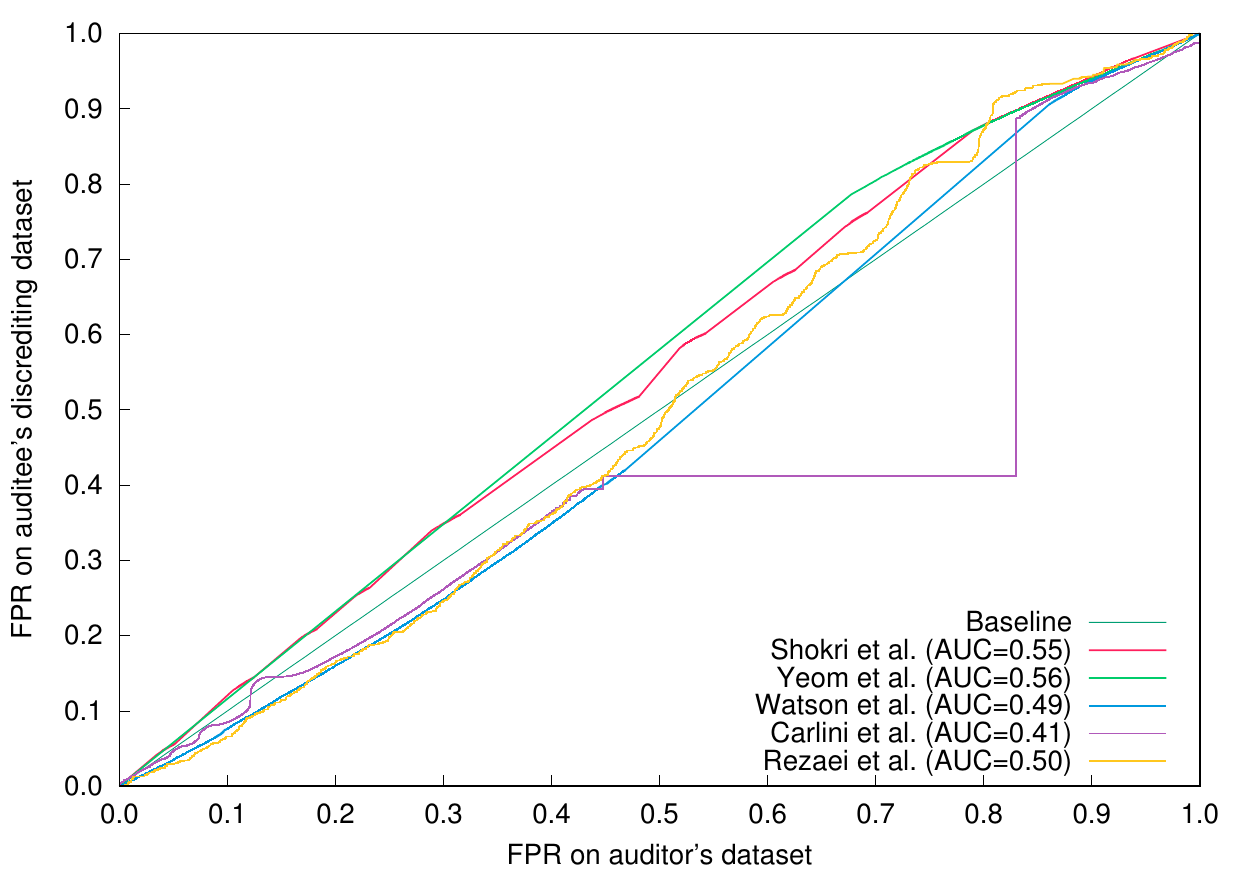}} 
& \subfloat[FPR/FPR logscale plot]{\includegraphics[width=0.3\linewidth]{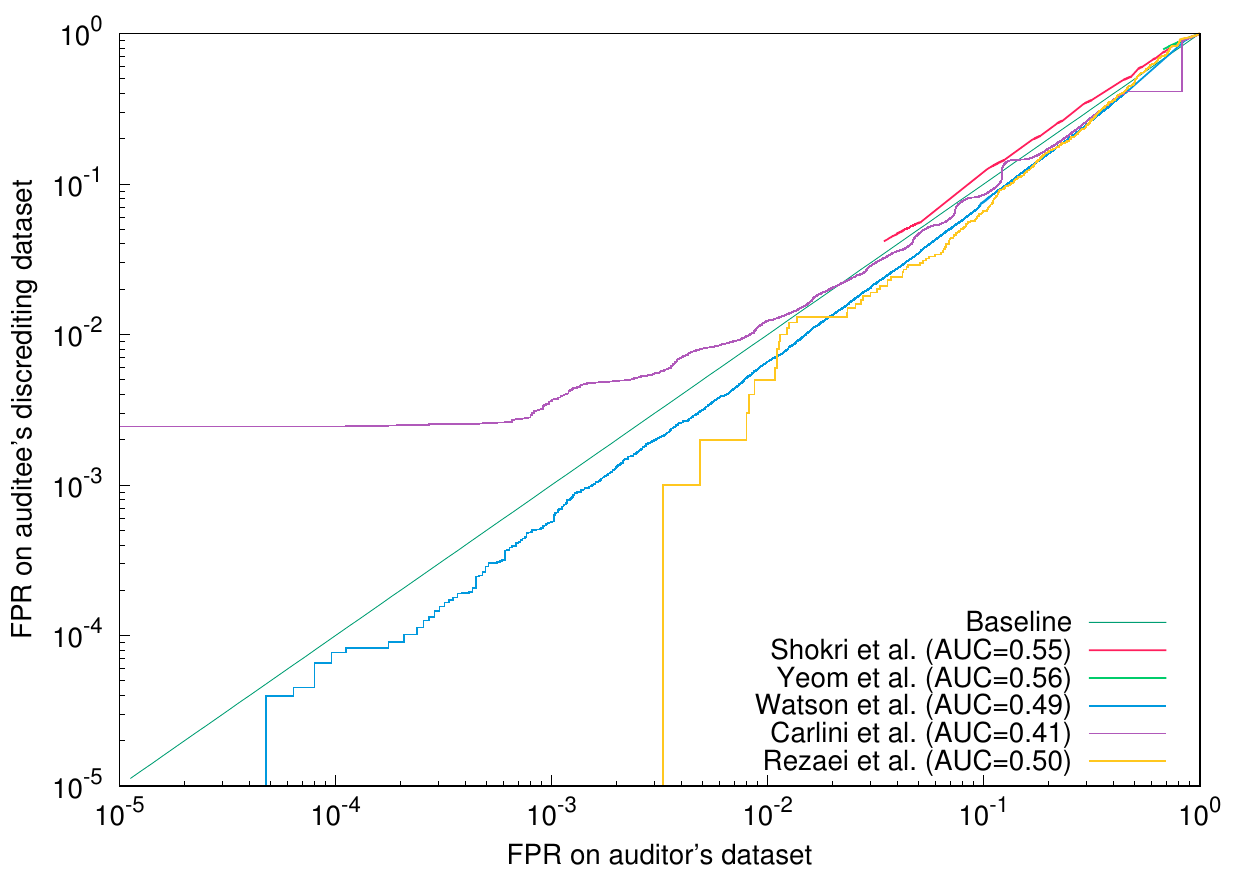}}\\
\end{tabular}
\caption{SVHN/LeNet model. Here, the discrediting dataset is the entire extra portion of the SVHN dataset.}
\label{c7-fig-fpr-fpr-svhn-lenet-dist-shift}
\end{figure*}

Figure \ref{fig-fpr-fpr-c10-resnet-dist-shift} and \ref{c7-fig-fpr-fpr-svhn-lenet-dist-shift} shows the false positive ratio to false positive ratio in case of domain shift on ResNet and LeNet, trained on CIFAR10 and SVHN, respectively.

\end{document}